\documentclass[11pt]{article}
\usepackage{color}		
\usepackage[dvips]{graphicx}	
\usepackage{amsbsy}		
\usepackage{amsfonts}		
\usepackage{amsmath}		
\usepackage{amssymb}		
\usepackage{upgreek}		
\usepackage{wasysym}
\usepackage{multirow}
\usepackage{mciteplus}
\usepackage{psfrag}
\usepackage{url}
\usepackage{longtable}
\usepackage{rotate}
\usepackage{cite}
\usepackage{soul}
\usepackage{afterpage}
\usepackage{dsfont}
\usepackage{mathrsfs}
\usepackage[percent]{overpic}

\include{paperdef}
\graphicspath{{finalfigs/}}

\DeclareMathOperator\sgn{sgn}
\newenvironment{Eqnarray}%
     {\arraycolsep 0.14em\begin{eqnarray}}{\end{eqnarray}}
\def\beq{\begin{equation}}
\def\eeq{\end{equation}}
\def\beqa{\begin{Eqnarray}}
\def\eeqa{\end{Eqnarray}}

\def\Ref#1{ref.~\cite{#1}}

\def\eq#1{Eq.~(\ref{#1})}

\def\eqs#1#2{Eqs.~(\ref{#1}) and (\ref{#2})}
\def\eqss#1#2#3{Eqs.~(\ref{#1}), (\ref{#2}) and (\ref{#3})}
\def\eqst#1#2{Eqs.~(\ref{#1})--(\ref{#2})}

\def\vev#1{\langle #1 \rangle}
\def\sb{s_\beta}
\def\cb{c_\beta}
\def\sbt{s^2_\beta}
\def\cbt{c^2_\beta}
\def\sbma{s_{\beta-\alpha}}
\def\cbma{c_{\beta-\alpha}}
\def\ssqbma{s^2_{\beta-\alpha}}
\def\csqbma{c^2_{\beta-\alpha}}
\def\cba{\cos(\beta-\alpha)}
\def\sba{\sin(\beta-\alpha)}

\def\ur{U_R}
\def\dr{D_R}
\def\anti{\overline}

\def\hl{h}
\def\ha{A}
\def\hh{H}
\def\hpm{{H^\pm}}
\def\mha{m_{\ha}}
\def\mhl{m_{\hl}}
\def\mhh{m_{\hh}}
\def\mhpm{m_{\hpm}}
\def\phm{\phantom{-}}
\def\ifmath#1{\relax\ifmmode #1\else $#1$\fi}
\def\ls#1{\ifmath{_{\lower1.5pt\hbox{$\scriptstyle #1$}}}}
\makeatletter

\newcommand{\Rmnum}[1]{\expandafter\@slowromancap\romannumeral #1@}
\makeatother

\newcommand{\HB}{{\tt HiggsBounds}}
\newcommand{\HS}{{\tt HiggsSignals}}

\newcommand{\thdmc}{{\tt 2HDMC}}
\newcommand{\HH}{{hybrid}}

\newcommand{\GeV}{\;\mathrm{GeV}}
\newcommand{\TeV}{\;\mathrm{TeV}}

\newcommand{\half}{\tfrac{1}{2}}
\newcommand{\nn}{\nonumber}
\def\thefootnote{\fnsymbol{footnote}}
\begin{document}
\begin{flushright}
\normalsize{
SCIPP-15/10 \\
July, 2015
}
\end{flushright}
\vspace{1cm}

\begin{center}
\Large\bf\boldmath
New LHC Benchmarks for the \cp-conserving\\
 Two-Higgs-Doublet Model
\unboldmath
\end{center}
\vspace{0.4cm}
\begin{center}
Howard E.~Haber$^{1,}$\footnote{Electronic address: haber@scipp.ucsc.edu} and
Oscar St{\aa}l$^{2,}$\footnote{Electronic address: oscar.stal@fysik.su.se}\\
\vspace{0.4cm}
 {\sl $^1$ Santa Cruz Institute for Particle Physics \\
 University of California, Santa Cruz, CA 95064 USA}\\[0.2cm]
 {\sl $^2$ The Oskar Klein Centre, Department of Physics\\
  Stockholm University, SE-106 91 Stockholm, Sweden}\\[0.2cm]
\end{center}
\vspace{0.2cm}

\renewcommand{\thefootnote}{\arabic{footnote}}
\setcounter{footnote}{0}

\begin{abstract}
We introduce a strategy to study the parameter space 
of the general, $\cp$-conserving, two-Higgs-doublet Model (2HDM) with a softly broken $\mathbb{Z}_2$-symmetry by means of a  new ``hybrid'' basis. In this basis the input parameters are the measured values of the mass of the observed Standard Model (SM)-like Higgs boson and its coupling strength to vector boson pairs, the mass of the second $\cp$-even Higgs boson, the ratio of neutral Higgs vacuum expectation values, and three additional dimensionless parameters. Using the hybrid basis, we present numerical scans of the 2HDM parameter space where we survey available parameter regions and analyze model constraints. From these results, we define a number of benchmark scenarios that capture different aspects of non-standard Higgs phenomenology that are of interest for future LHC Higgs searches.

\end{abstract}
\newpage


\section{Introduction}

The observation in 2012 \cite{ATLASDiscovery, CMSDiscovery} of a new boson with a mass close to 125 GeV \cite{Aad:2015zhl} has been widely viewed as the discovery of the long sought after Higgs boson \cite{Higgs:1964pj,Englert:1964et,Higgs:1964ia}.  Despite the limited data set, the detailed analyses of the Run 1 LHC data by the ATLAS and CMS Collaborations have confirmed that the observed signal strengths (cross section times branching ratio) of the Higgs boson candidate relative to that expected of the Standard Model (SM) Higgs boson are consistent with SM predictions to within the accuracy of the current measurements \cite{Aad:2013wqa,Khachatryan:2014jba}.  Nevertheless, the current precision is at best $20\%$ in the bosonic channels and considerably less accurate in the fermionic channels.  Thus, there is still considerable room for new physics beyond the SM to contribute to the properties of the already discovered scalar state.

The Standard Model posits that the dynamics of one complex hypercharge-one, weak doublet of scalar fields is solely responsible for electroweak symmetry breaking.  Three of the four degrees of freedom of this scalar doublet provide the longitudinal modes of the massive $W^\pm$ and $Z$ gauge bosons.  The remaining scalar degree of freedom is the SM Higgs boson.  But, why should the scalar sector of the SM be of minimal form?  The spin-1/2 quark and lepton degrees of freedom of the SM appear in three generations.  The origin of the non-trivial flavor structure of the SM is presently unknown.  By analogy, one might expect a replication in the scalar sector as well---a non-minimal Higgs sector consisting of multiple doublets.  Adding additional doublets yields new phenomena in the scalar sector---charged Higgs scalars, neutral Higgs scalars of opposite \cp\ quantum numbers (if \cp\ is conserved by the scalar potential) or neutral Higgs of indefinite \cp\ (if \cp\ is violated by the scalar potential).  The current Higgs data do not rule out the existence of an extended Higgs sector which, as we will discuss in some detail, could be realized at mass scales close to (or in the extreme case even below) $125$~GeV.  Thus, apart from any other theoretical motivation, it behooves us to devote a dedicated program at the LHC to search for evidence of a non-minimal Higgs structure.

Of course, theoretical arguments have been also advanced in support of a non-minimal Higgs sector.  Perhaps the most persuasive is based on the fact
that the SM is unnatural~\cite{Weisskopf:1939zz,Altarelli:2013lla}--namely, it is difficult to understand how the scale of electroweak symmetry breaking ($v=246$~GeV) arises in a more fundamental framework that includes gravity and its associated Planck scale, $M_{\rm PL}\simeq 10^{19}$~GeV.  In the context of the SM coupled to gravity, electroweak symmetry is achieved only by fine-tuning the squared-mass parameter of the scalar potential to an accuracy of 34 decimal places.  New physics beyond the SM that attempts to address this issue introduces new phenomena that enters at or near the TeV scale.  Many such approaches invoke non-minimal Higgs sectors.  The most well studied example of this kind is the minimal supersymmetric extension of the SM (MSSM), which employs a two-Higgs doublet scalar sector with quartic terms in the scalar potential that respect 
supersymmetry (SUSY)~\cite{Fayet:1974pd,Inoue:1982ej,Flores:1982pr,Gunion:1984yn}.

Since we do not presently know the precise nature of the new physics beyond the SM, it is prudent to be open minded about the possible structure of the non-minimal Higgs sector.  However, the observed value of the electroweak parameter $\rho=m_W^2/(m_Z^2\cos^\theta_W)\simeq 1$ strongly favors non-minimal Higgs sectors comprised only of singlet and doublet scalar fields~\cite{Ross:1975fq,Gunion:1989we}.\footnote{Including triplet scalar fields for example typically violates $\rho\simeq 1$~\cite{Tsao:1980em,Gunion:1989we} except in special cases that must be considered as fine-tuned~\cite{Gunion:1990dt}.}  Among theories in this category, we find the two-Higgs doublet model~\!\footnote{For a comprehensive review of the 2HDM, see \cite{Branco:2011iw}.  A review that treats the 2HDM in the formalism employed in this paper can be found in \cite{Asner:2013psa}.}  
(2HDM) particularly attractive as the minimal extension that yields new scalar phenomena, including new charged states ($H^\pm$) and neutral states with different (or mixed) \cp\ properties. Direct searches for the additional 2HDM Higgs states based on run-I LHC data have appeared both from ATLAS \cite{ATLAS2013027} and CMS \cite{Khachatryan:2014jya,Khachatryan:2016are}.  Interpretations of the recently discovered Higgs boson at 125 GeV in the context of the 2HDM and implications for future LHC searches have been presented in \cite{Craig:2013hca,Eberhardt:2013uba,Coleppa:2013dya,Chang:2013ona,Cheung:2013rva,Baglio:2014nea,Dumont:2014wha,Dumont:2014kna,Chowdhury:2015yja,Craig:2015jba,Bernon:2015qea}.

In this paper, we develop a set of benchmark scenarios for LHC Higgs searches that capture different aspects of 2HDM phenomenology.
Our scenarios are devised taking into account that the observed Higgs boson already possess properties that are close to those expected in the SM.
The experimental data already provide important constraints on the 2HDM framework.  For example, the absence of observable flavor changing neutral currents (FCNCs) mediated by tree-level exchange of neutral Higgs bosons requires that the structure of the Higgs-fermion Yukawa couplings must be of a special form~\cite{Glashow:1976nt,Paschos:1976ay}.  The simplest way to eliminate tree-level Higgs-mediated FCNCs is to impose a particular  discrete $\mathbb{Z}_2$ symmetry [cf.~Table~\ref{Tab:type} in Section~\ref{sect:Yukawa}] that is exactly respected by the dimension-four quartic terms of the Higgs 
Lagrangian~\cite{Donoghue:1978cj,Haber:1978jt}.\footnote{Dimension-two quadratic terms in the Higgs Lagrangian that softly break the $\mathbb{Z}_2$ discrete symmetry
are allowed since these terms will only generate Higgs-mediated FCNCs at the loop level, which typically are not large enough to be in conflict with experimental data.}
We make one further assumption in our analysis by imposing \cp\ symmetry on the scalar potential.  This assumption is not required based on data (even though a \emph{pure} \cp-odd state has been ruled out as explanation for the 125 GeV state \cite{Khachatryan:2014kca}).  The restriction to \cp-symmetry provides an additional simplification to the analysis.   In particular, the neutral scalar spectrum consists of two \cp-even states $h$ and $H$ with $m_h<m_H$, and the \cp-odd state $A$.
Indeed, the methods developed in this paper are rather easily generalized to a 2HDM
with a \cp-violating scalar potential or \cp-violating vacuum. This possibility will be addressed elsewhere.

In Section 2, we provide the theoretical background relevant to the 2HDM.  We introduce the Higgs basis~\cite{Branco:1999fs,Davidson:2005cw}, which is especially useful in our analysis, as it provides a very clear way of parametrizing the terms that yield deviations of the properties of the observed Higgs boson from those of the SM.  
Ultimately, we construct a ``hybrid basis'' of parameters, which we will employ in 2HDM parameter scans.   These parameters include the two \cp-even neutral Higgs boson masses ($m_h$ and $m_H$, where one of these masses is identified with the mass of the observed Higgs boson), 
the parameter $\cbma$ which parametrizes the deviation of the lightest \cp-even neutral Higgs boson $h$ from that of the SM, the ratio of neutral Higgs vacuum expectation values, $\tan\beta$, and three dimensionless quartic couplings of the Higgs basis.  Two of these three dimensionless couplings are related in a simple way to the masses of the charged Higgs and neutral \cp-odd Higgs boson of the 2HDM.

In Section 3, we present our numerical scans of the 2HDM parameter space using the hybrid basis of parameters.  Based on these scans, which are performed using the code \thdmc\ \cite{Eriksson:2009ws,Eriksson:2010zzb}, we develop seven different benchmark scenarios: (1) a SM-like $h$, with $\cbma$ small but non-zero as allowed by current data to yield interesting phenomenology of the heavier~$H$; (2)~a ``flipped'' scenario with a SM-like $H$ (which implies that $|\cbma|$ is near 1); (3)~overlapping \cp-even and \cp-odd scalars with masses around 125 GeV; (4) a SM-like $h$ and a heavy non-SM-Higgs mass spectrum with short cascade decays $H\to ZA$ or $H\to W^\pm H^\mp $; (5) a SM-like $h$ and a heavy non-SM Higgs mass spectrum with long cascade decays $H^\pm\to W^\pm A \to W^\pm ZH$ or $A\to W^\pm H^\mp \to W^\pm W^\mp H$; 
(6)~SM-like $hVV$ and $ht\bar t$ couplings ($V=W^\pm$ and $Z$), but with opposite sign $hb\bar b$ and $h\tau^+\tau^-$ couplings as compared with the SM; and (7) an MSSM-like scenario, in which the scalar potential parameters (with one exception) are fixed by the corresponding tree-level MSSM relations.  Finally, in Section~4, we present our conclusions.  The proposed benchmark scenarios are summarized in a set of tables presented in Appendix A.


\section{Theoretical background}

\subsection{The general two-Higgs doublet model (2HDM)}

The two-Higgs doublet model (2HDM) consists of two complex SU(2) doublet, hypercharge-one fields, $\Phi_1$ and $\Phi_2$, and an SU(2)$\times$U(1)-invariant scalar potential,
\beqa
\mathcal{V}&=& m_{11}^2 \Phi_1^\dagger \Phi_1+ m_{22}^2 \Phi_2^\dagger \Phi_2 -[m_{12}^2
\Phi_1^\dagger \Phi_2+{\rm h.c.}]
+\half \lambda_1(\Phi_1^\dagger \Phi_1)^2+\half \lambda_2(\Phi_2^\dagger \Phi_2)^2
+\lambda_3(\Phi_1^\dagger \Phi_1)(\Phi_2^\dagger \Phi_2)
\nn\\
&&\quad
+\lambda_4( \Phi_1^\dagger \Phi_2)(\Phi_2^\dagger \Phi_1)
 +\left\{\half \lambda_5 (\Phi_1^\dagger \Phi_2)^2 +\big[\lambda_6 (\Phi_1^\dagger
\Phi_1) +\lambda_7 (\Phi_2^\dagger \Phi_2)\big] \Phi_1^\dagger \Phi_2+{\rm
h.c.}\right\}\,,\label{genpot}
\eeqa
where the parameters $m_{12}^2$, $\lambda_5$, $\lambda_6$ and $\lambda_7$ are potentially complex.  All other scalar potential parameters are manifestly real.  We assume that
the parameters are chosen such that the minimum of the scalar potential spontaneously
breaks the SU(2)$\times$U(1) electroweak gauge symmetry to U(1)$_{\rm EM}$~\cite{Maniatis:2006fs,Ivanov:2006yq}.  That is,
at the minimum of the scalar potential, the neutral components of the complex doublet
scalar fields acquire vacuum expectation values (vevs),
\beq \label{phivevs}
\langle\Phi_i^0\rangle=\frac{v_i}{\sqrt{2}}\, e^{i\xi_i}\,,\qquad (i=1,2),
\eeq
where by convention $v_1$ and $v_2$ are real and non-negative. The combination $v^2\equiv v_1^2+v_2^2\simeq (246~{\rm GeV})^2$ is fixed by its relation to the Fermi constant and the 
$W$ boson mass, $v^2=1/(\sqrt{2}G_F)=4\,m_W^2/g^2$.

In the most general 2HDM, the fields $\Phi_1$ and $\Phi_2$ are indistinguishable.
Thus, it is always possible to define two orthonormal linear combinations of the two
scalar doublet fields without modifying any prediction of the model.  Performing such a
redefinition of fields (henceforth called \textit{a change of basis of the scalar doublet fields})
leads to a new scalar potential with the same form as \eq{genpot}
but with modified coefficients.  In this paper we shall focus on the case where the scalar potential and the vacuum state are \cp-conserving, leaving the more general case for future work.

The scalar potential is \textit{explicitly} \cp-conserving if and only if there exists a basis choice
for the scalar fields in which $m_{12}^2$, $\lambda_5$, $\lambda_6$ and $\lambda_7$ are simultaneously real.
The general conditions that guarantee the existence of such a basis (called a \textit{real basis}) were developed in \cite{Gunion:2005ja}.  Henceforth, we shall assume that all scalar potential parameters given in \eq{genpot} are real.   However, it is still possible that
the vacuum spontaneously breaks \cp.  In particular, spontaneous \cp-violation takes place
if and only if the scalar potential is explicitly \cp-conserving, but no real basis exists in which the scalar vacuum expectation values are simultaneously real.  Sufficient conditions for a \cp-conserving vacuum are easily obtained.   The minimization of the scalar potential fixes $v_1\equiv v\cos\beta$,
$v_2\equiv v\sin\beta$ (where $0\leq\beta\leq\half\pi$) and the relative phase of the two vevs, $\xi\equiv\xi_2-\xi_1$ through the equations,
\beqa
\!\!\!\!\!\!\!\! m_{11}^2&=& m_{12}^2\tan\beta\cos\xi-\half v^2\bigl[\lambda_1\cbt+(\lambda_{3}+\lambda_4+\lambda_5\cos 2\xi)\sbt
+3\lambda_6\sb\cb\cos\xi+\lambda_7\sbt\tan\beta\cos\xi\bigr],\label{m11}\\
\!\!\!\!\!\!\!\! m_{22}^2&=& m_{12}^2\cot\beta\cos\xi-\half v^2\bigl[\lambda_2\sbt+(\lambda_{3}+\lambda_4+\lambda_5\cos 2\xi)\cbt
+\lambda_6\cbt\cot\beta\cos\xi+3\lambda_7\sb\cb\cos\xi\bigr],\label{m22} \\
\!\!\!\!\!\!\!\! m_{12}^2\sin\xi&=&\half v^2\bigl[2\lambda_5\sb\cb\cos\xi+\lambda_6\cbt+\lambda_7\sbt\bigr]\sin\xi\,,
\label{m12}
\eeqa
where $\sb\equiv\sin\beta$ and $\cb\equiv\cos\beta$ and all scalar potential parameters are real by assumption.  If $\sin\xi\neq 0$, then \eq{m12} can be used to obtain
\beq \label{cosxi}
\cos\xi=\frac{m_{12}^2-\half \lambda_6 v_1^2-\half\lambda_7 v_2^2}{\lambda_5 v_1 v_2}\,.
\eeq
Moreover, this value of $\xi$ corresponds to the minimum [maximum] of the scalar potential if $\lambda_5>0$ [$\lambda_5<0$].
Thus, it follows (e.g., see Appendix B of \Ref{Gunion:2002zf}) that no scalar potential minimum occurs for $\sin\xi\neq 0$ if
\beq \label{cpcond}
|m_{12}^2-\half \lambda_6 v_1^2-\half\lambda_7 v_2^2|\geq\lambda_5 v_1 v_2\,,
\eeq
Thus if \eq{cpcond} is satisfied, then the scalar potential is minimized for $\sin\xi=0$ 
(i.e., $\xi=n\pi$ for integer~$n$).  One can then
perform a hypercharge gauge transformation, $\Phi_i\to e^{-i\xi_1}\Phi_i$, followed by the field redefinition,
$\Phi_2\to (-1)^n\Phi_2$, which yields real non-negative vevs
$\langle\Phi_i^0\rangle=v_i/\sqrt{2}$ (for $i=1,2$).

If \eq{cpcond} is not satisfied, then the scalar potential is minimized for $\sin\xi\neq 0$.  Nevertheless, one still must check
that there exists no basis transformation of the scalar fields such that the resulting vevs are real.
This can be accomplished without explicitly considering all possible basis choices for the scalar fields
by making use of the so-called Higgs basis.  In Section~\ref{sect:H}, we define the Higgs basis and explain how it can be used to verify that a 2HDM with an explicitly \cp-conserving scalar potential
also has a \cp-conserving vacuum.

\subsection{The Higgs basis}
\label{sect:H}

It is convenient to define new Higgs doublet fields,
\beq \label{newfields}
H_1=\begin{pmatrix}H_1^+\\ H_1^0\end{pmatrix}\equiv \frac{v_1 e^{-i\xi_1}\Phi_1+v_2 e^{-i\xi_2}\Phi_2}{v}\,,
\qquad\quad H_2=\begin{pmatrix} H_2^+\\ H_2^0\end{pmatrix}\equiv\frac{-v_2  e^{i\xi_2}\Phi_1+v_1 e^{i\xi_1}\Phi_2}{v}\,.
\eeq
It follows that $\vev{H_1^0}=v/\sqrt{2}$ and $\vev{H_2^0}=0$.
This is the \textit{Higgs basis}~\cite{Branco:1999fs,Davidson:2005cw}, which is uniquely defined
up to an overall rephasing, $H_2\to e^{i\chi} H_2$.

In the Higgs basis, the scalar potential takes the same form as \eq{genpot} but with new coefficients,
\beqa
\mathcal{V}&=& Y_1 H_1^\dagger H_1+ Y_2 H_2^\dagger H_2 +[Y_3
H_1^\dagger H_2+{\rm h.c.}]
+\half Z_1(H_1^\dagger H_1)^2\nn\\
&&\quad
+\half Z_2(H_2^\dagger H_2)^2
+Z_3(H_1^\dagger H_1)(H_2^\dagger H_2)
+Z_4( H_1^\dagger H_2)(H_2^\dagger H_1)\nn\\
&&\quad +\left\{\half Z_5 (H_1^\dagger H_2)^2 +\big[Z_6 (H_1^\dagger
H_1) +Z_7 (H_2^\dagger H_2)\big] H_1^\dagger H_2+{\rm
h.c.}\right\}\,,\label{hbasispot}
\eeqa
where $Y_1$, $Y_2$ and $Z_1,\ldots,Z_4$ are real and uniquely defined,
whereas $Y_3$, $Z_5$, $Z_6$ and $Z_7$ are complex and transform under
the rephasing of $H_2$,
\beq \label{chitrans}
 [Y_3, Z_6, Z_7]\to e^{-i\chi}[Y_3, Z_6, Z_7] \quad{\rm and}\quad
Z_5\to  e^{-2i\chi} Z_5\,.
\eeq
For an explicitly \cp-conserving scalar potential where all the coefficients given in
\eq{genpot} are real, the real coefficients of the scalar potential in the Higgs basis are given 
by
\beqa
Y_1&=& m_{11}^2\cbt+m_{22}^2\sbt-m_{12}^2 s_{2\beta}\cos\xi\,,\label{yone}\\
Y_2&=& m_{11}^2\sbt+m_{22}^2\cbt+m_{12}^2 s_{2\beta}\cos\xi\,,\label{ytwo}\\
Z_1&=&\lambda_1\cb^4+\lambda_2\sb^4+\half(\lambda_3+\lambda_4+\lambda_5\cos 2\xi)s^2_{2\beta}
+2s_{2\beta}(\lambda_6\cbt+\lambda_7\sbt)\cos\xi\,,\label{zone}\\
Z_2&=&\lambda_1\sb^4+\lambda_2\cb^4+\half(\lambda_3+\lambda_4+\lambda_5\cos 2\xi)s^2_{2\beta}
-2s_{2\beta}(\lambda_6\sbt+\lambda_7\cbt)\cos\xi\,,\label{ztwo}\\
Z_3&=&\tfrac{1}{4}(\lambda_1+\lambda_2-2\lambda_3-2\lambda_4-2\lambda_5\cos 2\xi)s^2_{2\beta}+\lambda_3-(\lambda_6-\lambda_7)s_{2\beta}c_{2\beta}\cos\xi\,,\\
Z_4&=&\tfrac{1}{4}(\lambda_1+\lambda_2-2\lambda_3-2\lambda_4-2\lambda_5\cos 2\xi)s^2_{2\beta}+\lambda_4 -(\lambda_6-\lambda_7)s_{2\beta}c_{2\beta}\cos\xi\,,
\eeqa
where $s_{2\beta}\equiv\sin 2\beta$, $c_{2\beta}\equiv\cos 2\beta$, etc.
The potentially complex coefficients of the scalar potential in the Higgs basis are given by
\beqa
Y_3&=&-e^{-i\xi}\bigl[\half(m_{11}^2-m_{22}^2)s_{2\beta}+m_{12}^2c_{2\beta}\cos\xi+im_{12}^2\sin\xi\bigr]\,,\label{ythree}\\
Z_5&=&e^{-2i\xi}\biggl\{\tfrac{1}{4}\bigl[\lambda_1+\lambda_2-2(\lambda_3+\lambda_4+\lambda_5\cos
2\xi)\bigr]s^2_{2\beta}
+\lambda_5 \cos 2\xi
-(\lambda_6-\lambda_7)s_{2\beta}c_{2\beta}\cos\xi \nonumber \\
&&\qquad\qquad +i\bigl[\lambda_5 c_{2\beta}\sin 2\xi-(\lambda_6-\lambda_7)s_{2\beta}\sin\xi\bigr]\biggr\}\,,\label{zfive}\\
Z_6&=&e^{-i\xi}\biggl\{-\half \bigl[\lambda_1\cbt-\lambda_2\sbt-(\lambda_3+\lambda_4+\lambda_5\cos 2\xi)c_{2\beta}\bigr]s_{2\beta}
+(\lambda_6\cb c_{3\beta}+\lambda_7\sb s_{3\beta})\cos\xi \nonumber \\
&&\qquad\qquad +i\bigl[\half \lambda_5 s_{2\beta}\sin 2\xi+(\lambda_6\cbt+\lambda_7\sbt)\sin\xi\bigr]\biggr\}\,,\label{zsix}
\eeqa
\beqa
Z_7&=&e^{-i\xi}\biggl\{-\half \bigl[\lambda_1\sbt-\lambda_2\cbt+(\lambda_3+\lambda_4+\lambda_5\cos 2\xi)c_{2\beta}\bigr]s_{2\beta}
+(\lambda_6\sb s_{3\beta}+\lambda_7\cb c_{3\beta})\cos\xi \nonumber \\
&&\qquad\qquad +i\bigl[-\half \lambda_5 s_{2\beta}\sin 2\xi+(\lambda_6\sbt+\lambda_7\cbt)\sin\xi\bigr]\biggr\}\,,\label{zseven}
\eeqa
One can check that by using \eqst{m11}{m12}, one recovers the expected scalar potential minimization conditions in the Higgs basis,
\beq \label{minconds}
Y_1=-\tfrac{1}{2}Z_1 v^2\,,\qquad\quad
Y_3=-\tfrac{1}{2} Z_6 v^2\,.
\eeq

Note that if $\sin\xi=0$ then $Y_3$, $Z_5$, $Z_6$ and $Z_7$ are all real and the scalar potential and the vacuum are \cp-conserving.  More generally, if all the coefficients of the Higgs basis are real for some choice of $\chi$ [cf.~\eq{chitrans}], then it follows that any basis related to this real Higgs basis
by a real orthogonal transformation of the two doublet fields is also a real basis with real vevs.
Thus, it follows that the scalar potential and vacuum are \cp-conserving if and only if~\cite{Lavoura:1994fv,Botella:1994cs,Davidson:2005cw,Gunion:2005ja}
\beq \label{cpconds}
{\rm Im}(Z_5^* Z_6^2)={\rm Im}(Z_5^* Z_7^2)={\rm Im}(Z_6^* Z_7)=0\,.
\eeq
No separate condition involving $Y_3$ is needed in light of \eq{minconds}.
Thus, if $\sin\xi\neq 0$, then it may still be possible that the vacuum is \cp-conserving if \eq{cpconds} is satisfied.  One can obtain the general conditions for a \cp-conserving vacuum by inserting
\eqst{zfive}{zseven} into \eq{cpconds}.   For example,
\beq \label{zees}
{\rm Im}(Z_6^* Z_7)=\tfrac{1}{4}\sin2\xi s^2_{2\beta}\bigl[\lambda_5(\lambda_1-\lambda_2)
-\lambda_6^2+\lambda_7^2\bigr]-\half\sin\xi s_{2\beta}c_{2\beta}\bigl[\lambda_1\lambda_7+\lambda_2\lambda_6
-(\lambda_3+\lambda_4+\lambda_5)(\lambda_6+\lambda_7)\bigr]\,.
\eeq
One noteworthy observation concerns the special case of $\lambda_1=\lambda_2$ and $\lambda_7=-\lambda_6$ (dubbed
an exceptional point of the 2HDM parameter space in Ref.~\cite{Davidson:2005cw}).  In fact, if both these relations hold
simultaneously in one basis then they hold simultaneously in \textit{all} bases.  Applying this observation to the Higgs basis, it follows that $Z_1=Z_2$ and
$Z_7=-Z_6$, which can be easily verified using eqs.~(\ref{zone}), (\ref{ztwo}), (\ref{zsix}) and (\ref{zseven}). As a result
${\rm Im}(Z_6^* Z_7)=-{\rm Im}(|Z_6|^2)=0$, which is confirmed by \eq{zees}.

The general expressions for ${\rm Im}(Z_5^* Z_6^2)$ and ${\rm Im}(Z_5^* Z_7^2)$ are much
more complicated and not particularly illuminating.
Nevertheless, if \eq{cpconds} is satisfied for some value of $\xi\neq 0$, then it follows that a real
Higgs basis exists.  Performing an O(2) basis transformation then yields a scalar potential of
the form given by \eq{genpot} with real vevs.   That is, without loss of generality, we can assume
that $\sin\xi=0$ in \eqst{ythree}{zseven}, which yields a real Higgs basis that is unique up to
a possible sign ambiguity that corresponds to redefining the Higgs basis field $H_2\to -H_2$.

Thus, we shall parametrize the general \cp-invariant 2HDM scalar potential by its real Higgs basis
form given in \eq{hbasispot}, where all the $Y_i$ and $Z_i$ are real.  The scalar potential minimum
conditions given in \eq{minconds} fix $Y_1$ and $Y_3$ in terms of $Z_1$ and $Z_6$, respectively.
Since $\vev{H_1^0}=v/\sqrt{2}$ and $\vev{H_2^0}=0$,
it follows that $Y_2$, $Z_1,\ldots,Z_5$ and the product $Z_6 Z_7$ are uniquely defined, whereas $Z_6$ and $Z_7$
separately change sign under the only possible transformation among real Higgs bases, $H_2\to -H_2$.
That is, $Z_6$ and $Z_7$ are pseudo-invariant quantities with respect to the real Higgs basis transformation $H_2\to -H_2$.  However,  the relative sign of $Z_6$ and $Z_7$ is meaningful.

Physical observables must be invariant with respect to any possible Higgs basis transformation. 
In the most general 2HDM, $\tan\beta$ is also unphysical~\cite{Davidson:2005cw,Haber:2006ue} since there is no physical significance
to an arbitrary real basis of scalar fields (apart from the Higgs basis and the basis defined
in terms of neutral Higgs mass eigenstates).  However, it is often the case that the form of the
Higgs-fermion Yukawa couplings pick out a special scalar field basis, in which case $\tan\beta$ is
promoted to a physical parameter.

It is convenient to employ the real Higgs basis to evaluate the spectrum of Higgs masses in the \cp-conserving model.
The physical charged Higgs boson is the charged component of the Higgs-basis doublet $H_2$, and its mass
is given by
\beq \label{chhiggsmass}
m_{H^\pm}^2=Y_{2}+\half Z_3 v^2\,.
\eeq
The three physical neutral Higgs boson mass-eigenstates
are determined by diagonalizing a $3\times 3$ real symmetric squared-mass
matrix that is defined in the Higgs basis~\cite{Branco:1999fs,Haber:2006ue}
\beq   \label{mtwo}
\mathcal{M}^2=
\begin{pmatrix}
Z_1 v^2&\quad Z_6 v^2 &\,\, 0\\
Z_6 v^2  &\quad Y_2+\half (Z_3+Z_4+Z_5)v^2 & \,\, 0 \\
0 &\quad 0 &\,\,
 Y_2+\half (Z_3+Z_4-Z_5)v^2\end{pmatrix}\,.
\eeq
We immediately identify the \cp-odd Higgs boson $A=\sqrt{2}\,{\rm Im}~H_2^0$ with
squared mass,
\beq \label{ma}
m_A^2=Y_2+\half (Z_3+Z_4-Z_5)v^2\,.
\eeq
Note that the real Higgs mass-eigenstate field $A$ is defined up to an overall sign change, which corresponds to the freedom to redefine $H_2\to -H_2$.

The upper $2\times 2$ matrix block given in \eq{mtwo} is the $\cp$-even Higgs squared-mass matrix,
\beq \label{Hmm}
\mathcal{M}_H^2=\begin{pmatrix} Z_1 v^2 & \quad Z_6 v^2 \\  Z_6 v^2 & \quad m_A^2+Z_5 v^2\end{pmatrix}\,,
\eeq
where we have used \eq{ma} to eliminate $Y_2$.
To diagonalize $\mathcal{M}^2_H$, we define
the \cp-even mass eigenstates, $h$ and $H$ (with $m_{h}\leq m_{H}$) by
\beq \label{hH}
\begin{pmatrix} H\\ h\end{pmatrix}=\begin{pmatrix} \cbma & \,\,\, -\sbma \\
\sbma & \,\,\,\phantom{-}\cbma\end{pmatrix}\,\begin{pmatrix} \sqrt{2}\,\,{\rm Re}~H_1^0-v \\ 
\sqrt{2}\,{\rm Re}~H_2^0
\end{pmatrix}\,,
\eeq
where $\cbma\equiv\cba$ and $\sbma\equiv\sba$ are defined in terms of the mixing angle $\alpha$ that diagonalizes the \cp-even Higgs squared-mass matrix when expressed in the original basis of scalar fields, $\{\Phi_1\,,\,\Phi_2\}$.  
The real Higgs mass-eigenstate fields $H$ and $h$ are defined up to an overall sign change.
This implies that $\beta-\alpha$ is defined modulo $\pi$.  In particular,
\beqa
&&\cbma\to -\cbma\,,\quad \sbma\to -\sbma\,,\quad H_2\to +H_2\,\qquad\Longrightarrow\qquad
H\to -H\,,\quad h\to -h\,,\label{sign1}\\
&&\cbma\to +\cbma\,,\quad \sbma\to -\sbma\,,\quad H_2\to -H_2\,\qquad\Longrightarrow\qquad
H\to +H\,,\quad h\to -h\,,\label{sign2}\\
&&\cbma\to -\cbma\,,\quad \sbma\to +\sbma\,,\quad H_2\to -H_2\,\qquad\Longrightarrow\qquad
H\to -H\,,\quad h\to +h\,.\label{sign3}
\eeqa
It follows that the product $\sbma\cbma$ is a pseudo-invariant quantity with respect to the real Higgs basis transformation $H_2\to -H_2$.

The squared masses of $h$ and $H$ are then given by,
\beq \label{mhH}
m_{H,h}^2=\tfrac{1}{2}\biggl\{\mha^2+(Z_1+Z_5)v^2\pm
\sqrt{\bigl[\mha^2+(Z_5-Z_1)v^2\bigr]^2+4Z_6^2 v^4}\,\biggr\}\,.
\eeq
The following identity therefore holds,
\beq \label{Hrelation}
|Z_6| v^2=\sqrt{\bigl(\mhh^2-Z_1 v^2)(Z_1 v^2-\mhl^2\bigr)}\,.
\eeq 
Hence, diagonalizing $\mathcal{M}^2_H$ yields the following expressions:
\beqa
Z_1 v^2&=&\mhl^2 \ssqbma+\mhh^2 \csqbma\,,\label{z1v}\\
Z_6 v^2&=&(\mhl^2-\mhh^2)\sbma\cbma\,,\label{z6v} \\
\mha^2+Z_5 v^2&=&\mhh^2 \ssqbma+\mhl^2 \csqbma\,,\label{z5v}
\eeqa
where $\sbma\equiv\sba$ and $\cbma\equiv\cba$.
Note that \eq{z6v} imply that 
\beq \label{z6cos}
Z_6\sbma\cbma\leq 0\,.
\eeq
Indeed, \eq{z6cos} is invariant with respect to the real Higgs basis transformation $H_2\to -H_2$, so the sign of the quantity $Z_6\sbma\cbma$ is physically meaningful.

Using the fact that $\beta-\alpha$ is defined modulo $\pi$, we shall establish a convention where
\beq \label{range2}
0\leq\beta-\alpha\leq\pi\,.
\eeq
In this convention, $\sbma$ is non-negative, the sign of the field $h$ is fixed and $\cbma$ is pseudo-invariant with respect to the real Higgs basis transformation $H_2\to -H_2$.
One can then derive expressions for $\cbma$ and $\sbma$ from \eqs{z1v}{z6v}, where the signs
of the corresponding quantities are fixed by \eqs{z6cos}{range2}:
\beqa 
\cbma&=&-\sgn(Z_6)\sqrt{\frac{Z_1 v^2-\mhl^2}{\mhh^2-\mhl^2}}=\frac{-Z_6 v^2}{\sqrt{(\mhh^2-\mhl^2)(\mhh^2-Z_1 v^2)}}\,,\label{c2exact}\\[6pt]
\sbma&=&\phm\sqrt{\frac{\mhh^2-Z_1 v^2}{\mhh^2-\mhl^2}}=\frac{|Z_6| v^2}{\sqrt{(\mhh^2-\mhl^2)(Z_1 v^2-\mhl^2)}}
\,.\label{s2exact}
\eeqa
Note that we have used \eq{Hrelation} to obtain the second form for $\cbma$ and $\sbma$ in \eqs{c2exact}{s2exact}, respectively.  

\subsection{The \cp-conserving 2HDM scalar potential with a softly-broken discrete $\mathbb{Z}_2$ symmetry}
\label{z2soft}

When we introduce the Higgs-fermion Yukawa couplings in Section~\ref{sect:Yukawa}, there will be some motivation
to restrict the parameter freedom of the most general scalar potential given in \eq{genpot} by
requiring the invariance of the scalar potential under
the discrete $\mathbb{Z}_2$ symmetry $\Phi_1\to +\Phi_1$ and $\Phi_2\to -\Phi_2$.  Imposing this discrete
symmetry implies that $m_{12}^2=\lambda_6=\lambda_7=0$ in \eq{genpot}.   In this case, $\lambda_5$ is the only potentially
complex scalar potential parameter, which can be rendered real by an appropriate rephasing of $\Phi_1$.
It then follows from \eqs{cosxi}{cpcond} that the scalar potential is minimized for $\sin\xi=0$
if $\lambda_5<0$ and $\cos\xi=0$ if $\lambda_5>0$.  In the latter case, $\vev{\Phi_2^0}/\vev{\Phi_1^0}=\pm i\tan\beta$.
However, a redefinition of $\Phi_1\to \mp i\Phi_1^0$ yields real vevs while $\lambda_5\to -\lambda_5$.  Thus,
the $\mathbb{Z}_2$--invariant scalar potential and the vacuum are \cp-invariant.

One can relax the discrete symmetry by allowing for $m_{12}^2\neq 0$ in \eq{genpot} which softly breaks the
$\mathbb{Z}_2$ symmetry.  The quartic terms in \eq{genpot} still respect the $\mathbb{Z}_2$ symmetry, so that $\lambda_6=\lambda_7=0$.   
However, the scalar potential is now \cp-violating unless
${\rm Im}(\lambda_5^*[(m_{12}^2]^2)=0$.  In what follows, we shall assume that the latter condition is satisfied.  In this
case, one can rephase $\Phi_1$ such that $m_{12}^2$ and $\lambda_5$ are both real.
That is, the \cp-conserving scalar potential of interest is given by
\beqa
\mathcal{V}&=& m_{11}^2 \Phi_1^\dagger \Phi_1+ m_{22}^2 \Phi_2^\dagger \Phi_2 -m_{12}^2[\Phi_1^\dagger \Phi_2+\Phi_2^\dagger\Phi_1]
+\half \lambda_1(\Phi_1^\dagger \Phi_1)^2+\half \lambda_2(\Phi_2^\dagger \Phi_2)^2
+\lambda_3(\Phi_1^\dagger \Phi_1)(\Phi_2^\dagger \Phi_2)
\nn\\
&&\quad
+\lambda_4( \Phi_1^\dagger \Phi_2)(\Phi_2^\dagger \Phi_1)
 +\half \lambda_5\Bigl[(\Phi_1^\dagger \Phi_2)^2 +(\Phi_2^\dagger \Phi_1)^2\bigr]\,,\label{z2genpot}
\eeqa
where all scalar potential parameters are real.  We denote this basis of scalar fields as the $\mathbb{Z}_2$-basis.

If $\lambda_5\leq |m_{12}^2|/(v_1 v_2)$, then the vevs are also real in
light of \eq{cpcond}, which implies that the vacuum is also \cp-conserving.
Otherwise, there is the potential for spontaneous \cp-violation.  To guarantee that
the vacuum is \cp-conserving, we check the conditions given by \eq{cpconds}.
Setting $\lambda_6=\lambda_7=0$ in \eq{zees} yields
\beq \label{cond1}
{\rm Im}(Z_6^* Z_7)=\tfrac{1}{4}\lambda_5(\lambda_1-\lambda_2) s^2_{2\beta}\sin2\xi\,.
\eeq
The other two conditions given in \eq{cpconds} simplify considerably when $\lambda_6=\lambda_7=0$,
\beqa
{\rm Im}(Z_5^* Z_6^2)&\!=\!&-\tfrac{1}{4}\lambda_5 s_{2\beta}^2\sin 2\xi\biggl\{(\lambda_1-\lambda_3-\lambda_4)^2 \cb^4-(\lambda_2-\lambda_3-\lambda_4)^2\sb^4
+\half \lambda_5(\lambda_1-\lambda_2)
s_{2\beta}^2\cos 2\xi-\lambda_5^2 c_{2\beta}\biggr\}\,,\nn \\
&&\phantom{line} \label{cond2}\\
{\rm Im}(Z_5^* Z_7^2)&\!=\!&-\tfrac{1}{4}\lambda_5 s_{2\beta}^2\sin 2\xi\biggl\{(\lambda_1-\lambda_3-\lambda_4)^2\sb^4-(\lambda_2-\lambda_3-\lambda_4)^2\cb^4
+\half\lambda_5(\lambda_1-\lambda_2)
s_{2\beta}^2\cos 2\xi+\lambda_5^2 c_{2\beta}\biggr\}\,,\nn  \\
\phantom{line}\label{cond3}
\eeqa
where $\cos\xi=m_{12}^2/(\lambda v_1 v_2)$ if $\lambda_5>|m_{12}^2|/(v_1 v_2)$ and $\sin\xi=0$ otherwise.

Although the vacuum is \cp-violating for generic values of the softly broken $\mathbb{Z}_2$-invariant 2HDM scalar
potential when $\sin\xi\neq 0$, special cases can arise in which $\sin\xi\neq 0$ and yet the vacuum is \cp-conserving.
We have already encountered one such example when
$m_{12}^2=\lambda_6=\lambda_7=0$ and $\lambda_5>0$, in which case the scalar potential is minimized for
$\xi=\half\pi$.  More generally, consider the case of $m_{12}^2\neq 0$, $\lambda_1=\lambda_2$,
$\lambda_5=|\lambda_1-\lambda_3-\lambda_4|$ and $\lambda_6=\lambda_7=0$.  In this case,
\eqst{cond1}{cond3} all vanish, which implies that a real Higgs basis exists.   Indeed, for this particular example,
\eqst{zfive}{zseven} yield
\beq
\frac{Z_6^2}{Z_5}=\pm\half \lambda_5 s_{2\beta}^2(1\pm\cos 2\xi)\,,\qquad\quad Z_6=-Z_7=\pm\half\lambda_5 s_{2\beta}\bigl[(1\pm\cos 2\xi)
c_{2\beta}+i\sin 2\xi\bigr]\,,
\eeq
for $\lambda_5=\pm(\lambda_1-\lambda_3-\lambda_4)$.  Thus, one can rephase the Higgs basis field $H_2$ to render
$Z_5$, $Z_6$ and $Z_7$ real.  Henceforth, we assume that a Higgs basis has been chosen such that all the $Z_i$ are real and $\sin\xi=0$.

The most general parametrization of the \cp-conserving 2HDM scalar potential and vacuum is easily specified in the Higgs basis.  To
determine the constraints on the Higgs basis parameters that guarantee the existence of a softly-broken
$\mathbb{Z}_2$-invariant, \cp-conserving scalar potential, one simply sets $\lambda_6=\lambda_7=\sin\xi=0$ in
\eqst{yone}{zseven}.  It is also convenient to redefine $\tan\beta$ as the ratio of vevs.  That is, 
henceforth we shall define
$\tan\beta\equiv\langle\Phi_2^0\rangle/\langle\Phi_1^0\rangle=e^{i\xi}v_2/v_1=\pm v_2/v_1$ for $\xi=0$ and $\xi=\pi$, respectively.
In particular, the $Z_i$ are now given by
\beqa
Z_1 & \equiv & \lambda_1 c^4_\beta+\lambda_2 s^4_\beta+\half\lambda_{345} s^2_{2\beta}\,,\label{zeeone}\\
Z_2 & \equiv & \lambda_1 s^4_\beta+\lambda_2 c^4_\beta+\half\lambda_{345} s^2_{2\beta}\,,\label{zeetwo}\\
Z_i & \equiv & \tfrac{1}{4} s^2_{2\beta}\bigl[\lambda_1+\lambda_2-2\lambda_{345}\bigr]+\lambda_i\,,\quad \text{(for $i=3,4$ or 5)}\,,\label{zeefive}\\
Z_6 & \equiv & -\half s_{2\beta}\bigl[\lambda_1 c^2_\beta-\lambda_2 s^2_\beta-\lambda_{345} c_{2\beta}\bigr]\,,\label{zeesix} \\
Z_7 & \equiv & -\half s_{2\beta}\bigl[\lambda_1 s^2_\beta-\lambda_2 c^2_\beta+\lambda_{345} c_{2\beta}\bigr]\,,\label{zeeseven}
\eeqa
where $\lambda_{345}\equiv\lambda_3+\lambda_4+\lambda_5$.
Since there are five nonzero $\lambda_i$ and seven nonzero $Z_i$, there must be two relations.   One can check that the following two
identities are satisfied:
\beqa 
Z_2&=& Z_1+2(Z_6+Z_7)\cot 2\beta\,,\label{zt}\\
Z_{345}&=&Z_1+2Z_6\cot 2\beta-(Z_6-Z_7)\tan 2\beta\,,\label{z345}
\eeqa
where 
\beq \label{z345def}
Z_{345}\equiv Z_3+Z_4+Z_5\,.
\eeq
Eliminating $\tan 2\beta$ yields the following relation among the $Z_i$,
\beq \label{hbasiscond}
(Z_1-Z_2)\bigl[Z_1 Z_7+Z_2 Z_6-Z_{345}(Z_6+Z_7)\bigr]+2(Z_6+Z_7)^2(Z_6-Z_7)=0\,.
\eeq
Indeed, in a general \cp-conserving 2HDM, if \eq{hbasiscond} is satisfied then there must exist a basis in which
$\lambda_6=\lambda_7=0$.  The corresponding value of $\tan\beta$ (which specifies the basis) can be determined from either
\eq{zt} or (\ref{z345}).  
At this stage, the parameter $\tan\beta$ is a pseudo-invariant, since its sign can be flipped by redefining $H_2\to -H_2$ 
[e.g., $\Phi_2\to -\Phi_2$ and $\xi_2\to \xi_2+\pi$ in \eq{newfields}].
It is convenient to work in a convention where the ratio of vevs is non-negative, in which case we
can take 
\beq \label{interval}
0\leq\beta\leq\half\pi\,.  
\eeq
In this convention, the signs of the pseudo-invariants $Z_6$ and $Z_7$ are now fixed, since we can no longer
flip their signs by redefining $H_2\to -H_2$.  This implies that the relative sign of $\sbma$ and $\cbma$ is fixed by \eq{z6cos}, in which case the sign of $\cbma$ is determined by combining \eqs{z6cos}{range2}.

Finally, in the \cp-conserving, softly-broken $\mathbb{Z}_2$-invariant 2HDM, it is convenient to introduce the squared-mass parameter,
\beq \label{mbar}
\overline{m}^{\,2}\equiv \frac{2m_{12}^2}{\sin 2\beta}=\mha^2+\lambda_5 v^2\,.
\eeq
One can express $\overline{m}^{\,2}$ in terms of $Y_2$, $Z_1$ and $Z_6$,
\beq \label{mbarid}
\overline{m}^{\,2}=Y_2+\half Z_1 v^2+Z_6 v^2\cot 2\beta\,.
\eeq
Then, combining \eq{mbarid} with \eqss{ma}{z6v}{z5v} yields
\beq \label{z7v}
Z_7 v^2=(\mhl^2-\mhh^2)\sbma\cbma+2\cot 2\beta\bigl[\mhh^2 \ssqbma+\mhl^2 \csqbma-\overline{m}^{\,2}\bigr]\,.
\eeq

\subsection{Special forms for the Higgs--fermion Yukawa couplings}
\label{sect:Yukawa}

We next turn to the Higgs-fermion Yukawa couplings.  One starts out initially with a Lagrangian
expressed in terms of the scalar doublet fields $\Phi_i$ ($i=1,2$) and the
interaction--eigenstate quark fields.  After electroweak symmetry breaking,
one can identify the $3\times 3$ quark mass matrices.  By redefining the left
and right-handed quark and lepton fields appropriately, the quark and charged lepton
mass matrices are transformed into diagonal form, where the diagonal elements are real
and non-negative.  The resulting Higgs--fermion interaction Lagrangian in terms of the
quark and lepton mass-eigenstate fields,
$U=(u,c,t)$, $D=(d,s,b)$, $N=(\nu_e,\nu_\mu,\nu_\tau)$, and
$E=(e,\mu,\tau)$, is given by
\beqa
-\mathcal{L}_{\rm Y}&=&\overline U_L \Phi_{a}^{0\,*}{{h^U_a}} \ur -\anti
D_L K^\dagger\Phi_{a}^- {{h^U_a}}\ur
+\overline U_L K\Phi_a^+{{h^{D\,\dagger}_{a}}} \dr
+\overline D_L\Phi_a^0 {{h^{D\,\dagger}_{a}}}\dr \nn \\
&&\qquad\quad +\overline N_L\Phi_a^+{{h^{E\,\dagger}_{a}}} E_R
+\overline E_L\Phi_a^0 {{h^{E\,\dagger}_{a}}}E_R +{\rm h.c.}\,,
\label{higgsql}
\eeqa
where $K$ is the CKM quark mixing matrix,
$h^{U,D,L}$ are $3\times 3$ Yukawa coupling matrices and there is an
implicit sum over $a=1,2$.  The diagonal quark and charged lepton mass matrices are given by
$M_F=(v_1h_1^F+v_2 h_2^F)/\sqrt{2}$, where $F=U,D,E$.
However, the couplings of the neutral Higgs bosons to the fermions are not flavor-diagonal.
Thus, \eq{higgsql} would yield large tree-level Higgs-mediated flavor changing neutral
currents (FCNCs) which is in conflict with observed data.

In a general extended Higgs model, tree-level Higgs mediated FCNCs are
absent if for some choice of basis of the scalar fields,
at most one Higgs multiplet is responsible for
providing mass for quarks or leptons of a given electric
charge, as first pointed out by Glashow, Weinberg and Pascos (GWP)~\cite{Glashow:1976nt,Paschos:1976ay}.  This GWP
condition can be imposed by a symmetry
principle, which guarantees that the absence of  tree-level Higgs-mediated FCNCs is natural.  By an appropriate choice of symmetry transformation
laws for the fermions and the Higgs scalars, the resulting
Higgs-fermion Yukawa interactions take on the required form in a specific
basis of scalar fields.  The symmetry also restricts the form of the Higgs scalar
potential in the same basis.  These considerations were first applied in
the 2HDM in Refs.~\cite{Donoghue:1978cj} and \cite{Haber:1978jt}.

The GWP condition can be implemented in four different ways~\cite{Hall:1981bc,Barger:1989fj,Akeroyd:1996he,Aoki:2009ha}:
\begin{enumerate}
\item Type-\Rmnum{1} Yukawa couplings: $h_1^U=h_1^D=h_1^L=0$,
\item Type-\Rmnum{2} Yukawa couplings: $h_1^U=h_2^D=h_2^L=0$.
\item Type-X Yukawa couplings: $h_1^U=h_1^D=h_2^L=0$,
\item Type-Y Yukawa couplings: $h_1^U=h_2^D=h_1^L=0$.
\end{enumerate}
The four types of Yukawa couplings can be implemented by a discrete
symmetry as shown in Table~\ref{Tab:type}.
\vskip -0.05in
\begin{table}[hb!]
\begin{center}
\begin{tabular}{|cl||c|c|c|c|c|c|}
\hline && $\Phi_1$ & $\Phi_2$ & $U_R^{}$ & $D_R^{}$ & $E_R^{}$ &
 $U_L$, $D_L$, $N_L$, $E_L$ \\  \hline
Type I  && $+$ & $-$ & $-$ & $-$ & $-$ & $+$ \\
Type II &(MSSM like)& $+$ & $-$ & $-$ & $+$ & $+$ & $+$ \\
Type X  &(lepton specific) & $+$ & $-$ & $-$ & $-$ & $+$ & $+$ \\
Type Y  &(flipped) & $+$ & $-$ & $-$ & $+$ & $-$ & $+$ \\
\hline
\end{tabular}
 \caption{Four possible $\mathbb{Z}_2$ charge assignments that forbid
tree-level Higgs-mediated FCNC effects in the 2HDM~\cite{Aoki:2009ha}.}
\label{Tab:type}
\end{center}
\end{table}
%
\noindent
The neutral Higgs Yukawa couplings (relative
to the corresponding couplings of the SM Higgs boson) are
conveniently summarized
in Table~\ref{yukawa_tab} for the four possible implementations of
the GWP condition.

\begin{table}[t!]
 \begin{center}
{\renewcommand\arraystretch{1.5}
\begin{tabular}{|c||ccccccccc|}\hline
&$h\overline{U}U$
&$h\overline{D}D$&$h\overline{E}E$&$H\overline{U}U$&$H\overline{D}D$&$H\overline{E}E$&$iA\overline{U}\gamma\ls{5}U$&$iA\overline{D}\gamma\ls{5}D$
& $iA\overline{E}\gamma\ls{5}E$\\
\hline\hline
Type I &$\frac{\cos\alpha}{\sin\beta}$&$\phm\frac{\cos\alpha}{\sin\beta}$&$\phm\frac{\cos\alpha}{\sin\beta}$&$\frac{\sin\alpha}{\sin\beta}$&$\frac{\sin\alpha}{\sin\beta}$&$\frac{\sin\alpha}{\sin\beta}$&$-\cot\beta$&$\phm\cot\beta$&$\phm\cot\beta$\\\hline
Type II &$\frac{\cos\alpha}{\sin\beta}$&$-\frac{\sin\alpha}{\cos\beta}$&$-\frac{\sin\alpha}{\cos\beta}$&$\frac{\sin\alpha}{\sin\beta}$&$\frac{\cos\alpha}{\cos\beta}$&$\frac{\cos\alpha}{\cos\beta}$&$-\cot\beta$&$-\tan\beta$&$-\tan\beta$\\\hline
Type X &$\frac{\cos\alpha}{\sin\beta}$&$\phm\frac{\cos\alpha}{\sin\beta}$&$-\frac{\sin\alpha}{\cos\beta}$&$\frac{\sin\alpha}{\sin\beta}$&$\frac{\sin\alpha}{\sin\beta}$&$\frac{\cos\alpha}{\cos\beta}$&$-\cot\beta$&$\phm\cot\beta$&$-\tan\beta$\\\hline
Type Y &$\frac{\cos\alpha}{\sin\beta}$&$-\frac{\sin\alpha}{\cos\beta}$&$\phm\frac{\cos\alpha}{\sin\beta}$&$\frac{\sin\alpha}{\sin\beta}$&$\frac{\cos\alpha}{\cos\beta}$&$\frac{\sin\alpha}{\sin\beta}$&$-\cot\beta$&$-\tan\beta$&$\phm\cot\beta$\\\hline
\end{tabular}
\caption{Neutral Higgs--fermion couplings in the 2HDM subject to the
$\mathbb{Z}_2$ symmetries given in Table~\ref{Tab:type}.  The
couplings listed above are normalized relative to the SM Higgs
couplings to $\overline{U}U$, $\overline{D}D$,
and $\overline{E}E$.
\label{yukawa_tab}}}
\end{center}
\vskip -0.2in
\end{table}

The imposition of the discrete symmetry also restricts the form of
the Higgs scalar potential given in \eq{genpot} by setting
$m_{12}^2=\lambda_6=\lambda_7=0$.  As discussed in Section~\ref{z2soft},
the condition $m_{12}^2=0$ can be relaxed.  In the case of
a softly-broken $\mathbb{Z}_2$-invariant 2HDM with $m_{12}^2\neq 0$, 
Higgs-mediated FCNCs are still absent at tree-level, although they are
generated at one-loop order.  Nevertheless, the size of these FCNCs
may be phenomenologically acceptable, depending on the region of the 2HDM
parameter space.  This motivates us to focus on the Higgs scalar potential
of the form given in \eq{z2genpot}.
Note that the parameter $\tan\beta\equiv\langle\Phi_2^0\rangle/ \langle\Phi_1^0\rangle$ is defined in terms of the vevs
with respect to the $\mathbb{Z}_2$-basis of scalar fields, where the discrete $\mathbb{Z}_2$ symmetry
of the Higgs-fermion Yukawa interactions is manifest
(i.e., where three of the six Higgs-fermion Yukawa matrices vanish).
Indeed as previously advertised, the parameter $\tan\beta$ has been promoted to a physical
parameter of the theory.

\subsection{Parametrizing the softly-broken $\mathbb{Z}_2$-symmetric \cp-conserving 2HDM}
\label{parms}

The scalar potential in the $\mathbb{Z}_2$-basis [\eq{z2genpot}] is governed by three squared-mass parameters and five dimensionless quartic coupling parameters.  Minimizing the scalar potential, we re-express $m_{11}^2$ and $m_{22}^2$ in terms of $v=246$~GeV and $\tan\beta$.  Excluding $v$ (which determines the $W$ and $Z$ masses), we are left with seven real parameters: $m_{12}^2$, $\tan\beta$ and the $\lambda_i$ ($i=1,2,\ldots,5$).   In the Higgs basis, the counting is also straightforward: after imposing the scalar potential minimum conditions the relevant real parameters (excluding $v$) are $Y_2$ and $Z_i$ ($i=1,2,\ldots,7$).  Imposing one relation, \eq{hbasiscond}, 
to guarantee the existence of a basis where $\lambda_6=\lambda_7=0$,  we are again left with seven real parameters.   

A more physical choice of parameters would consist of $\alpha$, $\beta$ and the four scalar masses, $m_h$, $m_H$, $m_A$ and $m_{H^\pm}$.  This leaves one additional parameter, which is usually chosen to be $m_{12}^2$ or $\lambda_5$ (cf.~Appendix~ D of \cite{Gunion:2002zf}).  In many of the previous studies of the 2HDM parameter space~\cite{Craig:2013hca,Eberhardt:2013uba,Coleppa:2013dya,Baglio:2014nea,Dumont:2014wha,Dumont:2014kna,Craig:2015jba,Bernon:2015qea}, scans were performed over the parameters $m_h$, $m_H$, $m_A$, $m_{H^\pm}$, $m_{12}^2$, $\alpha$ and 
$\beta$.  
Acceptable points in the scan must satisfy unitarity and perturbativity constraints~\cite{Kanemura:1993hm,Ginzburg:2005dt}.  However, for random choices of $m_{12}^2$ and masses of the three non-SM-like Higgs bosons, one finds that the unitarity and perturbativity constraints are often violated.  This is easily understood by examining the \textit{decoupling limit} where $Y_2\gg v$.  In this limit, $m_H\gg v$ in which case
$\cbma\to 0$ [cf.~\eq{c2exact}] and the properties of $h$ approach that of the SM Higgs boson.  Moreover, 
eqs.~(\ref{chhiggsmass}), (\ref{ma}), (\ref{z5v}) and (\ref{mbar}) yield squared-mass differences,
\beq \label{diffs}
m_A^2-m_{H^\pm}^2=\half(Z_4-Z_5)v^2\,,
\qquad
m_H^2-m_A^2\simeq Z_5 v^2\,,\qquad
\overline{m}^{\,2}-m_A^2=\lambda_5 v^2\,.
\eeq
The conditions of unitarity and perturbativity constrain these squared-mass differences to be of 
$\mathcal{O}(cv^2)$, where $c\lsim 10$.

To maximize the efficiency of costly scans over the 2HDM parameter space (e.g.~in a global fit), one should define the input parameters of the model to include the mass of the observed SM-like Higgs boson and at most one additional Higgs mass.  The other two Higgs masses and $\overline{m}^{\,2}$ are determined in terms of the quartic Higgs self-coupling parameters.  As long as the magnitudes of the quartic parameters are constrained to be less than~$\mathcal{O}(10)$, the scan over the other Higgs masses and $\overline{m}^2$ will not violate the unitarity and perturbativity constraints.

\subsection{The hybrid basis of parameters}
\label{sect:H2}

In light of the discovery of a SM-like Higgs boson, we have information on two of the parameters of the 2HDM.  Since the observed boson is not $\cp$-odd, we shall identify it with either $h$ or $H$.  Given that the $hVV$ and $HVV$ couplings ($V=W^\pm$ or $Z$) relative to the SM Higgs boson $h_{\rm SM}$ are given by
\beq \label{HVVc}
g\ls{hVV}=g\ls{h_{\rm SM}VV}\,\sbma\,,\qquad g\ls{HVV}=g\ls{h_{\rm SM}VV}\,\cbma\,,
\eeq
the values of the SM-like Higgs boson mass ($m_h$ or $m_H$) and $\cbma$ are already known with some precision.

We therefore propose a ``hybrid'' strategy for specifying the input parameters for the softly-broken $\mathbb{Z}_2$-invariant 2HDM.
The masses of the physical \cp-even Higgs scalars are given directly as input parameters together with $\cbma$, which determines the phenomenologically important couplings of the \cp-even scalars to the $W^\pm$ and $Z$ bosons [cf.~\eq{HVVc}],
and $\tan\beta$ which specifies the basis of scalar fields where the discrete symmetry
of the Higgs-fermion Yukawa interactions is manifest.
In addition, we specify the real Higgs basis self-coupling coefficients $Z_4$, $Z_5$, and $Z_7$ as input parameters.
We henceforth designate the input parameter set $\{m_h, m_H, \cbma, \tan\beta, Z_4, Z_5, Z_7\}$ as the hybrid basis of parameters as indicated in Table~\ref{tab:param}.
In the convention adopted in \eq{range2}, we have\footnote{Different convention choices for $\beta-\alpha$ appear in the 2HDM literature.   For example,
the convention $0\leq\cbma \leq 1$ and $-1\leq\sbma\leq 1$ is employed internally by the \thdmc\ code \cite{Eriksson:2009ws,Eriksson:2010zzb}.
Another common choice in the literature is to take $-\half\pi\leq\alpha\leq\half\pi$.  
Indeed, one is always free to define either $\alpha$ or $\beta-\alpha$ modulo $\pi$.   It is a simple matter to translate among the various conventions.
For any 2HDM parameter point $(\alpha, \beta)$ where $0\leq\beta\leq\half\pi$,
compute the values of $\sbma$ and $\cbma$.  Then, to convert to the convention
of \eq{sbmacon}, simply replace $(\sbma,\cbma)\to (-\sbma,-\cbma)$ if $\sbma$ is initially negative.  Only the \textit{relative} sign of $\sbma$ and $\cbma$ is physical under the independent convention of 
non-negative $\tan\beta$ (the latter is universally employed in the 2HDM literature), as explained below \eq{interval}.}
\beq \label{sbmacon}
0\leq\sbma\leq 1\,,\quad {\rm and}\quad  -1\leq \cbma \leq 1\,,
\eeq
in addition to the convention of \eq{interval} which implies that $\tan\beta$ is non-negative.


\begin{table}[h!]
\centering
\begin{tabular}{cl}
\hline
Parameter & Description\\
\hline
$\mh$ & Mass of the light $\cp$-even Higgs boson \\
$\mH$ & Mass of the heavy $\cp$-even Higgs boson \\
$\cbma$ & $\cos(\beta-\alpha)$, which determines the couplings of the \cp-even Higgs bosons to
$VV$\\
$\tan\beta$ & Ratio of vacuum expectation values in the basis with manifest $\mathbb{Z}_2$ symmetry \\
$Z_4, Z_5, Z_7$ & Quartic couplings in the Higgs basis of $\mathcal{O}(1)$\\
\hline
\end{tabular}
\caption{2HDM input parameters in the hybrid basis. By convention, we take $\tan\beta$ to be non-negative.  For a more detailed description of their phenomenological relevance, see the text.}
\label{tab:param}
\end{table}

With these seven input parameters, one may compute the real Higgs basis self-coupling coefficients
$Z_1$, $Z_2$, $Z_3$ and $Z_6$ as well as the \cp-odd Higgs and charged Higgs scalar masses $m_{A}$ and $m_{H^\pm}$ as follows.
First, we employ \eqs{z1v}{z6v}, 
\beqa
Z_1&=&\frac{\sbma^2\mh^2+\cbma^2\mH^2}{v^2}\,,\label{z1d}\\
Z_6&=&\frac{(m_h^2-m_H^2)\sbma\cbma}{v^2}\,.\label{z6d}
\eeqa
In order to ensure that unitarity and perturbativity constraints are not violated,
one should choose the input parameters $\mhh$ and $\cbma$ such that $\mhh^2\cbma\sim\mathcal{O}(cv^2)$, where $c\lsim\mathcal{O}(10)$
in order to avoid a potentially large value of $Z_6$.
Next, we use \eq{zt} to compute $Z_2$,
\beq
Z_2=\frac{\sbma^2\mh^2+\cbma^2\mH^2}{v^2}+2\cot 2\beta\left(\frac{(\mh^2-\mH^2)\sbma\cbma}{v^2}+Z_7\right)\,.\label{z2d}
\eeq
There is a danger that $Z_2$ may become too large if $\beta$ is near 0 or $\half\pi$.  
This is not a true singularity since in the formal limit of $\tan 2\beta=0$, it follows that $Z_6=Z_7=0$ [cf.~\eqs{zeeone}{zeetwo}] corresponding to the inert 2HDM limit discussed briefly at the end of this Section.  In practice, we restrict  our scans over regions of $\tan\beta$ such that $Z_2$ is never too large.

The parameter $Z_3$ is determined from \eq{z345},
\beq \label{z3d}
Z_3=\frac{\sbma^2\mh^2+\cbma^2\mH^2}{v^2}+(2\cot 2\beta-\tan 2\beta)\left(\frac{(\mh^2-\mH^2)\sbma\cbma}{v^2}\right)+Z_7\tan 2\beta-Z_4-Z_5\,.
\eeq
For this quantity, values of $\beta$ near 0, $\tfrac{1}{4}\pi$ or $\half\pi$ appear to be problematical, yielding potentially large values of $Z_3$.  The case of $\beta$ near 0 or $\half\pi$ has already been mentioned below \eq{z2d}.  The case of $\beta=\tfrac{1}{4}\pi$ is not a singular limit, as it corresponds to $Z_1=Z_2$ and $Z_6=Z_7$ with $Z_{345}$ an independent parameter [cf.~\eqst{zeeone}{zeeseven}].  In particular, for $\beta=\tfrac{1}{4}\pi$, we must replace $Z_7$ with $Z_3$ as an input parameter in the hybrid basis.  We can sidestep this special parameter regime by avoiding values of $\tan\beta$ too close to 1. 

Observe that \eqss{z6d}{z2d}{z3d} are consistent with the pseudo-invariant nature of $\tan\beta$, $\sbma\cbma$, $Z_6$ and $Z_7$. 
Since we have established a convention in which $\sbma$ and $\tan\beta$ are both non-negative [cf.~\eqs{range2}{interval}], it follows that that full 
parameter space of the model requires the consideration of all possible non-negative values of $\tan\beta$ and $\sbma$ 
and all possible sign combinations of $\cbma$, $Z_6$ and $Z_7$, subject to the constraint of \eq{z6cos}.
 
The masses of the \cp-odd Higgs and charged Higgs scalar masses are determined from \eqss{chhiggsmass}{ma}{z5v}
\beqa
m_A^2&=& \mH^2\ssqbma+\mh^2\csqbma- Z_5v^2\,,\label{ma2}\\
m_{H^\pm}^2&=& m_A^2-\half(Z_4-Z_5)v^2\,.\label{mhpm2}
\eeqa
As a side note, one can also determine $\overline{m}^{\,2}$ from \eq{z7v},
\beq \label{z7d}
\overline{m}^{\,2}=\mhh^2\ssqbma+\mhl^2\csqbma+\half\tan 2\beta(Z_6-Z_7)v^2\,,
\eeq
where $Z_6$ is given by \eq{z6d}.

In practice, taking the parameters $Z_4$ and $Z_5$ as input parameters is not very intuitive.  Clearly a more physical approach
is to take $\mha$ and $\mhpm$ as input parameter and then compute $Z_4$ and $Z_5$ using \eqs{ma2}{mhpm2}.  Then these parameters
can be employed in evaluating $Z_3$ via \eq{z3d}.  The only danger with such an approach is that a poor choice of $\mha$ and $\mhpm$
will yield values of $Z_4$ and $Z_5$ that violate unitarity and perturbativity constraints, as discussed in Section~\ref{parms}.  In practice, we first employ our hybrid
basis to perform our parameter scans.  Once we have identified suitable parameter regimes, we will then take a more physical approach by  
employing $\mha$ and $\mhpm$ to define the corresponding benchmark scenarios.

The parameter $Z_7$ is also not particularly intuitive, so one could advocate taking $\overline{m}^{\,2}$ as the input parameter
and then compute $Z_7$ using \eq{z7d}.  However, $\overline{m}^{\,2}$ is not a physical squared-mass parameter, so there is no particular advantage for
adopting this strategy.  A poor choice of $\overline{m}^{\,2}$ would yield a value of $Z_7$ that violates unitarity and perturbativity 
constraints.  Indeed, the \cp-conserving, softly-broken $\mathbb{Z}_2$-invariant 2HDM always requires one extra parameter beyond the four physical Higgs masses,
$\alpha$ and $\beta$.  There is an advantage to choosing this parameter to be dimensionless for the reasons noted above.
Another possible choice for this parameter that is often found in the literature is $\lambda_5$.  Using \eqst{zeefive}{zeeseven}, one can check that
\beq \label{L5}
\lambda_5=Z_5+\half(Z_6-Z_7)\tan 2\beta\,,
\eeq
so one can always trade in $\lambda_5$ for $Z_7$ and vice versa.
  
The case of $\overline{m}^{\,2}=0$ is special and corresponds to a 2HDM with an exact $\mathbb{Z}_2$ symmetry.  In this case,
$Z_7$ is determined by \eq{z7d} in terms of $\mhl$, $\mhh$ and $\cbma$.  It is noteworthy that one cannot take $\mhh$ arbitrarily large
in the $\mathbb{Z}_2$-invariant 2HDM without violating unitarity and perturbativity bounds.\footnote{In light of \eq{L5}, the quantity
$(Z_6-Z_7)\tan 2\beta\sim\mathcal{O}(1)$ for \textit{all} values of $\tan\beta$ as long as the $\lambda_i\sim\mathcal{O}(1)$.}
Thus, there is no consistent decoupling limit where the heavy Higgs states
$\hh$, $\ha$ and $\hpm$ decouple from the low-energy theory.  In contrast, in the softly-broken $\mathbb{Z}_2$-invariant 2HDM with $\overline{m}^{\,2}\neq 0$,
the decoupling limit exists in which $\hh$, $\ha$ and $\hpm$ are all very heavy with squared-masses of $\mathcal{O}(\overline{m}^{\,2})$
[cf.~\eq{mbarid}], while keeping all Higgs self-coupling parameters bounded.

Finally, the inert 2HDM~\cite{Deshpande:1977rw,Barbieri:2006dq,LopezHonorez:2006gr}, corresponding to $Z_6=Z_7=0$ and Type-I Yukawa couplings~\cite{Asner:2013psa}, must be treated separately.  In the inert 2HDM, we have either $\beta=0$ or $\beta=\half\pi$ and either $\cbma=0$ (in which case $h$ is identified as the SM-like Higgs boson)
or $\sbma=0$ (in which case $H$ is identified as the SM-like Higgs boson).  In both cases,
the SM-like Higgs boson is denoted by $h_{\rm SM}$ and the other inert neutral
Higgs bosons will be denoted by $H_I$ and $A_I$.  The states $H_I$ and $A_I$ are relatively \cp-odd, although (despite the notation) the individual \cp-quantum numbers of these two states are not well defined.  By convention, we shall define $H_I$ to be the heavier of the two neutral inert Higgs scalars.
Then, using \eqs{chhiggsmass}{mtwo}, the charged and neutral inert Higgs masses are given by
\beq
m_{H^\pm}^2=Y_2+\half Z_3 v^2\,,\qquad\quad
m_{H_I,A_I}^2=Y_2+\half(Z_3+Z_4\pm |Z_5|) v^2\,. 
\eeq
The relevant input parameter set in this
case is $\{m_{h_{\rm SM}}^2=Z_1 v^2, m_{H^\pm}^2, Z_2, Z_3, Z_4, Z_5\}$.  Note that parameters $Z_2$ and $Z_3$, which
govern the self-couplings of the inert scalar doublet, must be provided separately since \eqs{z2d}{z3d} are not applicable in this case. We shall not consider the inert 2HDM further here, but phenomenological constraints and benchmark scenarios for this model have been recently discussed in \cite{Ilnicka:2015sra}.

\section{Benchmark scenarios}
\label{bench}

\subsection{Numerical analysis}
To capture the 2HDM phenomenology that is interesting for upcoming LHC Higgs searches, we have performed numerical scans over the softly broken $\mathbb{Z}_2$-symmetric 2HDM parameter space using the hybrid basis of parameters for inputs for the scans.\footnote{The parameter scans presented in Section~\ref{bench} have employed either Type-I or Type-II Yukawa couplings.  Parameter scans with Type-X or Type-Y Yukawa couplings will be treated elsewhere.} 
Based on these results, we identify a set of useful \emph{benchmark scenarios} that we will now describe in more detail.  Most of the scenarios are presented in the form of 2-dimensional benchmark \emph{planes}, whereas some of them are in the form of \emph{lines} where only one parameter is varied. As a general rule, these scenarios could be made into benchmarks of higher dimensions by promoting additional parameters with fixed values to vary within certain ranges. Below we discuss a few examples of how this could be done.

For the numerical evaluations we use the code \thdmc\ \cite{Eriksson:2009ws,Eriksson:2010zzb} (v.~1.7.0), where the \HH\ basis of parameters has been implemented according to Section~\ref{sect:H2}. Constraints on the quartic couplings from (absolute) vacuum stability and $S$-matrix unitarity are evaluated at the input scale. For the case with $Z_2$ symmetry (as is implicit already in the definition of the hybrid basis), the condition that the Higgs potential is positive definite is equivalent to the well-known relations~\cite{Deshpande:1977rw},
\begin{equation}
\lambda_1 > 0,\quad \lambda_2>0, \quad \lambda_3 > -\sqrt{\lambda_1\lambda_2}, \quad
\lambda_3+\lambda_4-|\lambda_5|> -\sqrt{\lambda_1\lambda_2}\,.
\end{equation} 
For the unitarity constraints, we impose an upper limit which corresponds to saturation of the unitarity bound by the tree-level contribution. Using the formulation of \cite{Ginzburg:2005dt}, this is equivalent to the constraint $|\Lambda|<16\,\pi$ on individual matrix elements. 
\begin{figure}[t!]
\centering
\includegraphics[width=0.45\columnwidth]{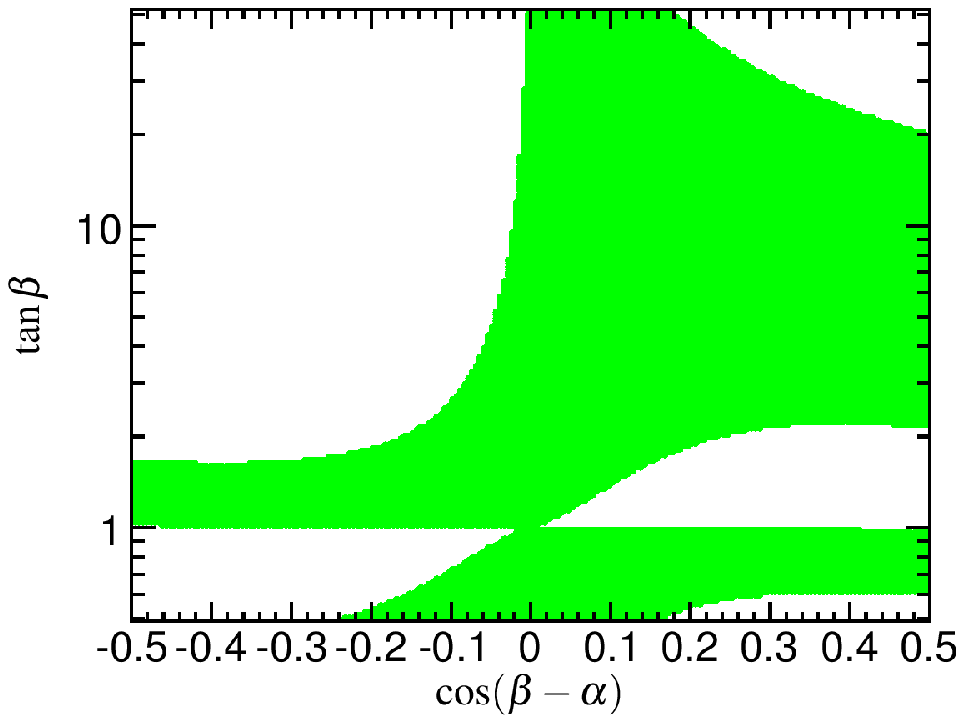}
\includegraphics[width=0.45\columnwidth]{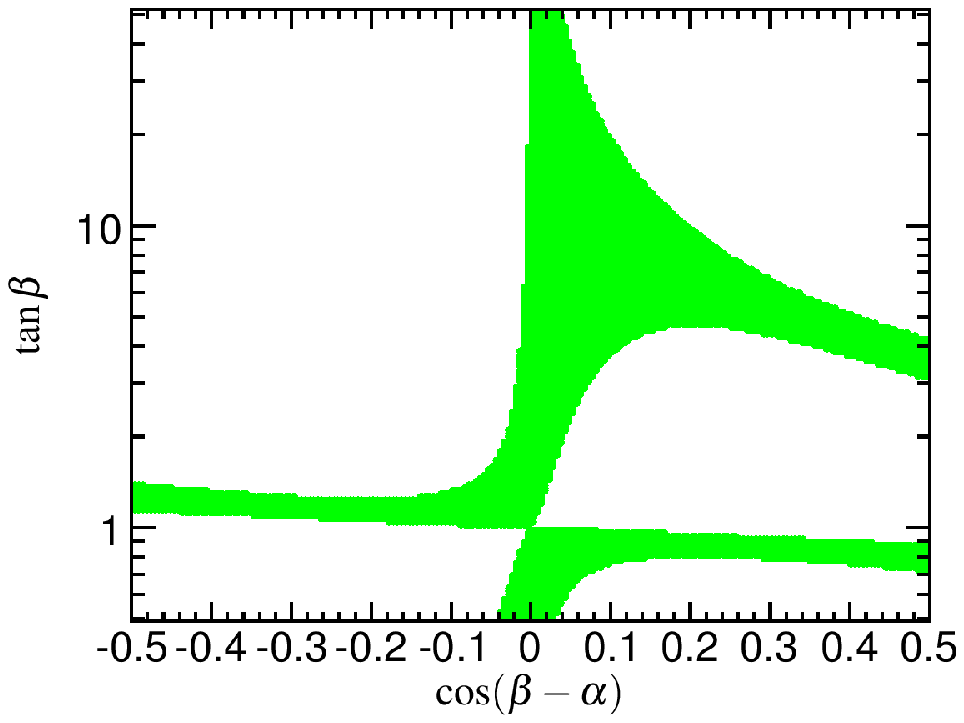}
\caption{Example 2HDM parameter regions respecting perturbative unitarity and stability constraints (green) for
$\mH=300\GeV$ (left) and $\mH=600\GeV$ (right), $Z_4=Z_5=-2$ and $Z_7=0$.}
\label{fig:Z}
\end{figure}
As an example, the regions in $(\cbma, \tanb)$ compatible with theoretical constraints from perturbative unitarity and positivity of the potential are shown in Fig.~\ref{fig:Z} for fixed values of the input parameters in the hybrid basis: $\mhl=125\GeV$, $\mhh=300\GeV$ (left) and $\mhh=600\GeV$ (right), $Z_4=Z_5=-2$ and $Z_7=0$. The theoretically allowed parameter space (shown in green) shrinks with increasing $\mhh$, owing to the existence of a proper decoupling limit where $\cbma$ is forced towards zero as $\mhh \gg v$ [cf.~eq.~\eqref{c2exact}].  As can be seen from this figure, there is an allowed (green) patch that survives for large values of $\cbma$ even for $\mhh=600\GeV$. This region is removed at higher masses by the unitarity constraint (which cuts in from the upper right corner). The constraints are invariant under the transformation $(\cbma,\tanb)\to (-\cbma,1/\tanb)$, reflecting the fact that these points are related by an interchange of the Higgs fields in the $\mathbb{Z}_2$-basis, $\Phi_1\longleftrightarrow \Phi_2$, or equivalently taking $H_2\to -H_2$ and $\cbma\to -\cbma$ in a convention where $\sbma$ is non-negative [cf.~\eq{sign3}].  We shall not consider values of $\tan\beta < 1$ in the following.

Branching ratios and LHC cross sections for Higgs production are evaluated using \thdmc\ and {\tt SusHi} \cite{Harlander:2012pb} in accordance with the recommendations of \cite{Harlander:2013qxa}. We recommend that numerical values used in an experimental analysis are also calculated with {\tt HIGLU} \cite{Spira:1995mt} and {\tt HDECAY} \cite{Djouadi:1997yw}, since such a comparison can provide a first estimate of the theoretical uncertainties in the treatment of (missing) higher-order corrections. These programs provide cross section predictions for the heavier 2HDM Higgs bosons, $H$ and $A$, in the dominant gluon fusion and $b\bar{b}$ associated production modes.  One of the benchmark scenarios discussed below (Scenario E) allows for the production of heavy charged Higgs bosons. The cross section for $pp\to t\bar{b}H^-/\bar{t}bH^+$ depends only on $\mHp$ and $\tan\beta$. A detailed numerical analysis of this process has recently been presented in \cite{Flechl:2014wfa}, with numbers that are applicable to our Scenario E.

Constraints from direct Higgs searches at LEP, the Tevatron, and the LHC are evaluated using \HB\ \cite{Bechtle:2008jh,Bechtle:2011sb,Bechtle:2013wla} (v.~4.2.0), which selects for each parameter point the most sensitive exclusion limit (at $95\%$ C.L.). The compatibility of the 2HDM with the observed $125$~GeV Higgs signal is calculated in terms of a $\chi^2$ value taking into account the full LHC run-I data using  \HS\ \cite{Bechtle:2013xfa} (v.~1.3.0). To determine the viable parameter regions in a benchmark scenario with two free parameters, we demand compatibility with the best-fit point, which usually is in very good agreement with the SM, within $2\,\sigma$ ($\Delta\chi^2=\chi^2-\chi^2_{\mathrm{min}} < 6.18$). The best-fit (minimal $\chi^2$ value) is reevaluated for each benchmark scenario. We make no quantitative statements about the relative degree of compatibility between different scenarios and the data, but it is checked explicitly that the best fit points obtained in all cases lie very close to the SM predictions.
\begin{figure}
\centering
\includegraphics[width=0.45\columnwidth]{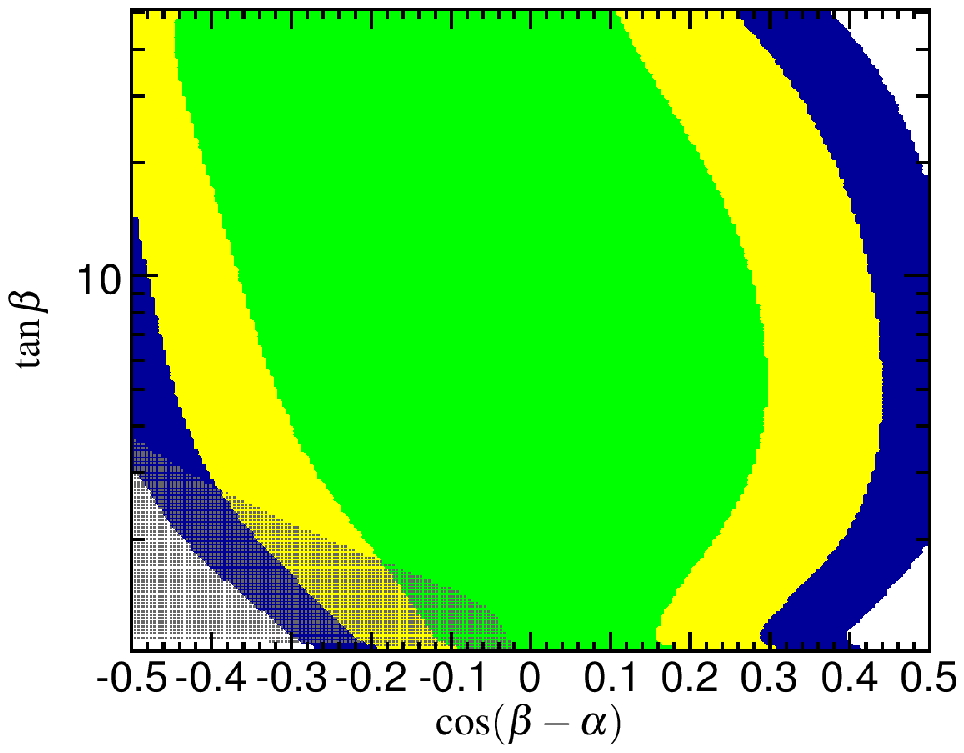}
\includegraphics[width=0.45\columnwidth]{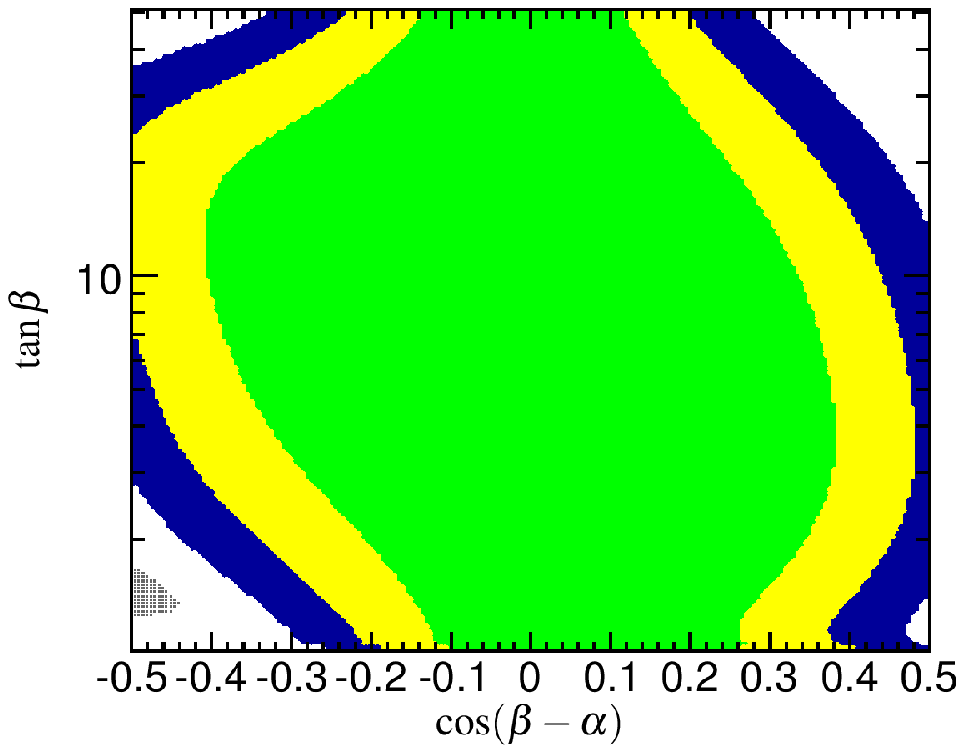}
\includegraphics[width=0.45\columnwidth]{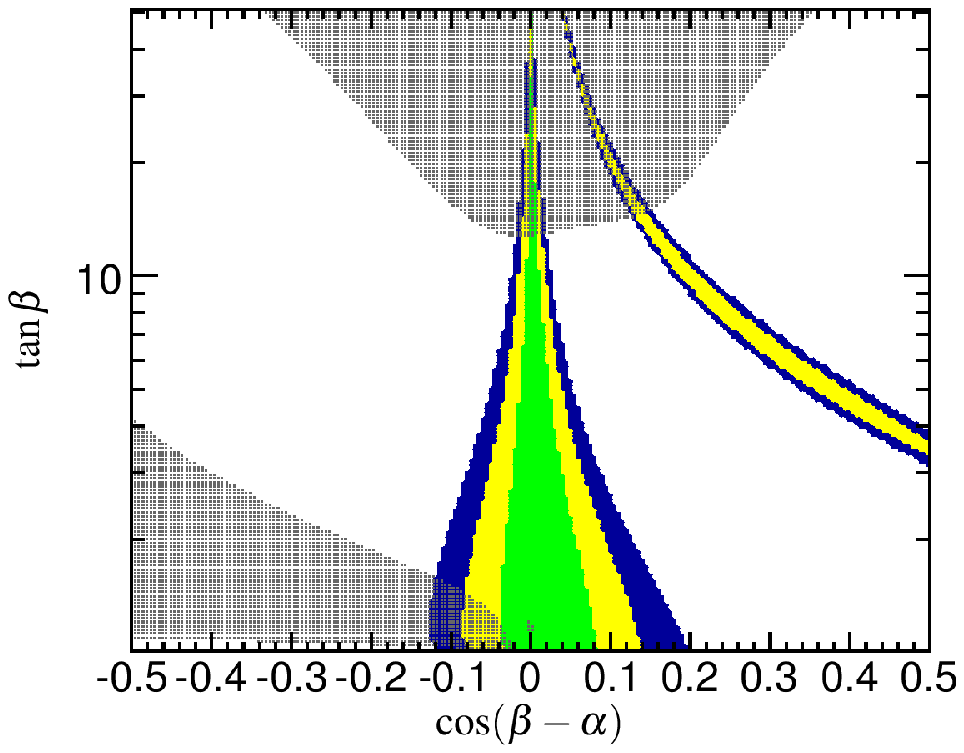}
\includegraphics[width=0.45\columnwidth]{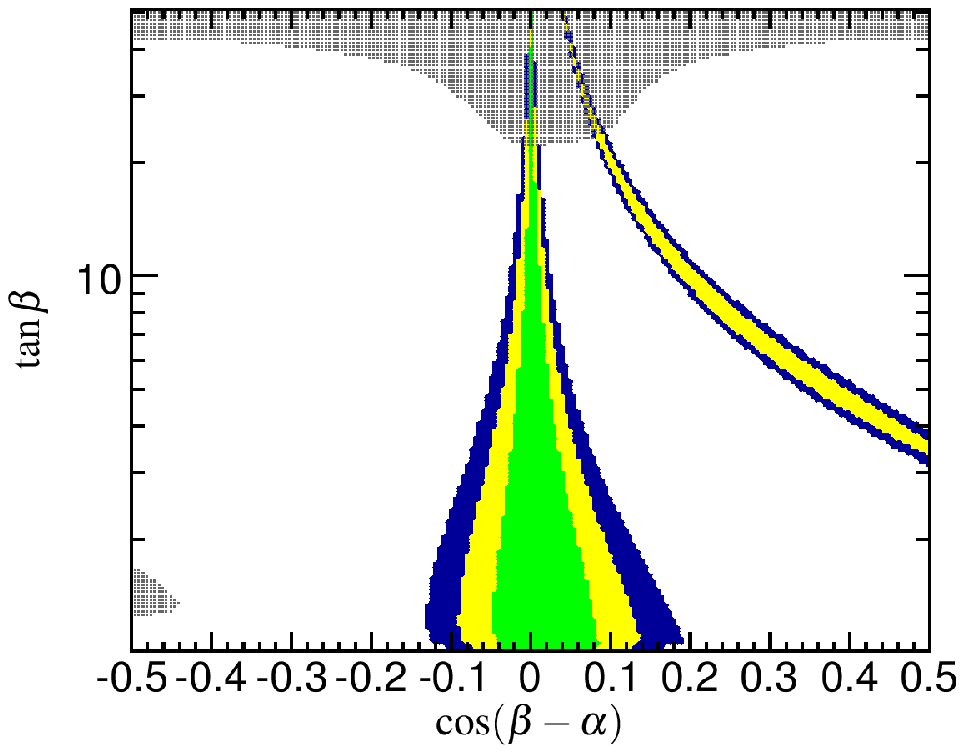}
\caption{Direct constraints from LHC Higgs searches on the parameter space for the 2HDM Type-I (top) and Type-II (bottom) with $\mH=300\GeV$ (left) and $\mH=600\GeV$ (right). In both cases $\mh=125\GeV$, $Z_4=Z_5=-2$ and $Z_7=0$. The colors indicate compatibility with the observed Higgs signal at $1\,\sigma$ (green), $2\,\sigma$ (yellow) and $3\,\sigma$ (blue). Exclusion bounds at $95\%$ C.L.~from the non-observation of the additional Higgs states are overlaid in gray.   Note that the constraints from Fig.~\ref{fig:Z} have \textit{not} been imposed here.
}
\label{fig:lhcconst}
\end{figure}
Fig.~\ref{fig:lhcconst} demonstrates how these constraints work on the 2HDM parameter space under the assumption that the observed LHC signal corresponds to the lightest 2HDM Higgs boson, $h$. Demanding that the Higgs rates are within $2\,\sigma$ of the measurements restricts $\cbma$ to be close to the SM limit ($\cbma\sim 0$) in the case of Type-II Yukawa couplings. For Type-I couplings, the deviation from the SM limit can be somewhat larger. The shape of the allowed region arises from the dependence of the production and decay rates on the mixing angles, most importantly the total width which in the Type-II model is dominated by the $hb\bar{b}$ coupling.

One additional set of constraints to keep in mind when designing viable benchmark scenarios are the electroweak precision tests, where in particular the oblique $T$ parameter \cite{Peskin:1991sw} can receive sizable contributions due to a large mass splitting between the non-SM-like charged and neutral Higgs states~\cite{Froggatt:1991qw,Froggatt:1992wt,Pomarol:1993mu,WahabElKaffas:2007xd,Haber:2010bw}.   In 
particular,
\beqa  \label{thdmtcp}
\alpha T &=&\frac{g^2}{64\pi m_W^2} \biggl\{
\mathcal{F}({m^2_{H^\pm}},m_{A^0}^2)+
\mathcal{F}({m^2_{H^\pm}},m_{H^0}^2)-\mathcal{F}({m^2_{A^0}},m_{H^0}^2)\nonumber \\
&&\quad +\csqbma\left[
\mathcal{F}({m^2_{H^\pm}},m_{h^0}^2)-\mathcal{F}({m^2_{A^0}},m_{h^0}^2)
+\mathcal{F}({m^2_{A^0}},m_{H^0}^2)-\mathcal{F}({m^2_{H^\pm}},m_{H^0}^2)\right]
\biggr\}+\mathcal{O}(g^{\prime\,2})\,,
\eeqa
where $\alpha\equiv e^2/(4\pi)$ with the electromagnetic coupling $e$ and the weak couplings $g$ and $g'$ defined in the $\overline{\rm MS}$ scheme evaluated at $m_Z$. The function $\mathcal{F}$ is defined by
\beq
\mathcal{F}(m_1^2,m_2^2) \equiv \half(m_1^2+m_2^2)-\frac{m_1^2m_2^2}{m_1^2-m_2^2}
\ln\left(\frac{m_1^2}{m_2^2}\right)\,,
\eeq
with $\mathcal{F}(m^2,m^2)=0$.
Typically, the $\mathcal{O}(g^{\prime\,2})$ contribution to $T$ is numerically small.  Hence, to ensure that the 2HDM contribution to $T$ is within the current bound, $T\leq 0.2$ at $95\%$~C.L.~\cite{Agashe:2014kda}, one 
must be close to the custodial limit.  In particular, in the decoupling limit of the 2HDM, Eq.~\eqref{thdmtcp} reduces to~\cite{Haber:2010bw}
\beq \label{alphaT}
\alpha T\simeq \frac{(m^2_{H^\pm}-m_A^2)(m^2_{H^\pm}-m_H^2)}{48\pi^2 v^2 m_H^2}
\left[1+\mathcal{O}\left(\frac{v^2}{m_H^2}\right)\right]\,,
\eeq
from which one can see that the 2HDM contribution to the $T$ parameter 
will be sufficiently small if either $\mhpm^2-\mha^2\lesssim \mathcal{O}(v^2)$ or $\mhpm^2-\mhh^2\lesssim \mathcal{O}(v^2)$.  Note that \eq{alphaT} can be rewritten as [cf.~\eq{diffs}]:
\beq
\alpha T\simeq \frac{(Z_4-Z_5)(Z_4+Z_5)v^2}{192\pi^2 m_H^2}
\left[1+\mathcal{O}\left(\frac{v^2}{m_H^2}\right)\right]\,.
\eeq
Consequently, the choice $Z_4=\pm Z_5$ is special in the sense that the corresponding Higgs sector contributions to the $T$ parameter vanish in the decoupling limit. Several of the benchmark scenarios that we propose below satisfy this property.

Finally, the general 2HDM (without supersymmetry) can also be constrained by various low-energy (flavor physics) processes (see, e.g., \cite{Mahmoudi:2009zx}). Since our main objective is to define scenarios capturing interesting LHC phenomenology, we will not be concerned with the details of these constraints, nor will they be explicitly applied in our numerical results. 
In all the following, it should be noted that there exist generic lower bound on the charged Higgs boson mass in the Type-II 2HDM, $\mhpm\gtrsim 480$~GeV at $95\%$~C.L., from measurements of the $\mathrm{BR}(B\to X_s\gamma)$~\cite{Misiak:2015xwa}.

\subsection{Scenario A (non-alignment)}
\begin{figure}[h!]
\centering
\includegraphics[width=0.48\columnwidth]{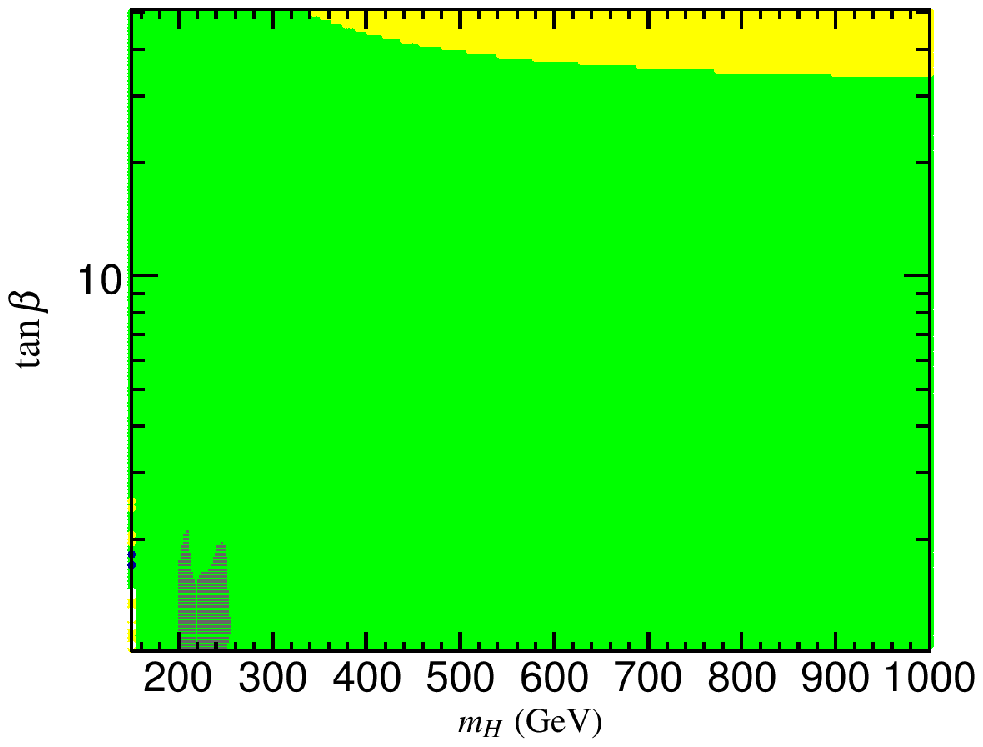}
\includegraphics[width=0.48\columnwidth]{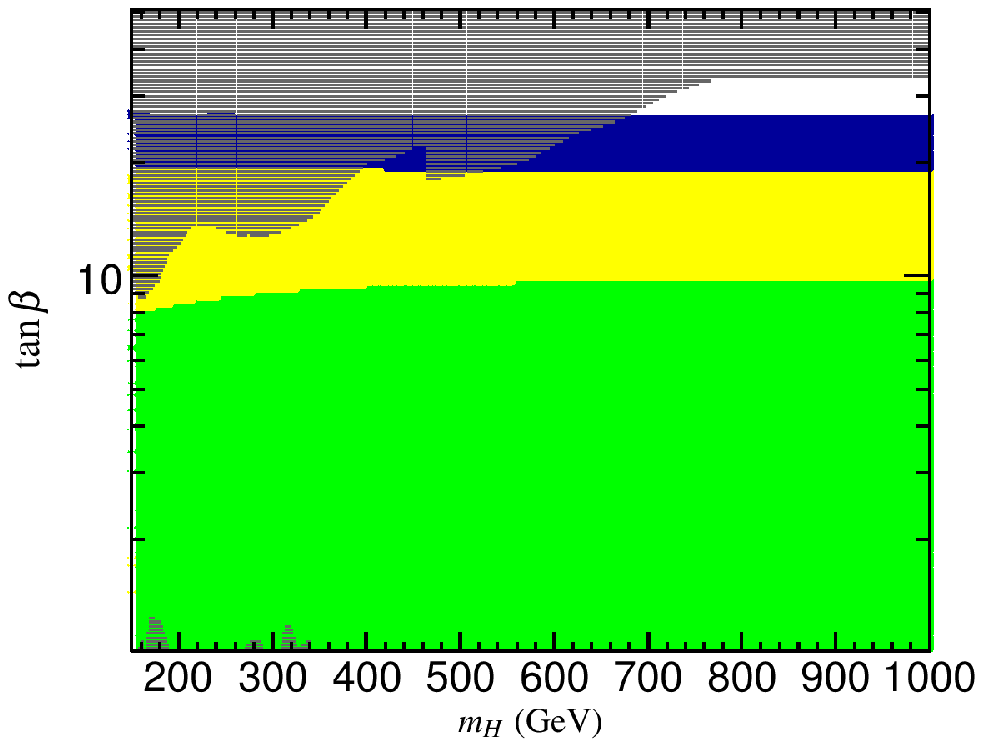}
\caption{Parameter space of the non-aligned benchmark Scenario A with Type-I couplings, $\cbma=0.1$ (left) and Type-II Yukawa couplings, $\cbma=0.01$  (right). The color coding is the same as in Fig.~\ref{fig:lhcconst}.}
\label{fig:Aconstr}
\end{figure}

Our first benchmark scenario has the ``normal'' interpretation of the $125\GeV$ signal as the lightest $\cp$-even Higgs boson, $h$, with SM-like properties. 
The $h$ is SM-like in the so-called alignment limit\footnote{In the alignment limit~\cite{Haber:2013mia,Craig:2013hca,Asner:2013psa,Carena:2013ooa,Carena:2014nza,Bernon:2015qea}, the SM-like Higgs boson is approximately aligned with the neutral component of the Higgs basis field $H_1$.
In particular, in the limit where $\cbma\to 0$, we have
 $h\simeq  \sqrt{2}\,{\rm Re}~\!H_1^0-v$, and $m_h^2\simeq Z_1 v^2$.  
In light of \eq{c2exact}, this limit can be realized if either $Z_6\to 0$ or if
$m_H\gg m_h$.  The latter is achieved in the decoupling limit, whereas the former can be achieved independently of the choice of the non-SM-like Higgs masses. \label{fn}}
where $\cbma\to 0$, in which case the $hVV$ coupling approaches the corresponding SM value.
On the other hand, to allow for some interesting phenomenology of the heavier $\cp$~even state, $H$, we define the scenario with a non-alignment ($\cbma\neq 0$) as allowed by the present constraints [cf.~Fig.~\ref{fig:lhcconst}]. The scenario focuses on searches for the heavier $\cp$-even state, $H$, in SM final states (including the $H\to hh$ decay). The remaining two Higgs bosons, $A$ and $H^\pm$ (which are kept mass-degenerate), are decoupled to a sufficient degree to create a small hierarchy
\vskip -0.05in
\[
\mhl=125\GeV < \mhh < \mha=\mhpm.
\]
For $\mH>150\GeV$, this can be achieved by setting $Z_4=Z_5 = -2$, which leads to values of $\mhpm$ satisfying the $b\to s\gamma$ constraint for Type-II models. The value of $\cbma$ is fixed close to the maximum allowed by the LHC Higgs constraints, $\cbma=0.1$ for Type-I and $\cbma=0.01$ for Type-II couplings. Consequently, we keep $\mhh$ and $\tanb$ as free parameters. 
As shown in Fig.~\ref{fig:Aconstr},
these parameter choices 

\noindent
lead to a very good fit to the light Higgs signal rates over a large fraction of the ($\mhh, \tanb$) plane.
An alternative to choosing a fixed value for 
$\cbma$ is to realize this scenario with a value that decreases with $\mH$ to emulate decoupling [cf.~\eq{c2exact}]. For example, in the case of Type-I couplings, we have performed scans
with $\cbma=0.1\times(150\GeV/m_H)^2$.

\begin{figure}[t!]
\centering
\includegraphics[width=0.45\columnwidth]{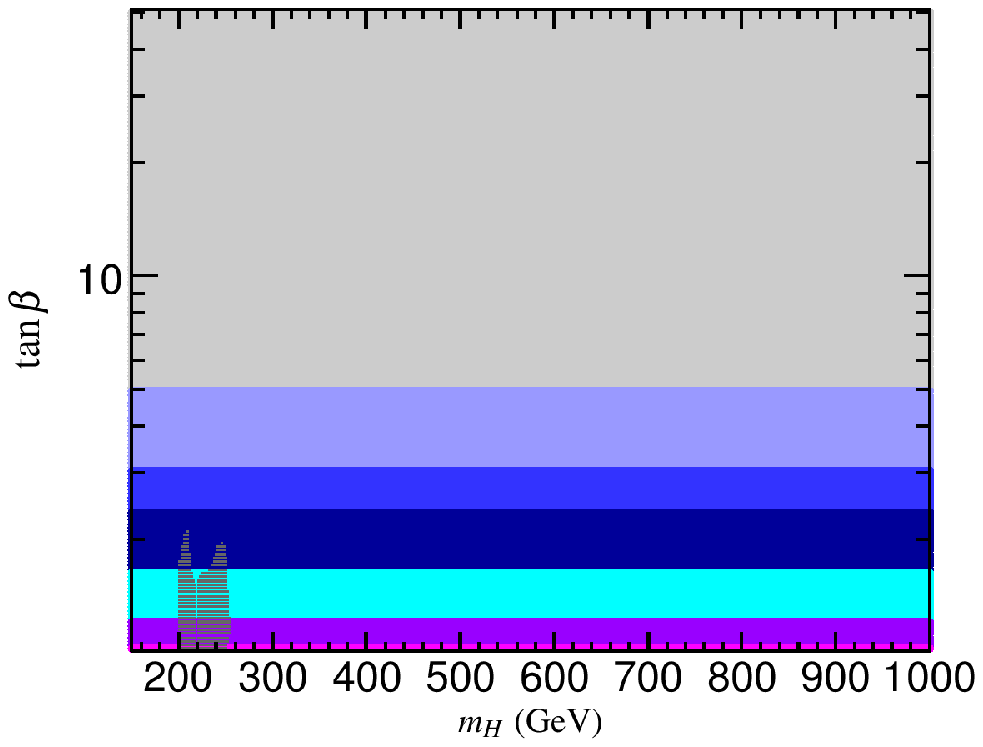}
\includegraphics[width=0.45\columnwidth]{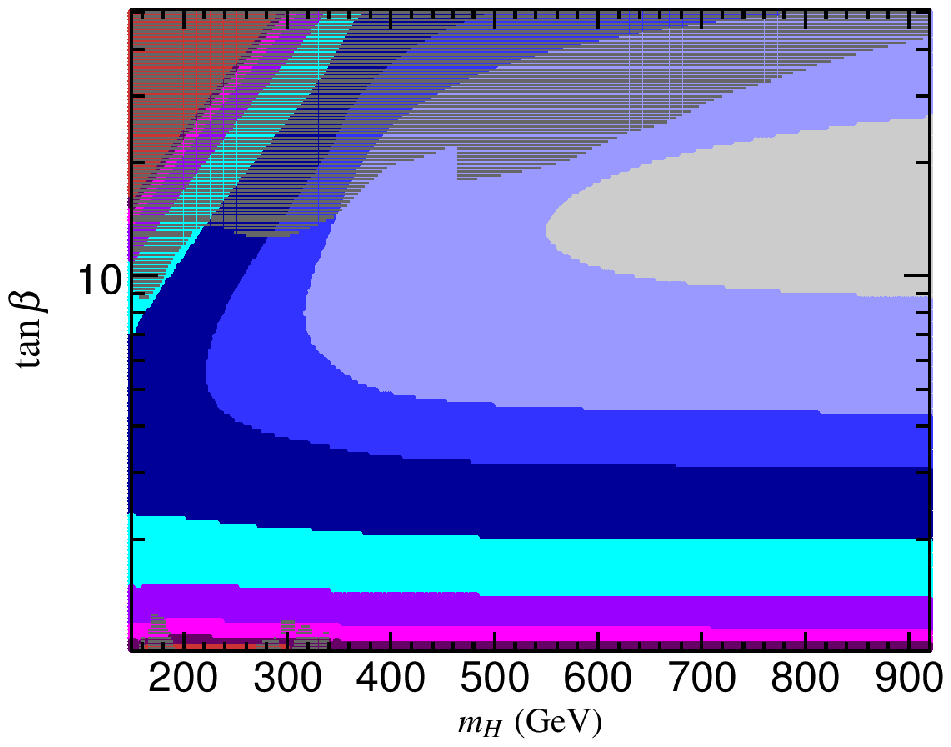}
\includegraphics[width=0.08\columnwidth]{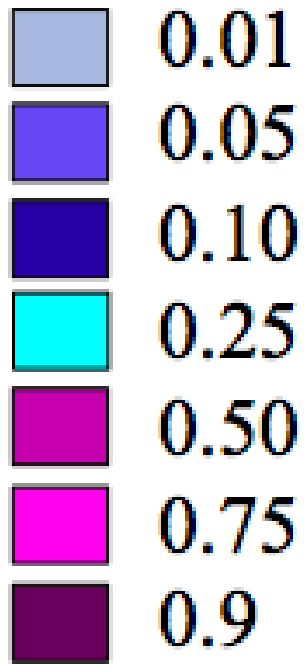}
\caption{Cross sections for gluon fusion $gg\to H$ at $\sqrt{s}=13$~TeV in the 2HDM relative to the prediction for a SM Higgs boson with the same mass in the alignment limit $\cbma\to 0$ with Type-I (left) and Type-II Yukawa couplings (right).}
\vskip 0.1in
\label{fig:A_ggHH}
\end{figure}

We now look at the predictions for the production and decay modes of the heavy $\cp$-even Higgs in this scenario.  Fig.~\ref{fig:A_ggHH} shows contours of the cross section for the dominant gluon fusion production mode, $gg\to H$ versus the free parameters $(\mhh, \tanb)$. As can be seen from this figure, the cross section is maximized as low values of $\tanb$ due to the suppression of the $Ht\bar{t}$ coupling by $1/\tanb$. This holds for both Type-I and Type-II couplings due to the universality of the couplings to up-type quarks. In the Type-II case, we also see an enhancement of the cross section at high values of $\tanb$ where the bottom loops become dominant, although in this scenario the parameter space regions with a significant enhancement is already disfavored by LHC constraints. In the Type-I case, the decoupling property of the $H$ couplings to fermions sets in at high $\tanb$, which can lead to arbitrarily small values of the cross section. This scenario can therefore provide a useful benchmark to gauge and interpret experimental progress in probing smaller and smaller values of $\sigma\times \mathrm{BR}$ in particular channels. In the Type-II model, at high $\tanb$, the cross section for associated production with $b$ quarks, $b\bar{b}\to H$, can become important since it scales as $\tan^2\beta$, whereas for Type-I couplings this process is suppressed. 

\begin{figure}[t!]
\centering
\begin{overpic}[width=0.48\columnwidth]{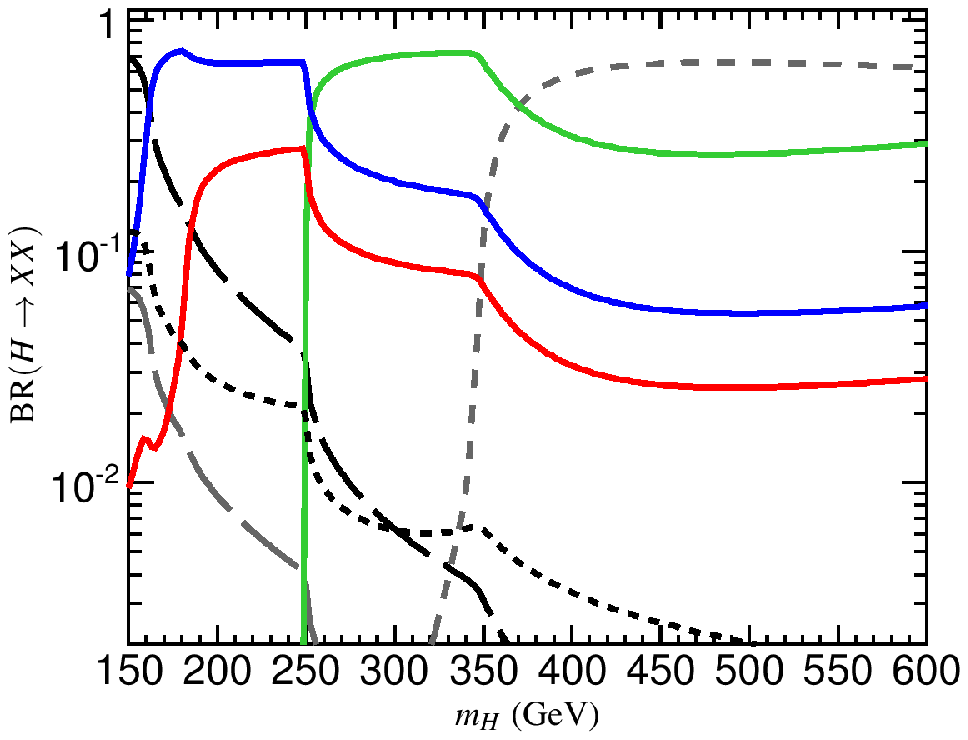}
\put (68,65) {\small $t\bar{t}$}
\put (68,55) {\small $hh$}
\put (68,40) {\small $W^+W^-$}
\put (68,32) {\small $ZZ$}
\put (66,16) {\small $gg$}
\put (38,28) {\small $b\bar{b}$}
\put (17,18) {\small $\tau^+\tau^-$}
\end{overpic}
\begin{overpic}[width=0.48\columnwidth]{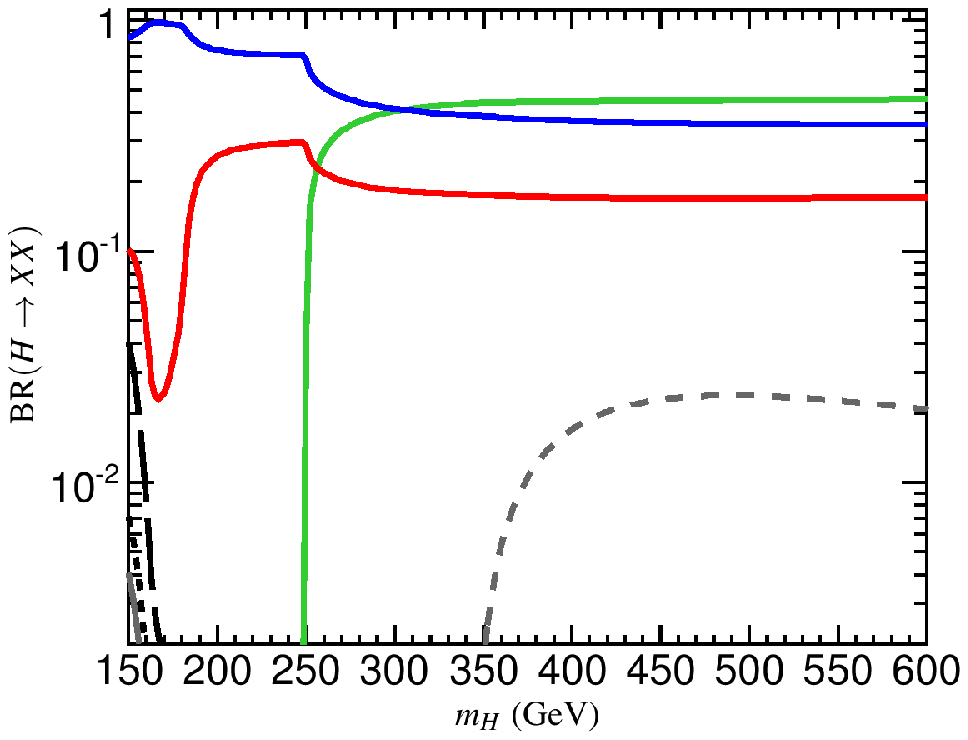}
\put (68,30) {\small $t\bar{t}$}
\put (68,68) {\small $hh$}
\put (68,59) {\small $W^+W^-$}
\put (68,51) {\small $ZZ$}
\put (17,28) {\small $b\bar{b}$}
\end{overpic}
\caption{Branching ratios of the Heavy Higgs boson, $H$, in scenario A with Type-I couplings for $\cbma=0.1$, $\tan\beta=1.5$ (left) and $\tan\beta=7$ (right). Colors: $H\to W^+W^-$ (blue, solid), $H\to ZZ$ (red, solid), $H\to hh$ (green, solid), $H\to t\bar{t}$ (gray, short dash), $H\to b\bar{b}$ (black, long dash), $H\to \tau\tau$ (gray, long dash) and $H\to gg$ (black, short dash).}
\label{fig:A_BRH_I}
\end{figure}
\begin{figure}[h!]
\centering
\begin{overpic}[width=0.48\columnwidth]{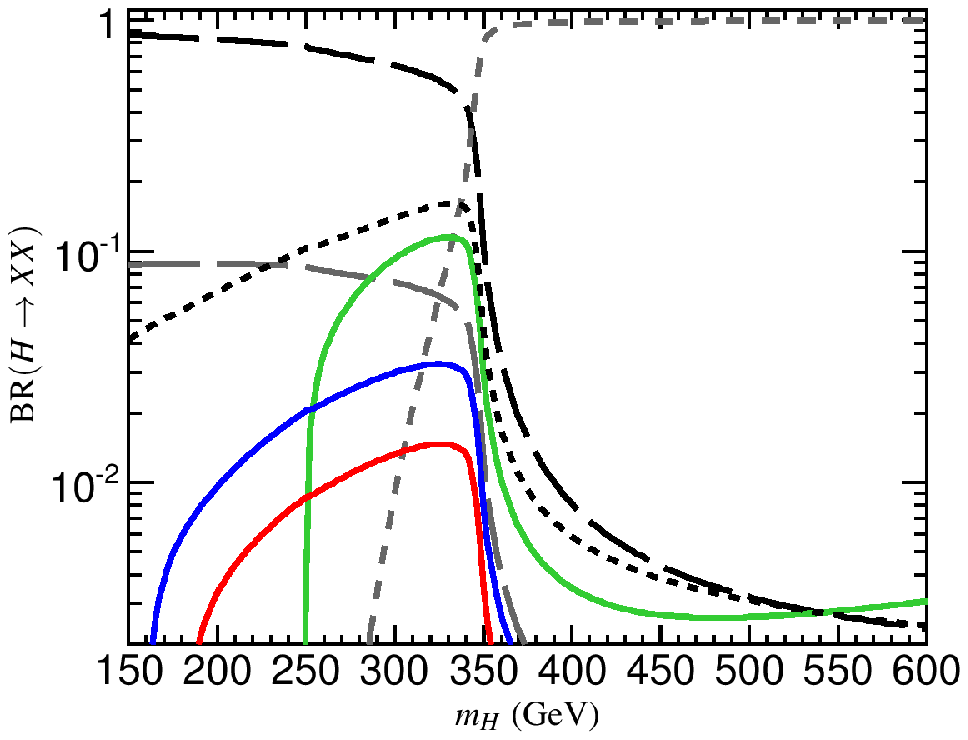}
\put (68,70) {\small $t\bar{t}$}
\put (27,44) {\small $hh$}
\put (17,35) {\small $W^+W^-$}
\put (25,15) {\small $ZZ$}
\put (32,56) {\small $gg$}
\put (17,68) {\small $b\bar{b}$}
\put (17,52) {\small $\tau^+\tau^-$}
\end{overpic}
\begin{overpic}[width=0.48\columnwidth]{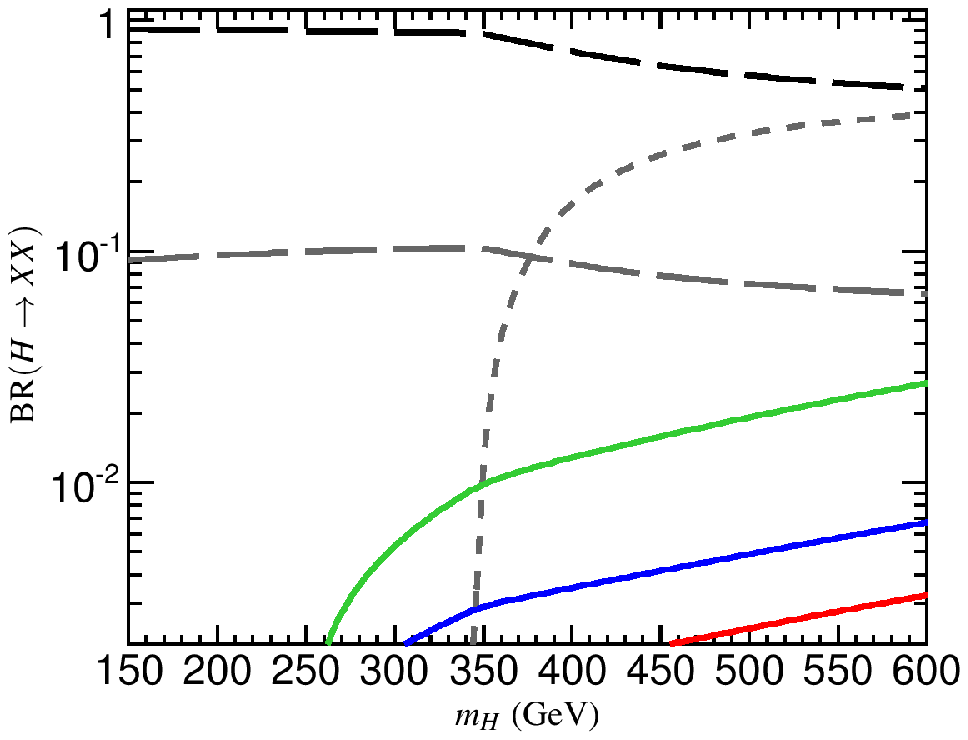}
\put (68,56) {\small $t\bar{t}$}
\put (68,35) {\small $hh$}
\put (68,23) {\small $W^+W^-$}
\put (72,14) {\small $ZZ$}
\put (17,68) {\small $b\bar{b}$}
\put (17,52) {\small $\tau^+\tau^-$}
\end{overpic}
\caption{Branching ratios of $H$ in scenario A with Type-II couplings for $\cbma=0.01$, $\tan\beta=1.5$ (left) and $\cbma=0.01$, $\tan\beta=7$ (right). The color coding is the same as in Fig~\ref{fig:A_BRH_I}.}
\label{fig:A_BRH_II}
\end{figure}

Turning now to the decay modes, 
the case of Type-I Yukawa couplings is shown in Fig.~\ref{fig:A_BRH_I} for $\tan\beta=1.5$ and $7$. The Type-I couplings gives a very distinct pattern of dominant decay modes, where the vector bosons dominate below $250\GeV$. Above this value, the cascade $H\to hh$ opens up and quickly becomes dominant, while there are still substantial contribution from the decays to $WW$ and $ZZ$. At low $\tan\beta$ this changes above the $t\bar{t}$ threshold, but for higher values of $\tan\beta$ this mode (like all fermionic decays) is suppressed. Scenario~A with Type-I Yukawa couplings is a suitable benchmark to replace the SM as the standard parametrization of $2\ell$ and $4\ell$ searches at higher masses, where the rates can basically be adjusted to anything below the present limit. Note that arbitrarily high values of $\mH$ are not allowed in this scenario for all $\tan\beta$ values, owing to the restrictive unitarity and stability constraints when $\cbma$ is non-zero [cf.~Fig.~\ref{fig:Z}].

With Type-II couplings, the restriction on $\mH$ from theoretical constraints is much less severe, since the LHC data already forces the scenario to be defined with $\cbma$ much closer to the alignment limit. This also has the effect of suppressing bosonic decay modes, which can be seen in Fig.~\ref{fig:A_BRH_II} where fermionic modes ($b\bar{b}$, $\tau\tau$ and $t\bar{t}$) are clearly dominant over the full mass range. This scenario can be used for combinations of searchers utilizing different fermionic decay modes. It is also useful to gauge performance of e.g.~$\tau\tau$ searches in a setting where only a single $\cp$-even Higgs boson is present at a given mass (in contrast to the MSSM case where typically $H$ and $A$ are nearly mass-degenerate). It also avoids other complications associated with SUSY interpretations, such as the non-holomorphic $\Delta_b$ corrections to the bottom Yukawa coupling and/or the presence of SUSY decay modes~\cite{Carena:2002es,Djouadi:2005gj}.


\subsection{Scenario B (low-$\mhh$)}
Scenario B corresponds to a ``flipped'' 2HDM benchmark scenario. In this scenario both $h$ and $H$ are light, but it is the heavier of the two which has $m_H = 125$ GeV and is SM-like. Since $m_h < m_H$, the lighter of the two $\cp$-even Higgs bosons must have strongly suppressed couplings to vector bosons to be compatible with direct search limits which forces $\sbma\to 0$ [cf.~\eq{HVVc}].\footnote{In light of \eq{s2exact}, the alignment limit of $\sbma\to 0$ is achieved in the limit of $Z_6\to 0$, in which case the heavier of the two $\cp$-even Higgs scalars is approximately aligned with the Higgs basis
field, $H\simeq \sqrt{2}\,{\rm Re}~\!H_1^0-v$, and $m_H\simeq Z_1 v^2$ [cf.~footnote~\ref{fn}].  In contrast to Scenario A, the decoupling limit can never be realized in Scenario B since at least one of the Higgs states is lighter than the SM-like Higgs boson.}

Similarly to Scenario A, the hybrid basis quartic parameters are chosen to decouple the other two Higgs states, $A$ and $H^\pm$, to a sufficient degree not to affect the phenomenology (here: $Z_4=Z_5=-5$). This is to ensure compatibility with the strong indirect constraints (from flavor physics) on a light charged Higgs boson with mass of order $m_H$, in particular for Type-II Yukawa couplings. In the parameter space region $90<\mh<120 \gev$ LHC constraints (from $h\to bb, \tau\tau$) apply, which leads to an upper limit on $\tan\beta$. This limit depends only weakly on $\cbma$ in the limit $\cbma\ll 1$. We focus here on the benchmark line with $\tan\beta=1.5$ as a function of $m_h$, although it is straightforward to generalize this analysis to a benchmark plane by varying $\tan\beta$. 

\begin{figure}[h!]
\centering
\includegraphics[width=0.4\columnwidth]{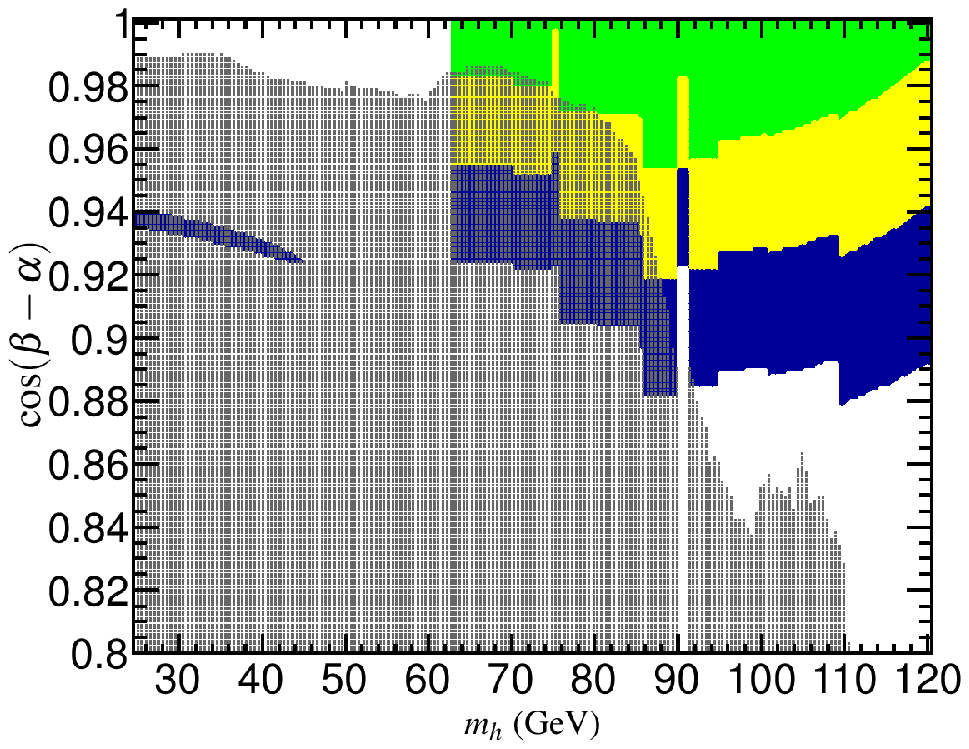}
\includegraphics[width=0.4\columnwidth]{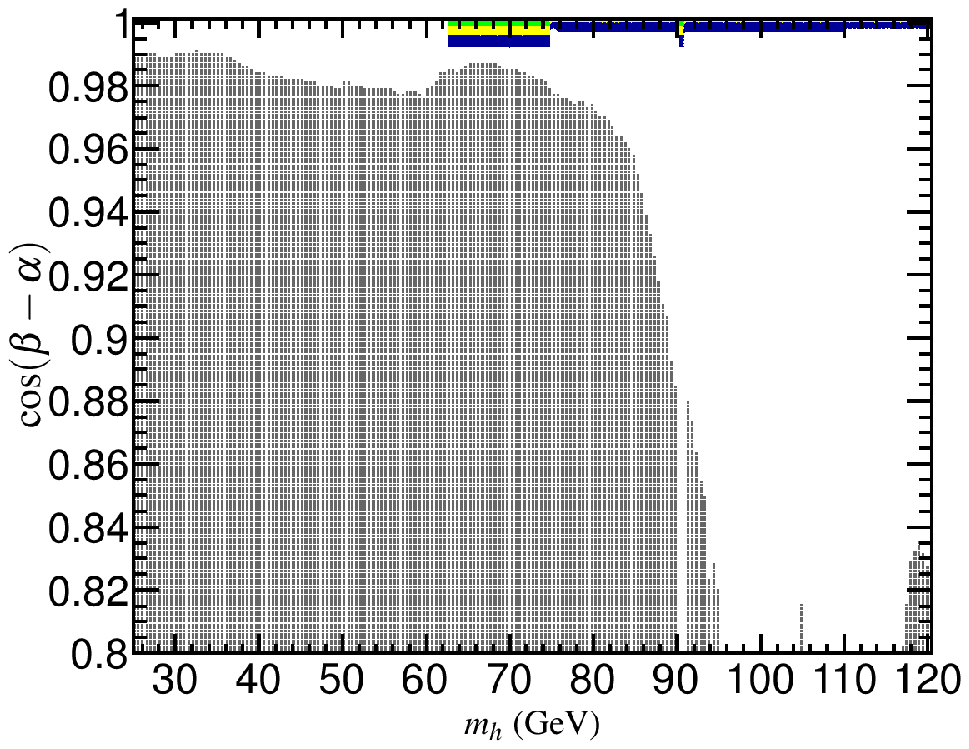}
\vskip -0.1in
\caption{Allowed parameter regions for the lightest 2HDM Higgs boson in Scenario B with Type-I Yukawa couplings (left) and Type-II couplings (right). The colors indicate statistical compatibility with the 125 GeV signal at $1\,\sigma$ (green), $2\,\sigma$ (yellow) and $3\,\sigma$ (blue).  The gray (dashed) region is excluded at $95\%$ C.L.~by constraints from direct searches at LEP and the LHC. }
\label{fig:lhcconst_B}
\end{figure}

Since this scenario implements a completely different interpretation of the $125\GeV$ signal, the assumptions for the parameter space fit presented in Fig.~\ref{fig:lhcconst} are not fulfilled. In Fig.~\ref{fig:lhcconst_B} we therefore show the corresponding results with the assumption of $m_H\simeq 125\gev$ for a region of parameter space near the SM limit, $|\cbma|\to 1$.
From this figure it can also be seen that the region where $H\to hh$ is open is very hard to reconcile with LHC measurements, we will therefore not consider this possibility any further.

As suitable benchmark scenarios we select the case of exact alignment, $\cbma=1$, with either Type-I or Type-II Yukawa couplings. As a third scenario, we choose a non-aligned value of $\cbma=0.9$, which however is only relevant for Type-I couplings. 
The dominant decay modes of the light, non-SM-like, Higgs bosons are similar for all the three different scenarios,  with $\mathrm{BR}(h\to b\bar{b})\sim 75$--$80\%$ and $\mathrm{BR}(h\to \tau^+\tau^-)\sim 8\%$ the two most interesting from the phenomenological point of view. 

\subsection{Scenario C (\cp-overlap)}
In this work we have restricted ourselves to benchmarks for a 2HDM Higgs sector with $\cp$-conservation, while a detailed analysis of the hybrid basis extended to the \cp-violating case is postponed to future work. Nevertheless, we now consider a scenario where overlapping \cp-odd and \cp-even Higgs bosons simultaneously have mass close to $125$~GeV \cite{Ferreira:2012nv}. Since the \cp-odd Higgs boson does not couple to vector bosons at tree level, there are surprisingly few channels where it is possible to distinguish this scenario from the case with a single light Higgs, $h$. The most important channel where the \cp-odd contribution to the total rate could reach $\mathcal{O}(1)$ is through gluon ($b\bar{b}$) fusion, followed by the decay $h/A\to \tau^+\tau^-$. We shall therefore analyze this process in more detail.

Requiring that $\mha=\mhl\,(=125\gev)$ in the hybrid basis of parameters, \eq{ma2} yields  
\begin{equation} \label{hy1}
 Z_5=\frac{(m_H^2-m_h^2)\ssqbma}{v^2}.
\end{equation}
Requiring in addition that $m_{H^\pm}=m_H$, \eq{mhpm2} implies that
\begin{equation} \label{hy2}
 Z_4=-Z_5-\frac{2(m_H^2-m_h^2)\csqbma}{v^2}.
\end{equation}
The remaining quartic parameter, $Z_7$, does not enter in the mass determination. To maintain a SM-like $h$, we focus on the alignment limit and set $\cbma=0$. In this case,
\eqs{hy1}{hy2} reduce to
\begin{equation}
Z_5=-Z_4=\frac{m_H^2-m_h^2}{v^2}.
\end{equation}
We fix $Z_7$, which has only minor impact on the phenomenology, such that $m_{12}^2 = \mA^2\sb\cb$, or equivalently $\lambda_5=0$. Setting $\mH=300\GeV$, this leaves $\tanb$ as the only remaining free parameter in Scenario C.  Other choices for $m_H$ (or even varying its value continuously) would lead to a benchmark plane generalization of Scenario C. However, varying $m_H$ has no impact on the properties of the overlapping $\cp$ states ($h$ and $A$).

\begin{figure}[t!]
\centering
\includegraphics[width=0.45\columnwidth]{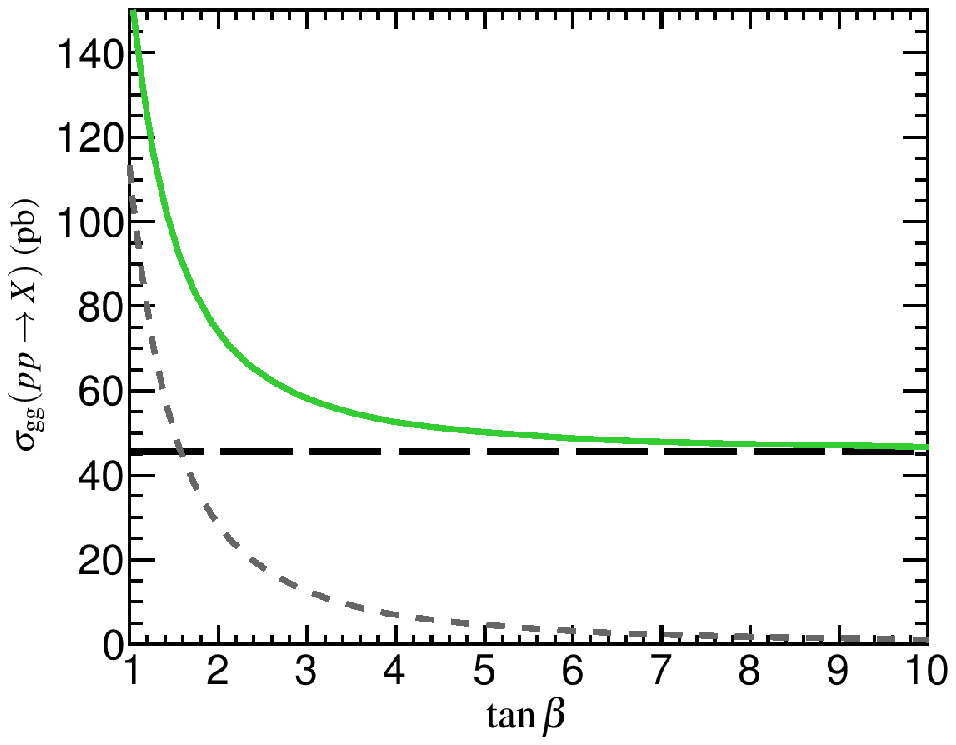}
\includegraphics[width=0.45\columnwidth]{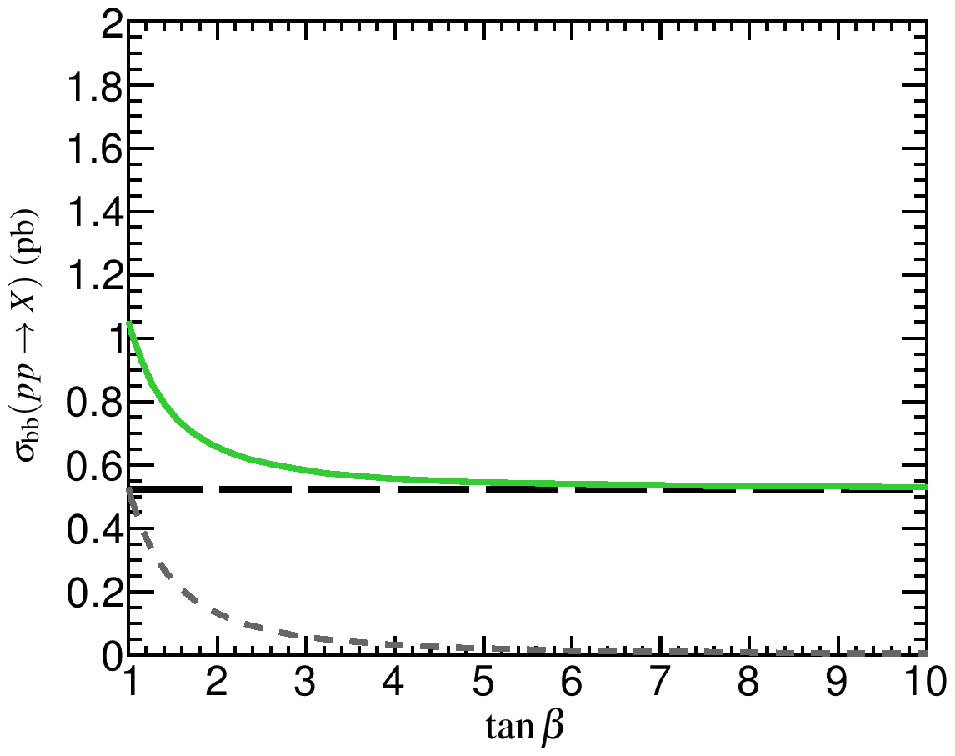}
\caption{Hadronic cross sections at $\sqrt{s}=13\TeV$ for production of the light \cp-even Higgs boson (long dashes), the \cp-odd Higgs $A$ (short dashes), and their sum (green, solid) in Scenario C with Type-I Yukawa couplings.}
\label{fig:C_ggh_I}
\end{figure}

\begin{figure}[t!]
\centering
\includegraphics[width=0.45\columnwidth]{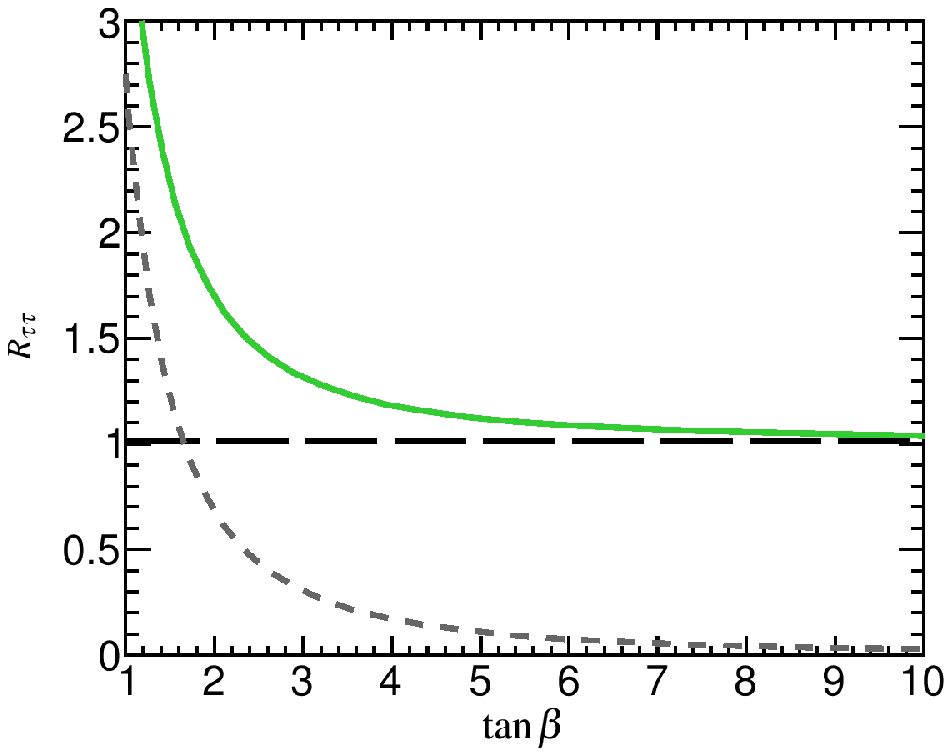}
\includegraphics[width=0.45\columnwidth]{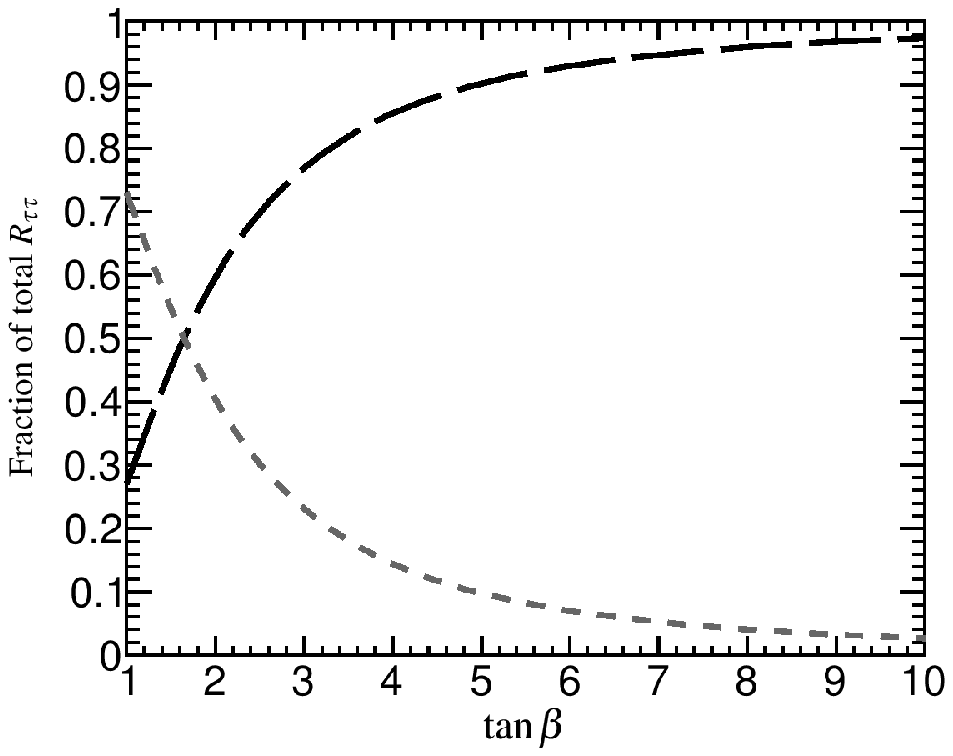}
\caption{The total $\tau\tau$ rate (adding $gg$ and $b\bar{b}$ production modes), relative to the SM, from $h$ (long dashes), $A$ (short dashes) and their sum (green, solid) in Scenario C with Type-I Yukawa couplings. Right: the respective fractions of the inclusive $\tau\tau$ rate resulting from $h$ (long dashes) and $A$ (short dashes).}
\label{fig:C_Rhtautau_I}
\end{figure}

Scenario C can be considered with both Type-I and Type-II Yukawa couplings. In Fig.~\ref{fig:C_ggh_I}, we show the production cross sections in the hadronic modes $gg\to h/A$ and $b\bar{b}\to h/A$ (and their sum) for Type-I couplings. In this case the \cp-even production dominates, except at low values of $\tanb\lesssim 1.6$. Since there is no $\tanb$ enhancement of the bottom Yukawa, the $b\bar{b}$-induced production remains small and can be neglected. Combining the (total) production cross sections ($gg+b\bar{b}$) with the branching ratios $\mathrm{BR}(h/A\to \tau\tau)$, we form the inclusive quantity 
\begin{equation}
R^{h/A}_{\tau\tau}=\frac{\sigma(pp\to h/A)\times \mathrm{BR}(h/A\to \tau\tau)}{\left[\sigma(pp\to h/A)\times \mathrm{BR}(h/A\to \tau\tau)\right]_{\mathrm{SM}}}
\end{equation}
for the $\tau\tau$ rate relative to the SM. Predictions for $R_{\tau\tau}$ in Scenario C are shown in Fig.~\ref{fig:C_Rhtautau_I}. The figure shows that the total rate (green curve) approaches the SM value ($R_{\tau\tau}=1$) in the limit of high $\tanb$ due to the decoupling property for the $A$ contribution with Type-I couplings. In the opposite limit, the \cp-odd contribution becomes increasingly important and dominates for $\tanb\lesssim 1.6$. In effect, the total predicted $\tau\tau$ rate significantly exceeds the SM prediction. Experimental constraints limit the maximal total rate, with the currently measured rates being $\mu_{\tau\tau}^{\mathrm{ATLAS}}=2.1^{+0.9}_{-0.8}$ \cite{Aad:2015vsa} and $\mu_{\tau\tau}^{\mathrm{CMS}}=0.34\pm 1.09$ \cite{Chatrchyan:2014nva}.\footnote{These results correspond to the ``boosted'' category of ATLAS and to the ``0 jet'' results of CMS, both of which are expected to be dominated by gluon fusion production.} To determine the fractions of the total rate which can be due to the \cp-even and \cp-odd 2HDM components, we show in the right panel of Fig.~\ref{fig:C_Rhtautau_I} their relative contributions $R_{\tau\tau}^i/R_{\tau\tau}^{\mathrm{tot}}$. Due to the decrease in the cross section, the \cp-odd component can be seen to decrease monotonously with increasing $\tan\beta$. This scenario therefore provides a consistent model to parametrize an arbitrarily small \cp-odd admixture in the $125\GeV$ signal.

\begin{figure}[h!]
\centering
\includegraphics[width=0.45\columnwidth]{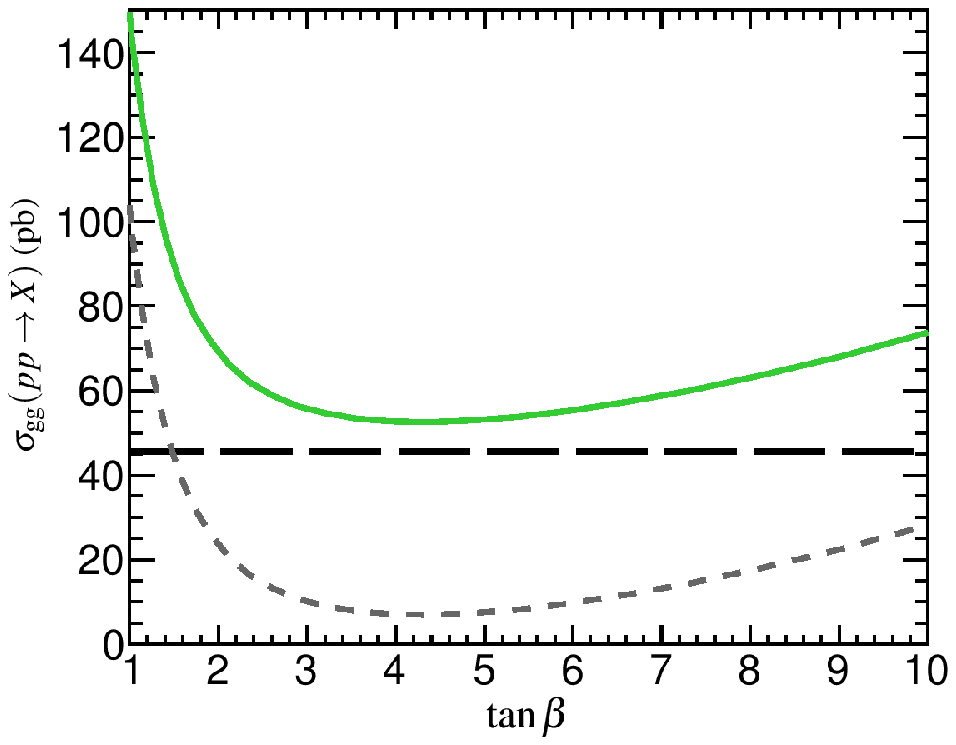}
\includegraphics[width=0.45\columnwidth]{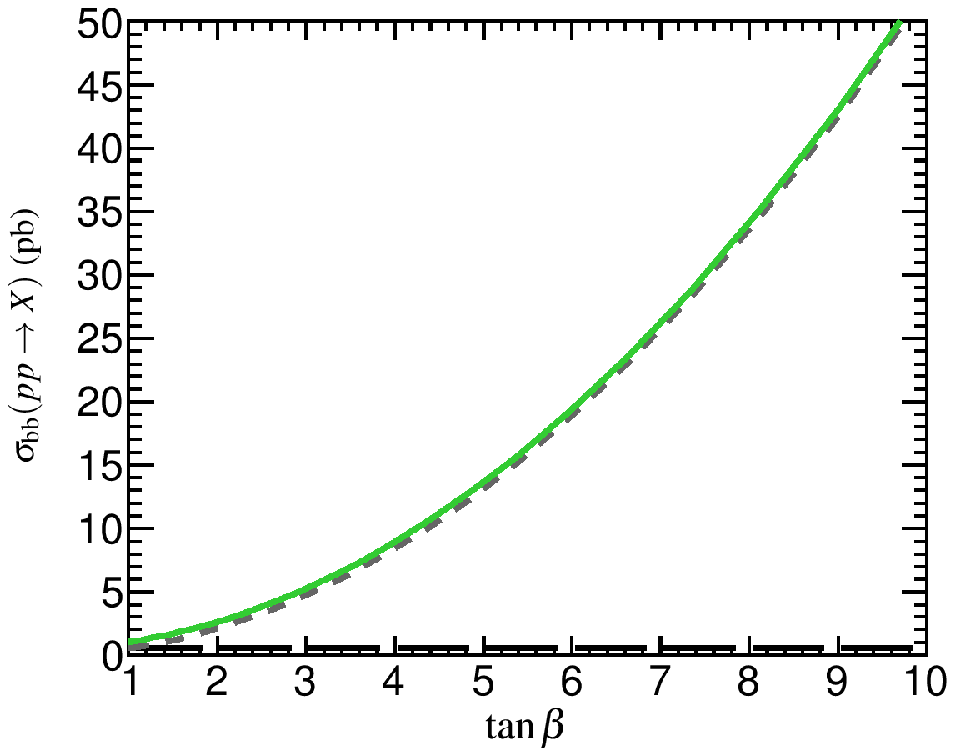}
\caption{Hadronic cross sections at $\sqrt{s}=13\TeV$ for production of the light \cp-even Higgs boson (long dashes), the \cp-odd Higgs $A$ (short dashes), and their sum (green, solid) in Scenario C with Type-II Yukawa couplings.}
\label{fig:C_ggh_II}
\end{figure}
\begin{figure}[t!]
\begin{center}
\includegraphics[width=0.45\columnwidth]{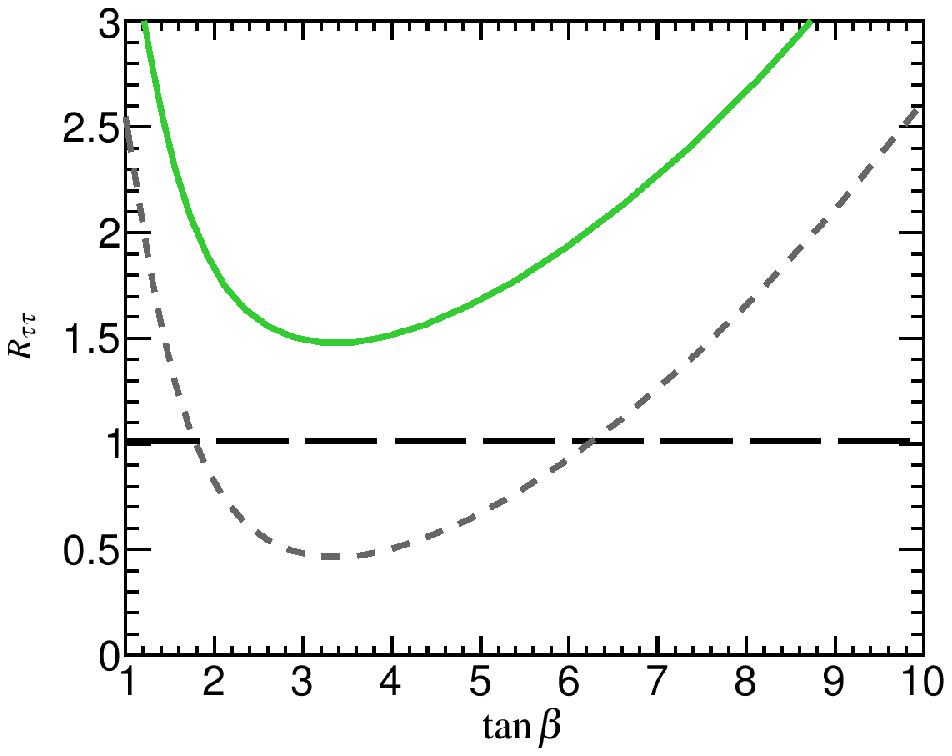}
\includegraphics[width=0.45\columnwidth]{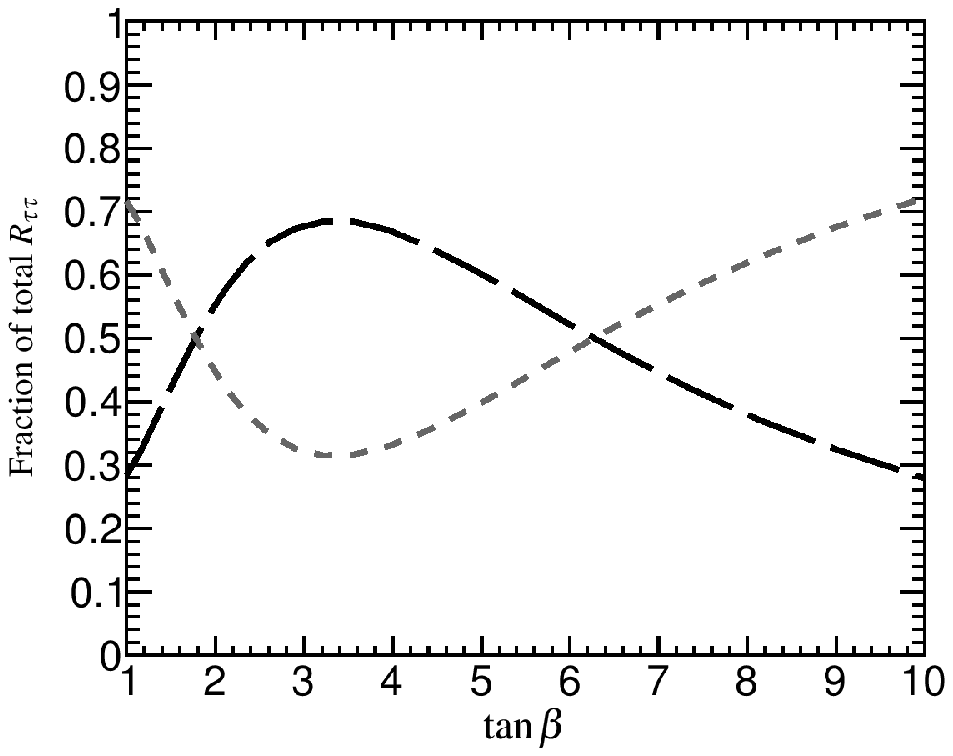}
\caption{The total $\tau\tau$ rate (adding $gg$ and $b\bar{b}$ production modes), relative to the SM, from $h$ (long dashes), $A$ (short dashes) and their sum (green, solid) in Scenario C with Type-II Yukawa couplings. Right: the respective fractions of the inclusive $\tau\tau$ rate resulting from $h$ (long dashes) and $A$ (short dashes).}
\label{fig:C_Rhtautau_II}
\end{center}
\end{figure}

With Type-II Yukawa couplings, the production cross sections for the \cp-odd state, $A$, have a different dependence on $\tanb$. The effect of this change is shown in Fig.~\ref{fig:C_ggh_II}. As can be seen from this figure the \cp-odd contribution now becomes much more important at high $\tanb$, both due to an increase in gluon fusion which has a minimum around $\tan\beta\sim 4$ and due to a contribution from the $b\bar{b}\to A$ process with a cross section increasing as $\tan^2\beta$. In the case of Type-II couplings the total cross section does not approach the SM value for \emph{any} value of $\tanb$. This also has consequences for the total $\tau\tau$ rate, as can be seen in Fig.~\ref{fig:C_Rhtautau_II}. In this scenario the minimum rate, obtained for $\tan\beta\sim 3.5$, is  $R_{\tau\tau}\simeq 1.5$. While $50\%$ above the SM expectation, this is not experimentally excluded. Interestingly, the similar magnitude of the \cp-even and \cp-odd contributions leads to fractions that are nearly equal over a large range in $\tan\beta$ (Fig.~\ref{fig:C_Rhtautau_II}, right). This scenario is therefore particularly suitable as a ``best-case'' test of the possibility to distinguish \cp-properties of the $125\GeV$ Higgs boson in the $\tau\tau$ channel.

\subsection{Scenario D (Short cascade)}

This scenario is constructed with a SM-like $h$ by fixing $\cbma$ to be zero (exact alignment). The mass hierarchy can be modified allow for either one---or both---of the decay modes $H\to ZA$ or $H\to W^\pm H^\mpÊ$ to be open, thus resulting in a ``small cascade'' of Higgs-to-Higgs decays.\footnote{Such decay modes have also been considered recently in \cite{Coleppa:2014hxa}.}
These decays can typically be made dominant in the mass window $250\GeV < \mH < 350\GeV$ (below $t\bar{t}$ threshold).   A recent search for $H\to ZA$ by the CMS 
Collaboration\cite{Khachatryan:2016are} 
already places interesting constraints on the parameter space of this type of scenario.
Other modes that can be potentially of simultaneous interest is $H\to hh$ and $H\to AA$ (when $A$ is very light). 

We present realizations of Scenario D for all the interesting cases below. We define these scenarios with a single free parameter by fixing $\tan\beta=2$, but this assumption could easily be relaxed.  For simplicity, we choose two of the three non-SM-like Higgs masses to be equal.  In the hybrid basis of parameters, these mass degeneracies can be implemented by an appropriate choice of $Z_4$ and $Z_5$ as follows [cf.~\eqs{ma2}{mhpm2}]:
\beqa
m_H=m_{H^\pm}\,\,\text{and}\,\,\cbma=0 \quad & \Longrightarrow & \quad Z_4=-Z_5\,,\\
m_H=m_A\,\,\text{and}\,\,\cbma=0 \quad & \Longrightarrow & \quad Z_5=0\,,\\
m_A=m_{H^\pm}\quad & \Longleftrightarrow & \quad Z_4=Z_5\,.
\eeqa

In addition to the short cascade that we are investigating, additional ``exotic'' decay modes may be accessible, such as e.g.~$A\to W^\pm H^\mp $ (when $H^\mp$ is light).  However, with the possibility of non-degeneracy between the heavy Higgs bosons ($H$ and $A$ in this case) there is in principle no guarantee for these other exotic decay modes to co-exist at any appreciable rate. We therefore choose to discuss only the $H$ decays here, but there could be additional modes of similar type to exploit.

\begin{figure}[t!]
\centering
\begin{overpic}[width=0.48\columnwidth]{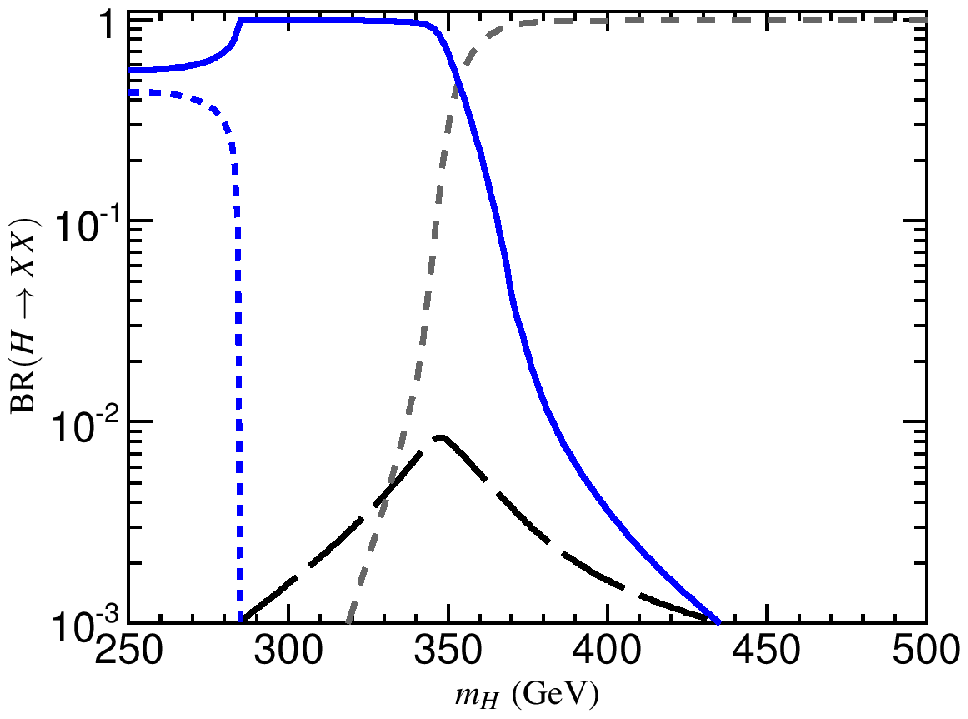}
\put (68,67) {\small $t\bar{t}$}
\put (30,67) {\small $ZA$}
\put (28,40) {\small $AA$}
\put (44,22) {\small $b\bar{b}$}
\end{overpic}
\begin{overpic}[width=0.48\columnwidth]{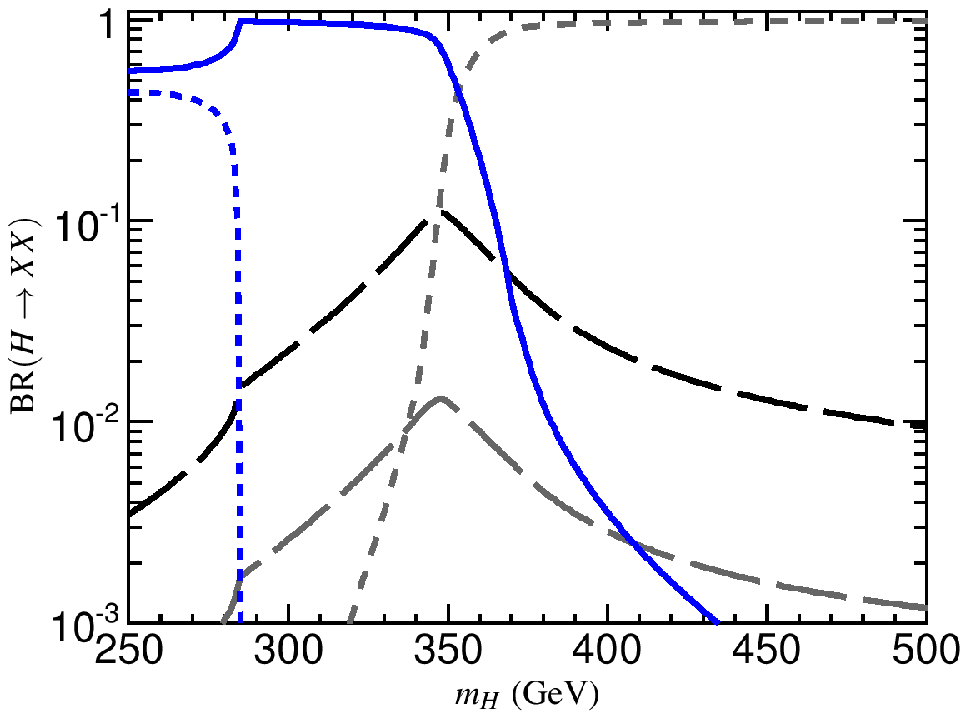}
\put (68,67) {\small $t\bar{t}$}
\put (30,67) {\small $ZA$}
\put (28,50) {\small $AA$}
\put (80,36) {\small $b\bar{b}$}
\put (80,18) {\small $\tau^+\tau^-$}
\end{overpic}
\caption{Branching ratios of $H$ in Scenario D (with low $\mA$) for $\tan\beta=2$ with Type-I (left) and Type-II Yukawa couplings. The colors show $H\to Z A$ (blue, solid), $H\to AA$ (blue, short dash), $H\to t\bar{t}$ (gray, dash) and $H\to b\bar{b}$ (black, long dash) and $H\to \tau\tau$ (gray, long dash).}
\label{fig:D_MA_BRH}
\end{figure}

Starting with the case of only a low $\mA$, this can be realized in the \HH\ basis of parameters by choosing $Z_4=-1$, $Z_5=1$. For $\mH$ close to $250\GeV$, the decay $H\to AA$ can be open, with a rate that can be adjusted by varying $Z_7$. Here we choose $Z_7=-1$, which satisfies stability requirements in the whole mass range.\footnote{In contrast, the opposite sign choice, $Z_7=1$ leads to problems with positivity of the scalar potential.} Fig.~\ref{fig:D_MA_BRH} shows the branching ratios of the heavy $\cp$-even Higgs boson, $H$, into the interesting final states. As this figures shows, there is nearly no difference in the decay pattern for the two Yukawa types (although, of course, the production cross sections are different). One thing that can be noteworthy is that the cascade decay, in this case $H\to Z A$ maintains an appreciable branching ratio also beyond the top threshold, with percent-level rates up to $\mhh\sim 380\GeV$. In Type-II models it is possible to suppress $H\to t\bar{t}$ further by going to higher $\tanb$, but we find for the chosen set of parameters that this requires adjusting $Z_7$ to maintain positivity of the potential.
\begin{figure}[h!]
\centering
\begin{overpic}[width=0.48\columnwidth]{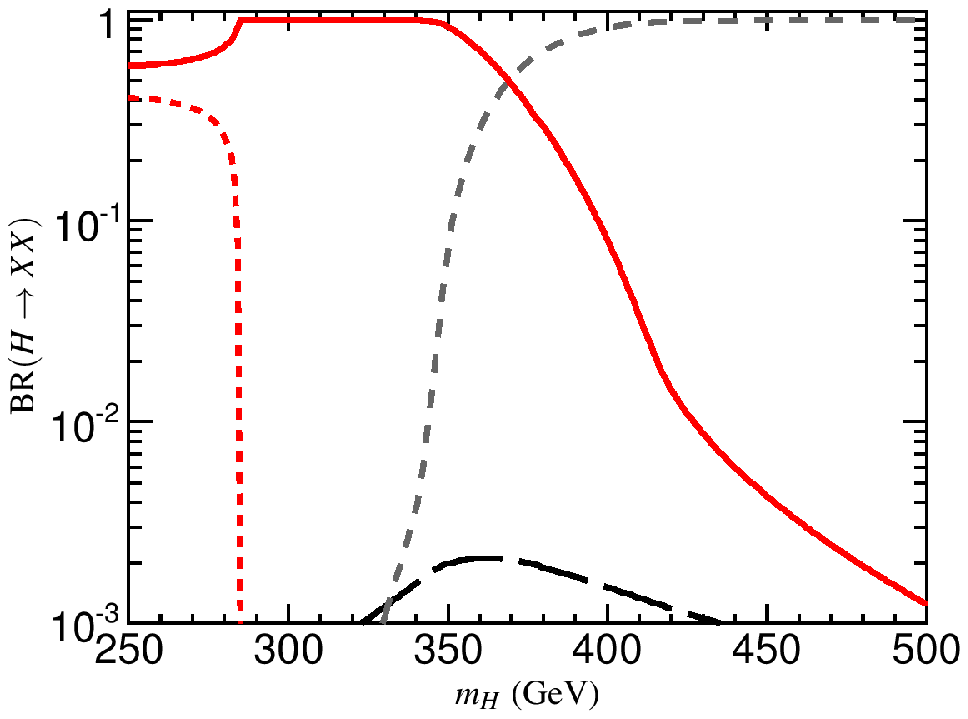}
\put (75,67) {\small $t\bar{t}$}
\put (30,67) {\small $W^\pm H^\mp$}
\put (28,40) {\small $H^+H^-$}
\put (48,20) {\small $b\bar{b}$}
\end{overpic}
\begin{overpic}[width=0.48\columnwidth]{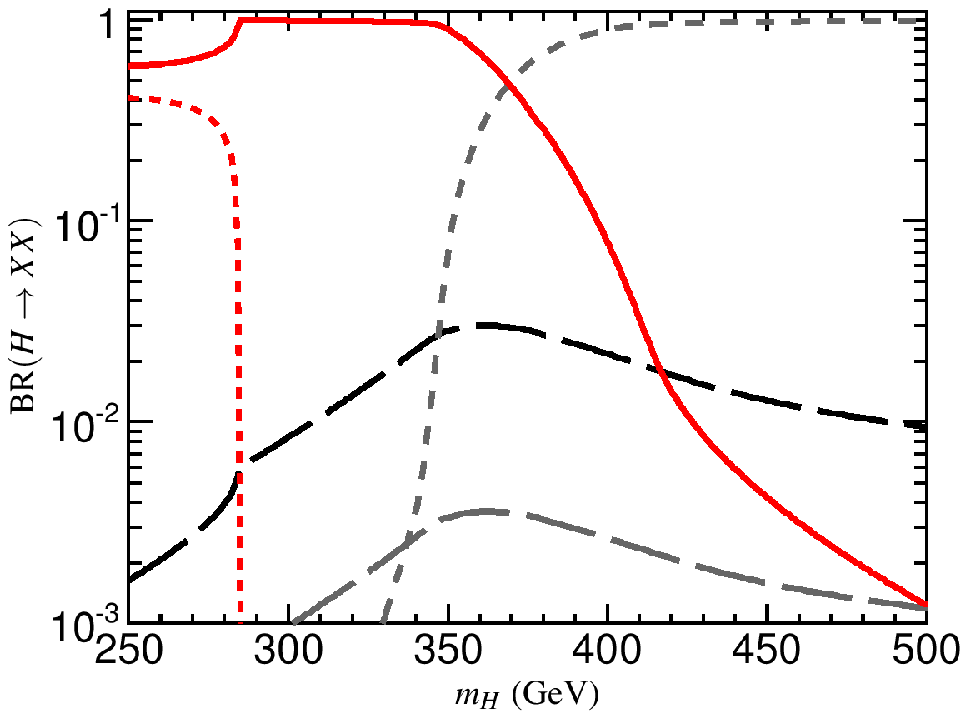}
\put (68,67) {\small $t\bar{t}$}
\put (30,67) {\small $W^\pm H^\mp$}
\put (28,50) {\small $H^+H^-$}
\put (80,36) {\small $b\bar{b}$}
\put (65,22) {\small $\tau^+\tau^-$}
\end{overpic}
\caption{Branching ratios of $H$ in Scenario D (with low $\mHp$) for $\tan\beta=2$ with Type-I (left) and Type-II Yukawa couplings. The colors show $H\to W^\pm H^\mp$ (red, solid), $H\to H^+H^-$ (red, short dash), $H\to t\bar{t}$ (gray, dash) and $H\to b\bar{b}$ (black, long dash) and $H\to \tau\tau$ (gray, long dash).}
\label{fig:D_MHp_BRH}
\end{figure}

An alternative mass hierarchy with $\mHp < \mA=\mH$ can be arranged by setting $Z_4=2$ and $Z_5=0$ while keeping the remaining parameters fixed. This generates the new possible decay modes $H\to W^\pm H^\mp$ (and $H\to H^+H^-$ for very low $\mHp$). Note that these low $\mHp$ might be in a region which is disallowed by flavor constraints. 
The decay branching ratios in this scenario are shown in Fig.~\ref{fig:D_MHp_BRH}. Similarly to the case with light $A$, the cascade decay $H\to W^\pm H^\mp$ is dominant over the range $250<\mH<350\GeV$, and with a sizable tail towards higher masses. In this case percent-level branching ratios are obtained for $\mH \simeq 420 \GeV$.

\begin{figure}[t!]
\centering
\begin{overpic}[width=0.48\columnwidth]{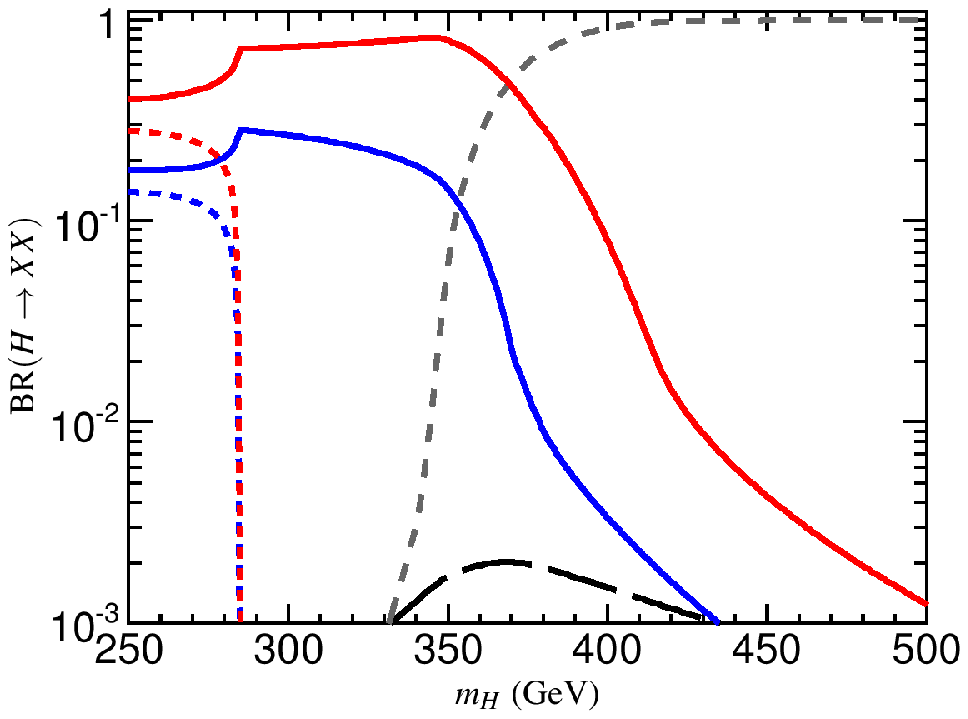}
\put (75,67) {\small $t\bar{t}$}
\put (30,64) {\small $W^\pm H^\mp$}
\put (30,54) {\small $ZA$}
\put (28,45) {\small $H^+H^-$}
\put (16,46) {\small $AA$}
\put (48,20) {\small $b\bar{b}$}
\end{overpic}
\begin{overpic}[width=0.48\columnwidth]{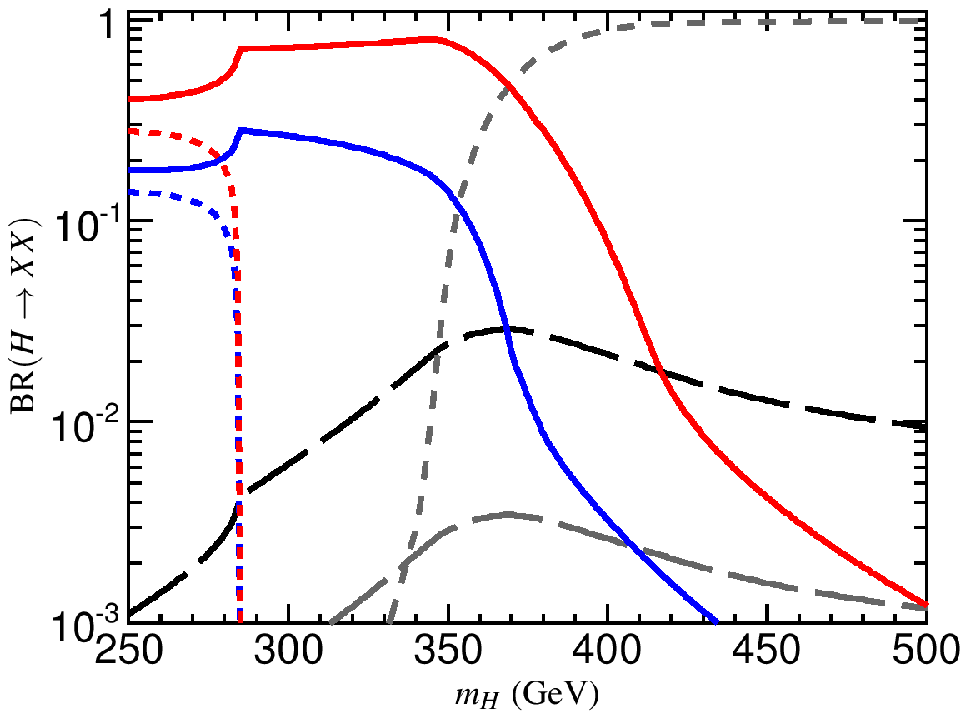}
\put (75,67) {\small $t\bar{t}$}
\put (30,64) {\small $W^\pm H^\mp$}
\put (30,54) {\small $ZA$}
\put (28,45) {\small $H^+H^-$}
\put (16,46) {\small $AA$}
\put (80,35) {\small $b\bar{b}$}
\put (65,22) {\small $\tau^+\tau^-$}
\end{overpic}
\caption{Branching ratios of $H$ in Scenario D with both $\mA$ and $\mHp$ low ($Z_4=Z_5$) for $\tan\beta=2$ with Type-I (left) and Type-II (right) Yukawa couplings. The color coding is the same as used in Figs.~\ref{fig:D_MA_BRH} and \ref{fig:D_MHp_BRH}.}
\label{fig:D_Mboth_BRH}
\end{figure}

Finally, by setting $Z_4=Z_5$ the hierarchy of masses becomes $\mA=\mHp < \mH$, with the light $\cp$-odd and charged Higgs degenerate in mass. In this case all the different cascade modes for $H$ can be open at the same time. The resulting branching ratios for the chosen scenario is shown in Fig.~\ref{fig:D_Mboth_BRH}. Note that the widths for the leading-order predictions for the decays involving one Higgs boson and one gauge bosons, e.g.~$H\to Z A$, are proportional to gauge couplings (and therefore relatively fixed), whereas the modes involving triple-Higgs couplings, such as $H\to AA$, are proportional to scalar couplings appearing in the Higgs potential.   For example,\footnote{A list of the triple Higgs couplings expressed in terms of the Higgs basis quartic parameters can be found in \cite{Bernon:2015qea}.}
\beq
 g\ls{\hh\ha\ha} =
   {-v}\bigl[(Z_3+Z_4-Z_5)\cbma-Z_7\sbma\bigr]\,.
\eeq
The strength of such modes is therefore more sensitive to the specific choice of the 2HDM parameters.

\subsection{Scenario E (Long cascade)}

The short cascade (Scenario D) was defined with a one-step decay involving two (or more) Higgs bosons. In Scenario E, we extend this to a ``long cascade'' where two-step decays involving all three non-SM-like Higgs bosons are possible, assuming that 
$H$ is always the lighter of the three. Assuming first (E1) that $\hpm$ is heavier than both $A$ and $H$ (realised in the hybrid basis by choosing $Z_4=-6$, $Z_5=-2$ (with $Z_7$ always zero in this scenario), can give rise to a ``long'' cascade and a second complementary direct decay that is always present~\cite{Coleppa:2014cca},
\begin{equation}
\begin{aligned}
H^\pm &\to W^\pm\, A  \to W^\pm\, Z\, H  \\
H^\pm &\to W^\pm\, H.
\label{eq:long1}
\end{aligned}
\end{equation}
The other hierarchy, with $\mA > \mHp > \mH$ is achieved by setting $Z_4=1$, $Z_5=-3$ (E2). This leads to long cascades which are inverted compared to Eq.~\eqref{eq:long1}~\cite{Coleppa:2014hxa}:
\begin{equation}
\begin{aligned}
A &\to  W^\pm\, H^\mp \to W^\pm  \, W^\mp \,H \\
A &\to Z\, H.
\end{aligned}
\end{equation}
Interesting signatures in this scenario include 
a heavy Higgs boson $H$ in the $\tau^+\tau^-$ or $b\bar{b}$ channels augmented with multiple leptons and/or additional jets. 

As in scenario D, the recent search for $A\to ZH$~\cite{Khachatryan:2016are} places constraints on the parameter space. It should be interesting to analyze the interplay between this mode and other available signatures in this scenario in more detail.
In Table~\ref{tab:ScenarioE} we present the branching ratios for the various cascade decay modes in the two different incarnations of Scenario E  for input masses $\mH=200\GeV$ and $\mH=300\GeV$ 
with Type-I Yukawa couplings.  
Note that for this low value of $\tan\beta$ there is no significant change in any of these branching ratios in a Type-II setting. The main difference between the two types are instead in the decay of $H$ at the end of the cascade, where the final states with down-type fermions ($b$, $\tau$) are 
more favored with Type-II couplings.
\begin{table}[t!]
\centering
\begin{tabular}{|c|ccc|cccc|}
\hline
 & \multicolumn{3}{|c|}{Masses (GeV)} &  \multicolumn{4}{|c|}{Branching ratios}\\
Scenario & $\mH$ & $\mA$ & $\mHp$ & $H^\pm \to W^\pm\, A$ & $H^\pm \to W^\pm\, H$ & $A \to Z\, H$  & $A \to W^\pm\, H^\mp$\\
\hline
E1.1 & $200$ & $402$ & $532$ & 0.053 & 0.79 & 0.62 & --  \\
     & $300$ & $460$ & $577$ & 0.041 & 0.74 &  0.39 & --  \\
     \hline
E1.2 & $200$ & $471$ & $317$ & -- & 0.27 & 0.56 & 0.25  \\
     & $300$ & $521$ & $388$ & -- & 0.026 &  0.50 & 0.20  \\
\hline
\end{tabular}
\caption{Mass spectrum and Type-I Higgs branching ratios to bosonic decay modes in Scenario E.}
\label{tab:ScenarioE}
\end{table}

There are several things of generic interest that can be noted from Table~\ref{tab:ScenarioE}. The heaviest (parent) Higgs boson has a rather high probability, $70$--$85\%$, to decay into a lighter Higgs and a vector boson. However, most of this decay goes directly into the lighter of the two states ($H$) due to the larger available phase space. The aggregated branching ratio for a ``long'' cascade is thus suppressed, reaching typical values at the (few) percent-level (up to $5$--$6\%$ in some cases). Nevertheless, to be able to study these types of final states would be an intriguing possibility, that would contain a lot of information about the structure of an extended Higgs sector.

\subsection{Scenario F (Flipped Yukawa)}
The flipped Yukawa scenario is characterized by SM-like couplings for the light Higgs, $h$, except for the couplings to down-type fermions which has a change of sign relative to the SM. This scenario is realized with Type-II Yukawa couplings for values of $(\cbma,\tanb)$ solving the equation \cite{Ferreira:2014naa}
\begin{equation}
\frac{g_{hdd}}{g_{hdd}^{\mathrm{SM}}}=-\frac{\sin\alpha}{\cos\beta}=\sba-\tanb\cba=-1.
\end{equation}
This unconventional solution is what generates the second branch of $2\,\sigma$-allowed parameter space in Fig.~\ref{fig:lhcconst} (right), with valid solution for not too small values of $\tan\beta$. We define Scenario F by simply fixing the value of $\cbma$ from the chosen $\tan\beta$ using this relation, while the remaining parameters are given the same values as in Scenario A: $Z_4=Z_5=-2$, $Z_7=0$. 

\begin{figure}[t!]
\includegraphics[width=0.42\columnwidth]{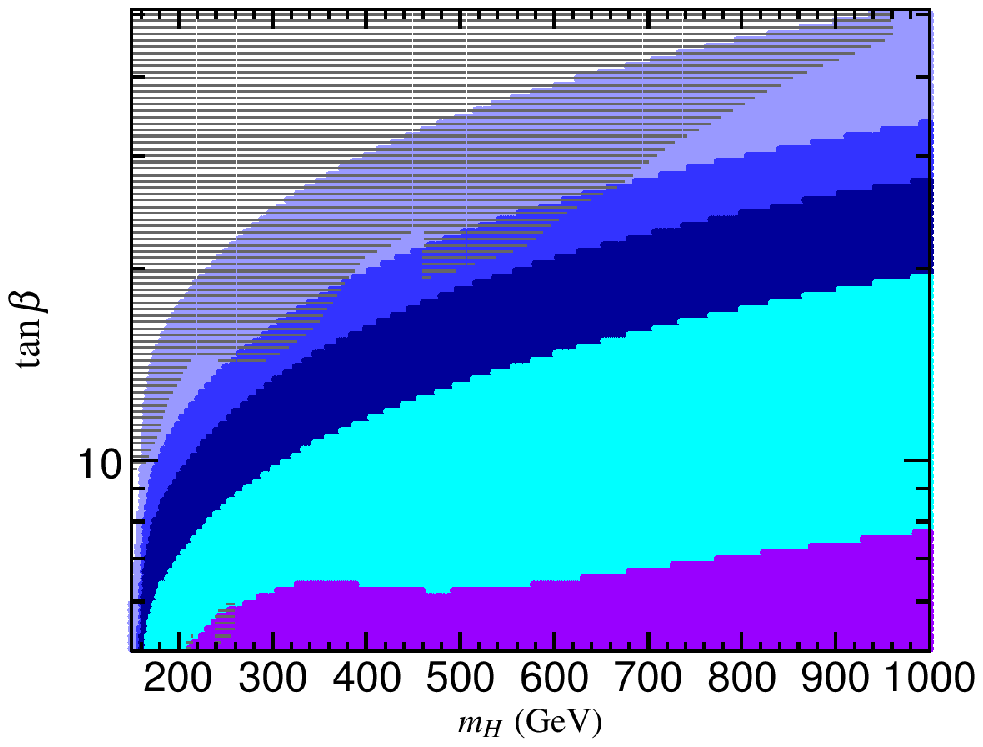}
\includegraphics[width=0.42\columnwidth]{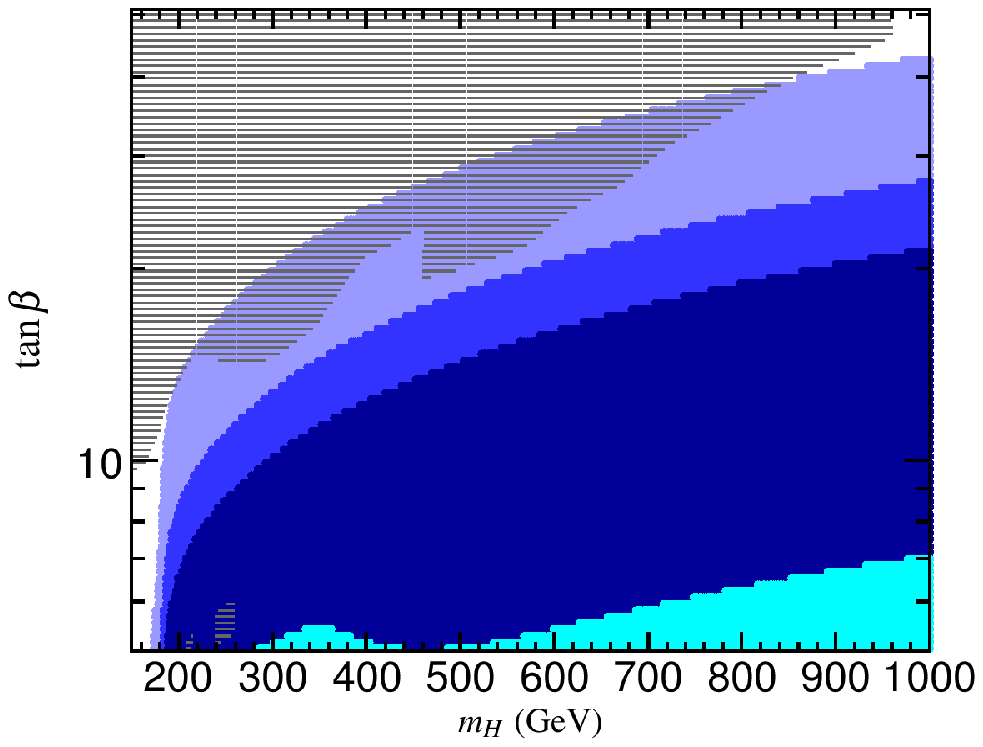}\\
\vspace{-0.75em}
\includegraphics[width=0.42\columnwidth]{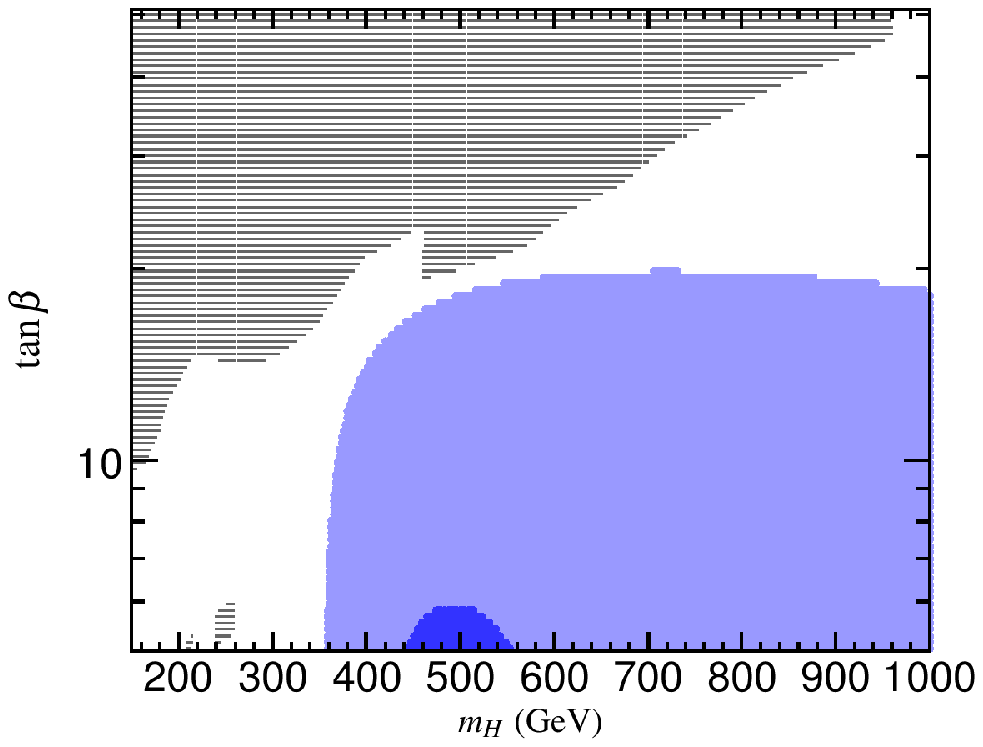}
\includegraphics[width=0.42\columnwidth]{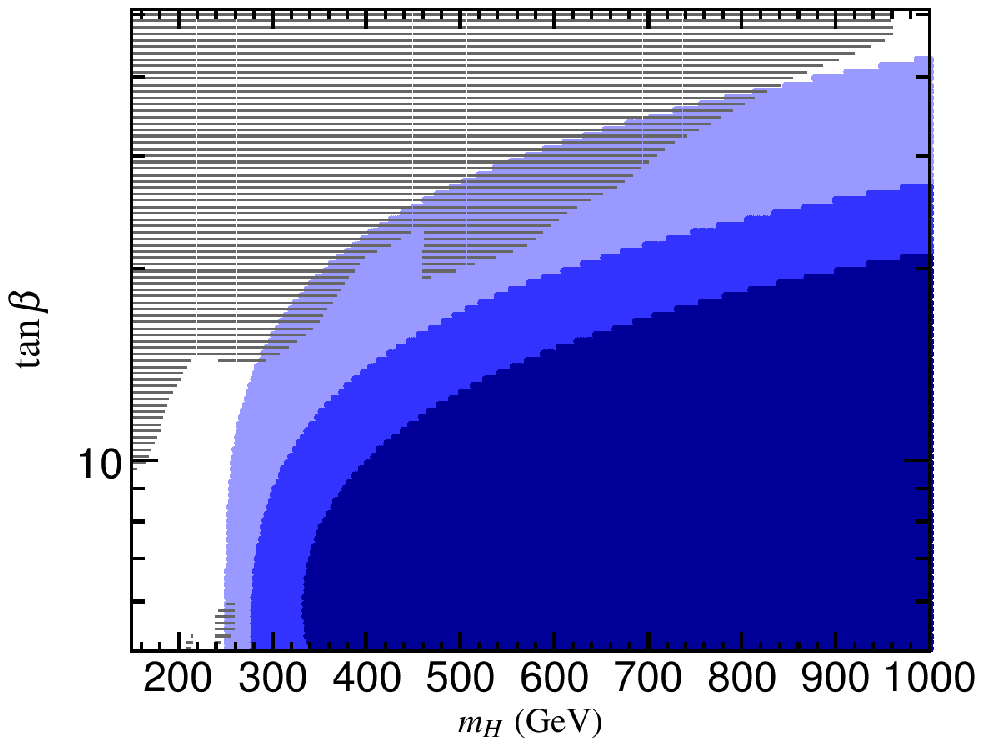}
\includegraphics[width=0.085\columnwidth]{legend}
\caption{Branching ratios of $H$ in Scenario F. The panels show $H\to WW$ (upper left), $H\to ZZ$ (upper right), $H\to t\bar{t}$ (lower left) and $H\to hh$ (lower right). Contours for each branching ratio are indicated with colors (see legend). Gray regions are excluded at $95\%$ C.L.~from direct searches.}
\label{fig:F_BRH}
\end{figure}

The (small) effects on the $h$ phenomenology induced by the opposite sign bottom Yukawa coupling have been studied in detail in \cite{Ferreira:2014naa}. However, the comparably large allowed values of $\cbma$ also open possibilities for the decay modes of the heavier Higgs boson beyond what can be achieved near the alignment limit as considered, for example, in Scenario A. To illustrate this point, we show in Fig.~\ref{fig:F_BRH} the decay branching ratios of $H$ as function of $\mH$ and $\tanb$. As this figure shows, the decays into vector bosons can be sizable over a large part of the remaining parameter space, clearly beyond what is possible in the alignment case. This observation holds even in the region above the top threshold, which makes the Flipped Yukawa scenario a suitable benchmark to replace the SM in interpretations of future heavy Higgs searches in the dilepton and $4\ell$ final states over a larger mass range.

\subsection{Scenario G (MSSM-like)}
This scenario is inspired by the Higgs potential of the MSSM. The tree-level MSSM is defined by the following values for the quartic couplings (in the ``SUSY-basis'' where supersymmetry is manifest),
\begin{equation}
\lambda_1=\lambda_2=\tfrac{1}{4}(g^2+g'^2),\quad \lambda_3=\tfrac{1}{4}(g^2-g'^2),\quad \lambda_4=-\half g^2,\quad\lambda_5=\lambda_6=\lambda_7=0.
\label{eq:MSSM}
\end{equation}
Following Eq.~\eqref{mbar}, the remaining parameter is given by $m_{12}^2 = \half\mA^2 \sin 2\beta$ 
in terms of the more usual MSSM inputs: the $\cp$-odd Higgs mass, $\mA$, and $\tan\beta$.
In the tree-level MSSM, the predicted mass of the lightest Higgs boson, $\mh<m_Z$, is not compatible with the LHC measurements. However, the $\cp$-even mass matrix receives corrections beyond leading order. To a good approximation, the leading radiative corrections can be parametrized as an additional contribution to $\lambda_2$ \cite{Haber:1993an}, corresponding to the shift $\lambda_2\rightarrow \lambda_2+\delta$, while (sub-leading) MSSM contributions to the other quartic couplings are neglected. This is reminiscent of the approach pursued in \cite{Djouadi:2013uqa}. Since the leading radiative corrections in the MSSM can most easily be specified in terms of the SUSY-basis, we use this instead of hybrid basis input for Scenario G.
Working in this approximation, we define our MSSM-inspired 2HDM scenario by three parameters:
$m_h$, $m_A$ and $\tan\beta$.
Using this input, together with the MSSM relations for the $\lambda_i$~($i\neq 2$) given by Eq.~\eqref{eq:MSSM}, we solve for the value of $\delta$ necessary to reproduce the desired value of $\mh$ for the chosen values of $\mA$ and $\tan\beta$. In practice, this is done with \thdmc\ using an iterative procedure.\footnote{This procedure can be accessed through the {\tt set\_hMSSM} method or by running the {\tt CalcHMSSM} example program.} To satisfy the constraints of unitarity and perturbativity, we impose the additional condition $\lambda_2<4\pi$. 

Although this scenario is \emph{inspired} by the MSSM, it is (like all our other benchmarks) completely defined within the $\cp$-conserving, softly-broken $\mathbb{Z}_2$-symmetric 2HDM. There is therefore no principal restriction to the Type-II structure for the Yukawa couplings, although we choose to stick to this familiar pattern here to keep the connection to the MSSM tree-level structure.

\begin{figure}[t!]
\centering
\includegraphics[width=0.5\columnwidth]{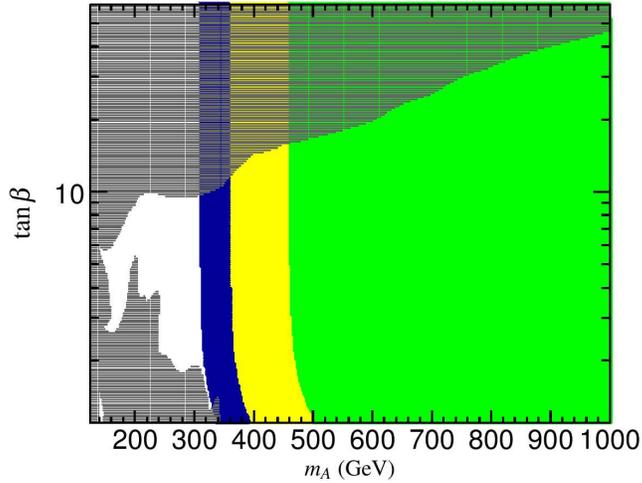}
\caption{Allowed parameter space by direct Higgs search constraints in the ``MSSM-like'' 2HDM with Type-II Yukawa couplings. The color coding is the same as in Fig.~\ref{fig:lhcconst}.}
\label{fig:E_mA_tanb}
\end{figure}
Using a fixed value $\mh=125\GeV$, we scan over the remaining parameters $\mA$ and $\tan\beta$ to determine the viable regions of parameter space. The results are shown in Fig.~\ref{fig:E_mA_tanb}, where green color indicates regions compatible with $\mh=125\GeV$ (and $\lambda_2<4\pi$). The shaded (gray) regions show the excluded regions at $95\%$ C.L.~from direct Higgs searches. In particular the limit $H/A\to \tau\tau$ plays a very important role to constrain this scenario both for high and low values of $\tan\beta$ for values $\mA<2\,m_t$. Constraints from heavy Higgs searches benefit in this scenario from the near-degeneracy of the heavy Higgs bosons $H$ and $A$. 

The resulting phenomenology of this scenario is very similar to the MSSM Higgs sector in the absence of additional low-energy degrees of freedom, which in the MSSM could provide additional decay channels for the heavy non-SM-like Higgs bosons. This can be seen, for example, from the region allowed by the LHC Higgs measurements in Fig.~\ref{fig:E_mA_tanb}. In the absence of large radiative corrections affecting the \cp-even Higgs mixing \cite{Carena:2014nza}, the LHC Higgs data forces the MSSM into the decoupling limit. Hence, we find $\mA>360\GeV$ at $95\%$  C.L. almost independently of $\tan\beta$.  A related analysis in the MSSM context at low values of $\tan\beta$ with a heavy supersymmetric spectrum that makes use of the framework described above has recently been presented in \cite{LHCHXSWG-2015-002}.

\subsection{Additional scenarios for consideration}

There are a number of additional scenarios worthy of consideration that we have not included in this work.  For example, one can consider the inert 2HDM with Type-I Yukawa couplings, which provides a plausible dark matter candidate~\cite{LopezHonorez:2006gr,LopezHonorez:2010tb,Gustafsson:2012aj,Goudelis:2013uca,Krawczyk:2013jta}.   As noted at the end of Section~\ref{sect:H2}, phenomenological constraints and benchmark scenarios for this model have been recently discussed in \cite{Ilnicka:2015sra}.

It is also possible to consider simplified models of 2HDM phenomena, where $h$ is SM-like and one of the non-SM-like Higgs states among $H$, $A$ and $H^\pm$ is significantly lighter than the two heaviest states (keeping in mind that the split spectrum must be consistent with constraints due to the $T$ parameter).   In this case, one can focus on the phenomenology of a single non-SM-like scalar.  Scenarios A and B already provide examples of this type, in which $A$ and $H^\pm$ are significantly heavier than $h$ and $H$. But other scenarios could be considered that would feature the production and decay of $A$ or $H^\pm$ alone.  In these latter scenarios, it would be appropriate to slightly modify the hybrid basis, which specifies two scalar masses as input parameters.  When focusing on either $A$ or $H^\pm$, it would be more appropriate to specify
the SM-like Higgs mass and one other scalar mass [$m_A$ or $m_{H^\pm}$] as the two input scalar masses.  Of course, one could then use \eqs{ma2}{mhpm2} to determine the masses of the remaining non-SM-like Higgs scalars.


\section{Conclusions}
We have introduced a new ``hybrid'' basis to define input parameters in the general, $\cp$-conserving, two-Higgs-doublet Model with a softly broken $Z_2$ symmetry in a way that naturally accommodates  constraints from perturbativity and unitarity. In this basis, the input parameters are the two $\cp$-even Higgs masses, $\mh$ and $\mH$, three quartic couplings defined in the Higgs basis, $Z_4$, $Z_5$, and $Z_7$, and the mixing angles $\cba$ and $\tan\beta$ which are very relevant for the phenomenology. In this hybrid approach, $\tan\beta$ can be interpreted as specifying the basis where the soft $Z_2$-breaking is manifest.

The hybrid basis of parameters has been implemented as an input option for the public computer code 2HDMC. Using this setup, we have performed a scan and a detailed numerical analysis of the parameter space constraints and predictions of cross sections and branching ratios for different 2HDM configurations. Using the results of this analysis, we have defined a set of 2HDM benchmark scenarios that we deem relevant for the design and interpretation of Higgs searches beyond the SM at LHC run-II and beyond. The scenarios are defined to cover different aspects of the 2HDM phenomenology, including interpretation of the 125 GeV Higgs signal as either the light or the heavy $\cp$-even Higgs boson. We also devise a new scenario with overlapping $\cp$-even and $\cp$-odd Higgses at 125 GeV, which could potentially be very useful for the interpretation of $\cp$-studies in the $\tau\tau$ final state, as well as a scenario where the Yukawa coupling of the Higgs to down-type fermions has the ``flipped'' sign (relative to the coupling to vector bosons). Another two sets of scenarios provide Higgs cascade decays as a prominent feature, giving rise to completely non-standard final states that may contain additional leptons and/or jets. Finally, we define the 2HDM equivalent of the decoupled MSSM, where the Higgs spectrum follows MSSM relations with the dominant radiative corrections to the Higgs masses completely captured in a shift of the quartic coupling $\lambda_2$ appearing in the Higgs potential. Results presented for this scenario should be immediately familiar to anyone following MSSM Higgs searches at the LHC, which could make it useful for comparison and communication of 
results.

We hope that the presented scenarios will be useful and serve to inspire new discoveries!


\section*{Acknowledgments}
We are grateful to the encouragement and interest in this work shown by the experimental LHC community.  We would also like to thank the participants in the LHC Higgs cross section working group, and in particular the members of WG3/Extended scalars, for many interesting discussions. 
We are especially thankful for the hospitality and support of Sven Heinemeyer during the Higgs Days meeting in Santander, Spain where this work was initiated, and to Georg Weiglein who made a number of useful suggestions that led to the development of Scenarios C, D and E.
In addition, H.E.H. gratefully acknowledges the hospitality of the Theory Group at CERN, where this work was completed.  H.E.H.~is supported in part by U.S. Department of
Energy grant number DE-FG02-04ER41286, and
O.S.\ is supported by the Swedish Research Council (VR) through the Oskar Klein Centre.

\newpage
\section*{A.~Parameter values for 2HDM benchmark lines and planes}
For convenience, this appendix contains a list of the input parameters that can be used to realize the 2HDM benchmark scenarios described in this paper.

\vspace{0.75em}

\centering
\begin{tabular}{ccccccccc}
\hline
\multicolumn{9}{c}{\bf Scenario A (Non-alignment)}\\
  & $\mh$ (GeV) & $\mH$ (GeV) & $\cbma$ & $Z_4$ & $Z_5$ & $Z_7$ & $\tan\beta$ & Type \\
\hline
A1.1 & $125$ & $150\ldots 600$ & $0.1$ & $-2$ & $-2$ & $0$ & $1\ldots 50$ & I\\
A1.2 & $125$ & $150\ldots 600$ & $0.1\times \left(\frac{150\GeV}{\mH}\right)^2$ & $-2$ & $-2$ & $0$ & $1\ldots 50$ & I\\
A2.1 & $125$ & $150\ldots 600$ & $0.01$ & $-2$ & $-2$ & $0$ & $1\ldots 50$ & II\\
A2.2 & $125$ & $150\ldots 600$ & $0.01\times \left(\frac{150\GeV}{\mH}\right)^2$ & $-2$ & $-2$ & $0$ & $1\ldots 50$ & II\\
\hline
\end{tabular}
\centering

\vspace{0.75em}
\begin{tabular}{ccccccccc}
\hline
\multicolumn{9}{c}{\bf Scenario B (Low-$\mH$)}\\
 & $\mh$ (GeV) & $\mH$ (GeV) & $\cbma$ & $Z_4$ & $Z_5$ & $Z_7$ & $\tan\beta$ & Type \\
\hline
B1.1 & $65\ldots 120$ & $125$ & $1.0$ & $-5$ & $-5$ & $0$ & $1.5$ & I\\
B1.2 & $80\ldots 120$ & $125$ & $0.9$ & $-5$ & $-5$ & $0$ & $1.5$ & I\\
B2 & $65 \ldots 120$ & $125$ & $1.0$ & $-5$ & $-5$ & $0$ & $1.5$ & II\\
\hline
\end{tabular}

\vspace{0.75em}
\centering
\begin{tabular}{ccccccccc}
\hline
\multicolumn{9}{c}{\bf Scenario C (\cp-overlap)}\\
 & $\mh$ & $\mH$ & $\mA$ & $\mHp$ & $\cbma$ &  $\lambda_5$ & $\tan\beta$ & Type \\
\hline
C1 & $125$ & $300$ & $125$ & $300$ & $0$ & $0$   & $1\ldots 10$ & I\\
C2 & $125$ & $300$ & $125$ & $300$ & $0$ & $0$   & $1\ldots 10$ & II\\
\hline
\end{tabular}

\vspace{0.75em}
\begin{tabular}{ccccccccccc}
\hline
\multicolumn{9}{c}{\bf Scenario D (Short cascade)}\\
  & $\mh$ (GeV) & $\mH$ (GeV) & $\cbma$ & $Z_4$ & $Z_5$ & $Z_7$ & $\tan\beta$ & Type \\
\hline
D1.1 & $125$ & $250\ldots 500$ & $0$ & $-1$ & $1$ & $-1$ & $2$ & I\\
D1.2 & $125$ & $250\ldots 500$ & $0$ & $2$ & $0$ & $-1$ & $2$ & I\\
D1.3 & $125$ & $250\ldots 500$ & $0$ & $1$ & $1$ & $-1$ & $2$ & I\\
D2.1 & $125$ & $250\ldots 500$ & $0$ & $-1$ & $1$ & $-1$ & $2$ & II\\
D2.2 & $125$ & $250\ldots 500$ & $0$ & $2$ & $0$ & $-1$ & $2$ & II\\
D2.3 & $125$ & $250\ldots 500$ & $0$ & $1$ & $1$ & $-1$ & $2$ & II\\
\hline
\end{tabular}

\vspace{0.75em}
\begin{tabular}{ccccccccccc}
\hline
\multicolumn{9}{c}{\bf Scenario E (Long cascade)}\\
  & $\mh$ (GeV) & $\mH$ (GeV) & $\cbma$ & $Z_4$ & $Z_5$ & $Z_7$ & $\tan\beta$ & Type \\
\hline
E1.1 & $125$ & $200\ldots 300$ & $0$ & $-6$ & $-2$ & $0$ & $2$ & I\\
E1.2 & $125$ & $200\ldots 300$ & $0$ & $1$ & $-3$ & $0$ & $2$ & I\\
E2.1 & $125$ & $200\ldots 300$ & $0$ & $-6$ & $-2$ & $0$ & $2$ & II\\
E2.2 & $125$ & $200\ldots 300$ & $0$ & $1$ & $-3$ & $0$ & $2$ & II\\
\hline
\end{tabular}

\vspace{0.75em}
\begin{tabular}{ccccccccccc}
\hline
\multicolumn{9}{c}{\bf Scenario F (Flipped Yukawa)}\\
  & $\mh$ (GeV) & $\mH$ (GeV) & $\cbma$ & $Z_4$ & $Z_5$ & $Z_7$ & $\tan\beta$ & Type \\
\hline
F2 & $125$ & $150\ldots 600$ & $\sin 2\beta$ & $-2$ & $-2$ & $0$ & $5\ldots 50$ & II\\
\hline
\end{tabular}

\vspace{0.75em}
\centering
\begin{tabular}{ccccccccc}
\hline
\multicolumn{5}{c}{\bf Scenario G (MSSM-like)}\\
  & $\mh$ (GeV) & $\mA$ (GeV) &  $\tan\beta$ & Type \\

\hline
G2 & $125$ & $90\ldots 1000$ & $1\ldots 60$ & II\\
\multicolumn{5}{c}{$\lambda_{i}$ ($i\neq 2$) given by their MSSM values [cf.~Eq.~\eqref{eq:MSSM}]; $\lambda_2$ determined by $m_h$}\\

\hline
\end{tabular}

\clearpage
\newpage
\bibliography{2hdm_lhc_benchmarks}
\bibliographystyle{JHEP}

\end{document}